\documentclass[aps,pra,10pt,amsmath,onecolumn,notitlepage,tightenlines,superscriptaddress,floatfix,eqsecnum,showpacs,longbibliography]{revtex4-1}

\usepackage{soul}
\usepackage[caption=false]{subfig}
\usepackage{amssymb, amsmath}
\usepackage{bm}
\usepackage{graphicx}
\usepackage{upgreek}
\usepackage{bbold}% for blackboard bold Arabic numerals

\usepackage{xcolor}
\definecolor{dark-red}{rgb}{0.4,0.15,0.15}
\definecolor{dark-blue}{rgb}{0.15,0.15,0.4}
\definecolor{blue-green}{RGB}{3,80,50}
\definecolor{green-yellow}{RGB}{100,160,18}
\definecolor{dark-yellow}{RGB}{140,114,6}
\definecolor{medium-blue}{rgb}{0,0,0.5}
\usepackage{hyperref}
\usepackage[all]{hypcap}%fix problems with figures and tables of the 'hyperref' package
\usepackage{cleveref}% Package hperref works wrongly within the 'subequations' environment without this.
\hypersetup{
    colorlinks, linkcolor={dark-red},
    citecolor={dark-blue}, urlcolor={medium-blue}
} %These setups replace the ugly and distracting red and green boxes of the 'hyperref' package with colored texts.

\newcommand{\ssp}{\hspace{0.4pt}}%small horizontal space
\newcommand{\nsp}{\hspace{-0.7pt}}%negative horizontal space
\newcommand{\norm}[1]{\lvert #1 \rvert}

\newcommand{\ket}[1]{\lvert\, #1\, \rangle}
\newcommand{\bra}[1]{\langle\, #1\, \rvert}
\newcommand{\braket}[2]{\langle\, #1\,\vert\, #2 \,\rangle}

\newcommand{\ketb}[1]{\big\lvert\, #1\, \big\rangle}
\newcommand{\brab}[1]{\big\langle\, #1\, \big\rvert}
\newcommand{\projb}[1]{\ketb{#1}\brab{#1}}
\newcommand{\dif}{d\ssp} %differential symbol
\newcommand{\vac}{\mathrm{vac}} %vacuum state
\newcommand{\half}{\frac{1}{2}}
\newcommand{\identity}{\mbox{\textbb{1}}}
\newcommand{\pa}[2]{\frac{\partial #1}{\partial #2}}
\newcommand{\Fthree}{F\big(\frac{3}{2},-n;\frac{1}{2}-n;z\big)}
\newcommand{\Fone}{F\big(\frac{1}{2},-n;\frac{1}{2}-n;z\big)}
\newcommand{\Fthreeoverone}{\frac{\Fthree}{\Fone}}

\DeclareMathOperator{\diag}{diag}
\DeclareMathOperator{\tr}{tr}

\renewcommand{\a}{a}
\renewcommand{\b}{b}
\renewcommand{\c}{c}

\newcommand{\rank}{\nu}%{{\mbox{\hspace{-0.4pt}\small\it s}}}
\newcommand{\notrank}{\mu}

\newcommand{\NFB}{\mathsf{N}} %normalization factor of TPS for bosons
\newcommand{\BesselI}{I}
\newcommand{\idu}{u}
\newcommand{\idv}{v}

\newcommand{\idG}{G}

\newcommand{\sA}{\mathcal{A}}
\newcommand{\sB}{\mathcal{B}}

\newcommand{\sF}{\mathcal{F}}

\newcommand{\sH}{\mathcal{H}}

\newcommand{\sL}{\mathcal{L}}
\newcommand{\sM}{\mathcal{M}}

\newcommand{\sP}{\mathcal{P}}

\newcommand{\sS}{\mathcal{S}}

\newcommand{\sU}{\mathcal{U}}
\newcommand{\sV}{\mathcal{V}}

\newcommand{\xbf}{\mathbf{x}}
\newcommand{\ybf}{\mathbf{y}}

\newcommand{\PCS}{\mbox{PCS}}
\newcommand{\bigO}{O}
\newcommand{\pcss}{s}
\newcommand{\exrho}{\sigma}
\newcommand{\normalized}{}

\begin{document}

\title{\bf Bosonic Particle-Correlated States: A Nonperturbative Treatment Beyond Mean Field}

\date{\today}

\author{Zhang Jiang}
\email{zhang.jiang@nasa.gov}
\affiliation{Center for Quantum Information and Control, University of New Mexico, MSC07-4220, Albuquerque, New Mexico 87131-0001, USA}
\affiliation{NASA Ames Research Center Quantum Artificial Intelligence Laboratory (QuAIL),  Moffett Field, CA 94035, USA}
\affiliation{Stinger Ghaffarian Technologies Inc., 7701 Greenbelt Road, Suite~400, Greenbelt, MD 20770, USA}

\author{Alexandre B. Tacla}
%\email{alexandre.tacla@strath.ac.uk}
\affiliation{Center for Quantum Information and Control, University of New Mexico, MSC07-4220, Albuquerque, New Mexico 87131-0001, USA}
\affiliation{Department of Physics and SUPA, University of Strathclyde, Glasgow G4 0NG, UK}

\author{Carlton M. Caves}
%\email{ccaves@unm.edu}
\affiliation{Center for Quantum Information and Control, University of New Mexico, MSC07-4220, Albuquerque, New Mexico 87131-0001, USA}
\affiliation{Centre for Engineered Quantum Systems, School of Mathematics and Physics,
University of Queensland, Brisbane, Queensland 4072, Australia}

\begin{abstract}
Many useful properties of dilute Bose gases at ultra-low temperature are predicted precisely by the (mean-field) product-state \emph{Ansatz\/}, in which all particles are in the same quantum state.  Yet, in situations where particle-particle correlations become important, the product \emph{Ansatz\/} fails.  To include correlations nonperturbatively, we consider a new set of states: the particle-correlated state of $N=l\times n$ bosons is derived by symmetrizing the $n$-fold product of an $l$-particle quantum state.  Quantum correlations of the $l$-particle state ``spread out'' to any subset of the $N$ bosons by symmetrization.  The particle-correlated states can be simulated efficiently for large $N$, because their parameter spaces, which depend on $l$, do not grow with $n$.  Here we formulate and develop in great detail the pure-state case for $l=2$, where the many-body state is constructed from a two-particle pure state.  These paired wave functions, which we call pair-correlated states (PCS), were introduced by A. J. Leggett [\href{https://doi.org/10.1103/RevModPhys.73.307}{Rev.\ Mod.\ Phys.\ {\bf 73}, 307 (2001)}] as a particle-number-conserving version of the Bogoliubov approximation.  We present an iterative algorithm that solves for the reduced (marginal) density matrices (RDMs), i.e., the correlation functions, associated with \PCS\ in time $\bigO(N)$.  The RDMs can also be derived from the normalization factor of \PCS, which is derived analytically in the large-$N$ limit.  To test the efficacy of \PCS, we analyze the ground state of the two-site Bose-Hubbard model by minimizing the energy of the \PCS~state, both in its exact form and in its large-$N$ approximate form, and comparing with the exact ground state.  For $N=1\,000$, the relative errors of the ground-state energy for both cases are within $10^{-5}$ over the entire parameter region from a single condensate to a Mott insulator.  We present numerical results that suggest that \PCS\ might be useful for describing the dynamics in the strongly interacting regime.
\end{abstract}

% Measurement theory (quantum mechanics), 03.65.Ta
% Quantum information, 03.67.-a
% Metrology, 06.20.-f
%% Interferometry, nonclassical, 42.50.St
%% Error theory, 06.20.Dk

\maketitle

\section{Introduction}
\label{sec:intro}

The mean-field Gross-Pitaevskii (GP) theory~\cite{gross_structure_1961, pitaevsk_vortex_1961} successfully predicts many useful properties of weakly interacting Bose gases at ultra-low temperatures, yet a number of interesting many-particle phenomena, such as quantum phase transitions from superfluids to Mott insulators~\cite{jaksch_cold_1998, greiner_quantum_2002}, cannot be explained by the mean-field GP approach.  This is because particle-particle correlations are neglected when one approximates the state of the system by the GP \emph{Ansatz\/}, i.e., by a product of identical single-particle states at all times.  The Bogoliubov approximation~\cite{bogoliubov_theory_1947, fetter_nonuniform_1972, huang_statistical_1987} is an attempt to include particle correlations by perturbing the GP mean-field solution with collective excitations.  Although useful for analyzing the stability and validity of the GP solution, the Bogoliubov approximation is only valid when depletion of the condensate mode is quite small, i.e., when the single-particle reduced density matrix has only one dominant eigenvalue~\cite{penrose_bose-einstein_1956}.

To tackle the case where the condensate is fragmented into two parts, many authors have adopted the double-Fock state (sometimes also called a twin-Fock state); more generally, for a condensate fragmented into many parts, one can use the many-Fock state,
\begin{align}
 \ketb{\varPsi_\mathrm{Fock}} = \prod_{j=1}^\rank \frac{1}{\sqrt{N_j!}}\big(\a^\dagger_j\big)^{N_j}\ketb{\vac}\;,
\end{align}
where $\rank$ is the number of fragments, $\ket{\vac}$ is the vacuum state, and $a_j^\dagger$ creates a particle in the single-particle state for the $j$th fragment.  This approach thus uses a product-state \emph{Ansatz\/} for each of the fragments.  Using this \emph{Ansatz\/}, Streltsov \emph{et al.}~\cite{streltsov_ground-state_2004} argued that fragmentation of the ground state of a condensate only happens when the total number of particles is finite; Mueller \emph{et al.}~\cite{mueller_fragmentation_2006} showed that as degeneracies multiply, so do the varieties of fragmentation; and Alon \emph{et al.}~\cite{alon_time-dependent_2007} generalized the GP equation to include the several single-particle wave functions for the various fragments.  In contrast, fragmented condensates are also treated by evolving the single-particle reduced density matrix in an approximation that includes the back-reaction from Bogoliubov excitations~\cite{vardi_bose-einstein_2001, tikhonenkov_quantum_2007}.

A powerful idea for including particle correlations in many-particle bosonic or fermionic wave functions is to construct them from two-particle states.  For instance, the Jastrow wave function of $N$ particles is a product of two-particle states of all $N(N-1)/2$ pairs,
\begin{align}
 \varPsi_\mathrm{Jast}(\xbf_1,%\xbf_2,
 \ldots,\xbf_N) \sim \prod_{\substack{j,k=1\\j< k}}^N f(\xbf_j-\xbf_k)\;,
\end{align}
with bosons (fermions) corresponding to $f$ having even (odd) parity.  Many famous wave functions are of Jastrow type, e.g., the Laughlin wave function~\cite{laughlin_anomalous_1983} and the Gutzwiller wave function~\cite{gutzwiller_effect_1963, rokhsar_gutzwiller_1991}.  The Jastrow wave function has found wide application in strongly interacting systems: It is used in variational quantum Monte Carlo as a trial wave function~\cite{umrigar_optimized_1988}; it is used to show that the single-particle reduced (marginal) density matrix of $^4\mathrm{He}$ is an extensive quantity, thus demonstrating that Bose-Einstein condensation (BEC) underlies superfluidity~\cite{reatto_bose-einstein_1969}; and it is used to investigate the effect of the interatomic correlations and the accuracy of the GP~equation~\cite{fabrocini_beyond_1999, dubois_bose-einstein_2001, cowell_cold_2002}.  The Jastrow wave function, however, suffers from increasing demand for computational power for large numbers of bosons and from the requirement of using quantum Monte Carlo.

Here we propose a new set of states, which constitute a natural generalization of the GP product-state \emph{Ansatz\/},
\begin{align}
 \rho\ssp\big(\xbf^{(l)}_1,%\xbf^{(l)}_2,
 \ldots,\xbf^{(l)}_n\, ;\, \ybf^{(l)}_1,%\ybf^{(l)}_2,
 \ldots,\ybf^{(l)}_n \big)
 = \frac{ \sP_S\,  \sigma (\xbf^{(l)}_1, \,\ybf^{(l)}_1)%\otimes \sigma(\xbf^{(l)}_2, \,\ybf^{(l)}_2)
 \otimes \cdots \otimes \sigma(\xbf^{(l)}_n, \,\ybf^{(l)}_n )\, \sP_S}{\tr\!\Big(\sP_S\,  \sigma(\xbf^{(l)}_1, \,\ybf^{(l)}_1)%\otimes \sigma(\xbf^{(l)}_2, \,\ybf^{(l)}_2)
 \otimes \cdots \otimes \sigma(\xbf^{(l)}_n, \,\ybf^{(l)}_n)\, \sP_S\Big)}\;.\label{eq:pcs}
\end{align}
Here $\xbf^{(l)}_{j} = \big(\xbf_{j,1},%\,\xbf_{j,2},
\ldots, \xbf_{j,l}\big)$ and $\ybf^{(l)}_{j} = \big(\ybf_{j,1},%\,\ybf_{j,2},
\ldots, \ybf_{j,l}\big)$ denote the co\"ordinates of blocks of $l$ particles, $\sigma$ is an arbitrary state (density matrix) of $l$ particles, and $\sP_S$ is the projection operator onto the symmetric subspace of all the $N=l\times n$ particles,
\begin{align}
 \sP_S\, \ketb{\psi_1,%\psi_2,
 \ldots,\psi_N} = \frac{1}{N!}\sum_{\pi\in S_N}\, \ketb{\psi_{\pi(1)},%\psi_{\pi(2)},
 \ldots,\psi_{\pi(N)}}\;,
\end{align}
where the sum is over the permutations $\pi$ in the symmetric group $S_N$.  The state~(\ref{eq:pcs}) is derived by symmetrizing the $n$-fold tensor product of the $l$-particle state $\sigma$; we call the resulting state a \emph{bosonic particle-correlated state\/}~(BPCS).  The $l$-particle states $\sigma$ can be restricted to symmetrized states, or they can be left arbitrary, with $\sP_S$ taking care of the symmetrization when the BPCS state is constructed.  Note that we can extend~(\ref{eq:pcs}) to the case of fermions simply by substituting the anti-symmetrizing operator $\sP_A$ for the symmetrizing operator $\sP_S$.  The resulting state can be called a \emph{fermionic particle-correlated state\/} (FPCS).  As a consequence of symmetrization, the quantum correlations existing in the $l$-boson state $\sigma$ ``spread out'' to any subset of the $N=l\times n$ bosons.  Moreover, the parameter space of the BPCS does not grow with $n$; it remains the same as that of the bosonic states for $l$ particles.

This article is devoted to the case that $\sigma$ is pure and $l=2$, which we refer to simply as \PCS, where now \PCS\ can be read as \emph{pair-correlated state}.  Despite being constructed from a two-particle wave function, PCS is different from Jastrow's wave function.

This article is organized as follows.  In Sec.~\ref{sec:pcs}, we express PCS, i.e., the case that $\sigma$ is pure and $l=2$, in second-quantized form, and we discuss its introduction by Leggett~\cite{leggett_bose-einstein_2001} to implement a particle-number-conserving version of the Bogoliubov approximation.  In Sec.~\ref{sec:pcs_and_the_bogoliubov_approximation}, we prove that PCS indeed reproduces the number-conserving Bogoliubov approximation as a special case and, hence, encompasses weakly interacting Bose gases with small condensate depletion.  In Sec.~\ref{sec:normalization_factor}, we show how to calculate the diagonal elements of reduced density matrices (RDMs) exactly from the PCS normalization factor.  In Sec.~\ref{sec:large_N}, we derive approximations for the RDMs that are valid in the case of a large number of bosons; we discuss exactly solvable examples of these approximations in Sec.~\ref{sec:examples}.  In Sec.~\ref{sec:allrdms}, we show that the off-diagonal elements of the RDMs can be determined exactly from the diagonal elements.  In particular, we show in Sec.~\ref{sec:2rdms} how to calculate exactly all the matrix elements of the two-particle RDM (2RDM), we give the large-$N$ limit of the 2RDM for two Schmidt orbitals in terms of modified Bessel functions in Sec.~\ref{sec:2RDMexamples}, and in Sec.~\ref{sec:higher_rdms}, we find the off-diagonal elements of all RDMs in the large-$N$ limit.  The analytical calculation of few-particle RDMs of PCS allows the direct evaluation of physical observables in various regimes.  An appealing feature of PCS is that they can represent quantum states with or without off-diagonal long-range order~\cite{yang_concept_1962}.  For example, both the superfluid and the Mott-insulating phases in the two-site Bose-Hubbard model can be described by the PCS \emph{Ansatz\/}.  An interesting question is whether the PCS \emph{Ansatz\/} can faithfully interpolate between the two phases.  We show in Sec.~\ref{sec:two_site_bose_hubbard_model} that the answer is yes: The PCS description, both in its exact form and in its large-$N$ approximate form, provides a remarkably good account of the ground state of the two-site Bose-Hubbard model across the entire parameter region from superfluid to Mott insulator.  Moreover, we present numerical results that suggest that PCS might be useful for describing the dynamics in the strongly interacting regime.  We conclude in Sec.~\ref{sec:conclusions}.

This paper is mainly devoted to developing the mathematical formalism for manipulating and using PCS and, in particular, to calculating the reduced (marginal) density matrices (RDMs)---these are the correlation functions---associated with PCS and to investigating the PCS ground state of the two-site Bose-Hubbard model.  The results we present at the end of Sec.~\ref{sec:two_site_bose_hubbard_model} on the dynamics of PCS for the two-site Bose-Hubbard model assume, as is always true, that the single-particle spatial wave functions (orbitals) are held fixed.  A subsequent article will explore the power of \PCS\ dynamics when the spatial mode functions are included in the dynamics~\cite{pcs2}.  This investigation will be based on deriving time-dependent equations for the \PCS~\emph{Ansatz\/} by evolving a state (initially in \PCS\ form) for an infinitesimal time and then projecting it back to the \PCS\ manifold.  In subsequent papers, we also plan to discuss the \PCS\ ground state for more general problems, e.g., the Bose-Hubbard model for more than two sites and fragmented spin-orbital coupled BECs in a trapping potential.

The material in this paper is taken mainly from Jiang's doctoral dissertation at the University of New Mexico~\cite{JiangPhD}.

\section{Pair-Correlated States}
\label{sec:pcs}

\subsection{Second quantization of pair-correlated states}
\label{sec:second_quantization}

It is difficult to do any calculation with the form~(\ref{eq:pcs}), because of the need for an explicit symmetrization.  This motivates going to a second-quantized picture, where the symmetrization is taken care of automatically.  For a pure PCS with $l=2$, the PCS is specified by a two-boson wave function $\varPsi^{(2)}(\xbf_1,\, \xbf_2)$, and the PCS wave function is given by
\begin{align}\label{eq:pcs_rank2_pure_first_quantization}
\varPsi_\mathrm{pcs}&(\xbf_1,%\xbf_2,
\ldots,\xbf_{2n})\,\propto\,
\sP_S\Bigl(\varPsi^{(2)}(\xbf_1,\, \xbf_2)\,\varPsi^{(2)}(\xbf_3,\, \xbf_4)\cdots \varPsi^{(2)}(\xbf_{2n-1},\,\xbf_{2n})\Bigr)\;.
\end{align}
Such a PCS can be regarded as constructed by a mapping of the two-boson Hilbert space into a submanifold of the Hilbert space of $2n$ bosons.  Note that the two-boson wave function always has a Schmidt decomposition of the form~\cite{paskauskas_quantum_2001}
\begin{equation}\label{eq:schmidt_first_quantization}
\varPsi^{(2)}(\xbf_1,\, \xbf_2)=\sum_{j=1}^\rank \normalized\lambda_j\, \psi_j(\xbf_1)\psi_j(\xbf_2)\;,
\end{equation}
where the single-particle wave functions $\{\psi_j(\xbf)\,\vert\,j=1,2,\ldots, \rank\}$ form orthonormal Schmidt bases for the two particles, $\rank$ is the Schmidt rank, and the (real and positive) $\normalized\lambda_j$s are the Schmidt coefficients.  They satisfy the normalization condition
\begin{align}\label{eq:lambda_normalization}
\ssp\sum_{j=1}^\rank \normalized\lambda_j^2=1\;.
\end{align}
The coincidence of the Schmidt bases of the two bosons is a consequence of the symmetry of the wave function.  Throughout this paper we order the $\normalized\lambda_j$s as
\begin{equation}\label{eq:lambda_convention}
\lambda_1\geq \lambda_2\geq \cdots \geq \lambda_\rank\;.
\end{equation}
Note also that we use the terms single-particle states, modes, and orbitals interchangeably throughout the paper, both for the single-particle states and for their wave functions.

To get into the second-quantized picture, we find it convenient to introduce a pair creation operator,
\begin{align}\label{eq:pair_op}
\sA^\dagger\equiv
\int\varPsi^{(2)}(\xbf_1,\,\xbf_2)\uppsi^\dagger(\xbf_1)\uppsi^\dagger(\xbf_2)\,d\xbf_1\,d\xbf_2
=\sum_{j=1}^\rank \normalized\lambda_j \ssp \big(\a^{\dagger}_j\big)^2\;.
\end{align}
Here $\uppsi^\dagger(\xbf)$ is the field operator that creates a particle at $\xbf$, and
\begin{equation}
\a_j^\dagger=\int\psi_j(\xbf)\,\uppsi^\dagger(\xbf)\,d\xbf
\end{equation}
is the operator that creates a particle in the single-particle state with wave function $\psi_j(\xbf)$.  The two-particle state,
\begin{equation}\label{eq:schmidt_second_quantization}
 \ket{\varPsi^{(2)}}
 =\frac{1}{\sqrt2}\,\sA^\dagger\,\ket{\vac}
 =\frac{1}{\sqrt 2}\,\sum_{j=1}^{\rank} \normalized\lambda_j \ssp \big(\a^{\dagger}_j\big)^2 \,\ket{\vac}\;,
\end{equation}
has the wave function $\braket{\xbf_1,\xbf_2}{\varPsi^{(2)}}=\varPsi^{(2)}(\xbf_1,\, \xbf_2)$ of Eq.~(\ref{eq:schmidt_first_quantization}).

It is instructive to derive the Schmidt decomposition directly in the second-quantized picture, where an arbitrary two-boson state is given by
\begin{equation}\label{eq:Psi2_notschmidt}
 \ket{\varPsi^{(2)}}=\frac{1}{\sqrt 2}\,\sum_{j,k=1}^{\notrank} \normalized\Lambda_{jk}\, \b^\dagger_k \b^\dagger_j\,\ket{\vac}\,.
\end{equation}
Here $\b^\dagger_j$ creates a particle in the $j$th single-particle state, and $\normalized\Lambda_{jk}=\normalized\Lambda_{kj}$ is a symmetric matrix [if $\normalized\Lambda$ is not symmetric, we can always make it so by redefining $\Lambda\rightarrow (\Lambda+\Lambda^T)/2$, without changing $\ket{\varPsi^{(2)}}$], satisfying $\mbox{tr}(\Lambda\Lambda^\dagger)=1$ to make $\ketb{\varPsi^{(2)}}$ normalized.  The Autonne-Takagi factorization theorem (see Corollary 4.4.4(c) of Horn and Johnson~\cite{horn_matrix_2013}) says that any complex symmetric matrix $\normalized\Lambda$ can be diagonalized by a unitary matrix $U$:
\begin{equation}\label{eq:Lambda_diagonalized}
 U \normalized\Lambda\, U^T = \diag\big(\normalized\lambda_1, %\normalized\lambda_2,
 \ldots, \normalized\lambda_\notrank\big) \;.
\end{equation}
The $\lambda_j$s are real and nonnegative and normalized according to Eq.~(\ref{eq:lambda_normalization});
they are the singular values of $\Lambda$, and the diagonalization~(\ref{eq:Lambda_diagonalized}) is a special case of the singular-value decomposition, specialized to symmetric matrices.  By introducing a new set of creation operators, $\a_j^\dagger=\sum_{k=1}^\rank U_{jk}^*\, \b_k^\dagger$, we bring the state~(\ref{eq:Psi2_notschmidt}) into Schmidt form:
\begin{equation}
 \ket{\varPsi^{(2)}}=\frac{1}{\sqrt 2}\,\sum_{j=1}^{\rank} \normalized\lambda_j \ssp \big(\a^{\dagger}_j\big)^2 \,\ket{\vac}\;.
\end{equation}
The number of nonzero singular values is the rank $\rank\le\notrank$ of $\Lambda$.  We only need the nonzero singular values in Eq.~(\ref{eq:schmidt_second_quantization}), and the sum is written so as to recognize this.

We are now prepared to give the second-quantized version of the \PCS.  We begin with
\begin{align}
\begin{split}
\Psi^\dagger(\xbf_1)\cdots\Psi^\dagger(\xbf_N)\,\ketb{\vac}
&=\sqrt{N!}\;\sP_S\ketb{\xbf_1,\ldots,\xbf_N}=\frac{1}{\sqrt{N!}}
\sum_{\pi\in S_N}\ketb{\xbf_{\pi(1)},\ldots,\xbf_{\pi(N)}}\;,
\end{split}
\end{align}
where the sum is over the permutations $\pi$ in the symmetric group $S_N$.  This gives immediately that
\begin{align}\label{eq:Anvac}
\big(\sA^\dagger\big)^n\,\ketb{\vac}
&=\sqrt{N!}\int\ketb{\xbf_1,\ldots,\xbf_{2n}}\,
\sP_S\Bigl(\varPsi^{(2)}(\xbf_1,\, \xbf_2) \cdots \varPsi^{(2)}(\xbf_{2n-1},\, \xbf_{2n})\Bigr) \,d\xbf_1\cdots d\xbf_{2n}\;,
\end{align}
where
\begin{align}
\sP_S\Bigl(\varPsi^{(2)}(\xbf_1,\, \xbf_2)\cdots \varPsi^{(2)}(\xbf_{2n-1},\, \xbf_{2n})\Bigr)
&=\frac{1}{N!}\sum_{\pi\in S_N}
\varPsi^{(2)}\big(\xbf_{\pi(1)},\,\xbf_{\pi(2)}\big) \cdots \varPsi^{(2)}\big(\xbf_{\pi(2n-1)},\, \xbf_{\pi(2n)}\big)\;.
\end{align}
Equation~(\ref{eq:Anvac}) is the relation between the first- and second-quantized pictures; it can be written in the equivalent form
\begin{equation}\label{eq:12rel}
 \hspace{-2em}\frac{1}{\sqrt{N!}}\,\brab{\xbf_1,\ldots,\xbf_{2n}}\big(\sA^\dagger\big)^n \ketb{\vac}=\sP_S\Bigl(\varPsi^{(2)}(\xbf_1,\, \xbf_2) \cdots \varPsi^{(2)}(\xbf_{2n-1},\, \xbf_{2n})\Bigr)\;,
\end{equation}
which states that one gets the PCS state~(\ref{eq:pcs_rank2_pure_first_quantization}) by applying the pair creation operator $n$ times to the vacuum.

The state in Eq.~(\ref{eq:12rel}) is not normalized.  The properly normalized PCS state for $N=2n$ particles is given by
\begin{align}\label{eq:pcs_rank2_pure_second_quantization}
\ket{\varPsi_\mathrm{pcs}}
\equiv \frac{1}{\sqrt{\NFB}}\, \big(\sA^\dagger\big)^n\,\ket{\vac}
=\frac{1}{\sqrt{\NFB}}\bigg(\sum_{j=1}^\rank \normalized\lambda_j \ssp \big(\a^{\dagger}_j\big)^{\!2}\bigg)^n\,\ket{\vac}\;,
\end{align}
where $\NFB$ is a normalization factor that plays an important role in our consideration of \PCS.  The PCS and this normalization factor are functions of $n$ (or $N$) and the $\lambda_j$s, but for brevity, we generally allow this functional dependence to remain implicit; when we want to be explicit, we denote the normalization factor as $\NFB_{\smash{\vec\lambda},n}$.  Henceforth, we abandon the normalization~(\ref{eq:lambda_normalization}) of the Schmidt coefficients, requiring only that the $\lambda_j$s be real and positive; we can do this because an overall scaling of the $\lambda_j$s is automatically absorbed into $\NFB$.

The second-quantized form~(\ref{eq:pcs_rank2_pure_second_quantization}) is convenient for calculations, but we can build some intuition by considering relative-state decompositions in the position basis of the first-quantized form~(\ref{eq:pcs_rank2_pure_first_quantization}) of the PCS.  The relative-state decomposition of the particle $\xbf_1$, relative to all the other particles, is
\begin{align}\label{eq:1q_1relativestate}
\varPsi_\mathrm{pcs}
\propto\,\sum_j\lambda_j\,\psi_j(\xbf_1)\,\sP_S\Bigl(\psi_j(\xbf_2)\,\varPsi^{(2)}(\xbf_3,\, \xbf_4)\cdots \varPsi^{(2)}(\xbf_{2n-1},\, \xbf_{2n})
\Bigr)\;.
\end{align}
This relative-state decomposition reveals a Schmidt decomposition, and the Schmidt basis of the particle $\xbf_1$ consists of all the single-particle wave functions $\psi_j(\xbf_1)$.  This means that the single-particle reduced density matrix (1RDM) is diagonal in the basis of the wave functions $\psi_j(\xbf_1)$.  The Schmidt coefficients of the decomposition (\ref{eq:1q_1relativestate}), i.e., the square roots of the eigenvalues of the 1RDM, are not, however, given by the $\lambda_j$s, because the norms of the relative states of the other particles are different for different values of~$j$.

More interestingly, we have the following relative-state decomposition for particles $\xbf_1$ and $\xbf_2$,
\begin{align}
\begin{split}
\varPsi_\mathrm{pcs}
&\,\propto\, \varPsi^{(2)}(\xbf_1,\, \xbf_2)\,\sP_S\Bigl(\varPsi^{(2)}(\xbf_3,\, \xbf_4)\cdots \varPsi^{(2)}(\xbf_{2n-1},\, \xbf_{2n})\Bigr)\\
&\qquad+(N-2)\sum_{j,k} \lambda_{\smash j}\lambda_{\smash k}\,\psi_j(\xbf_1)\psi_k(\xbf_2)\,\sP_S\Bigl(\psi_j(\xbf_3)\psi_k(\xbf_4)\,\varPsi^{(2)}(\xbf_5,\, \xbf_6)\cdots \varPsi^{(2)}(\xbf_{2n-1},\, \xbf_{2n})\Bigr) \;.
\end{split}
\end{align}
What this shows is that the two-particle reduced density matrix (2RDM) comes from two sorts of terms: In the first term, the two particles $\xbf_1$ and $\xbf_2$ can be perfectly correlated, whereas in the second, they are only partially correlated.  To determine the pairwise quantum correlations in the PCS, we need to find the 2RDM, and to do that, it turns out to be useful to investigate thoroughly the functional dependence of the normalization factor $\NFB$, which plays a role akin to a partition function.

Before getting to that in Sec.~\ref{sec:normalization_factor}, however, we detour into some historical remarks in Sec.~\ref{sec:pcs_history} and into showing how the Bogoliubov approximation arises from the PCS in Sec.~\ref{sec:pcs_and_the_bogoliubov_approximation}.

\subsection{Pairing wave functions}
\label{sec:pcs_history}

Inspired by the BCS wave function proposed by Bardeen, Cooper, and Schrieffer for superconductivity~\cite{bardeen_microscopic_1957, bardeen_theory_1957}, Valatin and Butler~\cite{valatin_collective_1958} introduced a similar pairing wave function for bosons,
\begin{align}\label{eq:valatin_butler_wave function}
\ketb{\varPsi_\mathrm{VB}}
=\frac{1}{\sqrt\sM}\, \exp \! \Big(\, \lambda_0\ssp \a_0^\dagger\ssp \a_0^\dagger
+ 2\sum_k \lambda_k\ssp \b^\dagger_k \b^\dagger_{-k}\Big)\,\ketb{\vac}\;,
\end{align}
where $\sM$ is a normalization factor. This state, with a quadratic form of the creation operators in the exponential, is very different from a coherent state.  A coherent state has a Poissonian number distribution, peaked around $N = \norm{\alpha}^2$, whereas the Valatin-Butler state~(\ref{eq:valatin_butler_wave function}) satisfies an exponential number distribution. Consequently, it might not be suited to situations where the total number of particles is conserved.  An easy way to see the exponential number distribution is by setting $\lambda_k = 0$ for all $k\neq 0$, which gives
\begin{align}
\begin{split}
 \ketb{\varPsi_\mathrm{VB}}
 &= \frac{1}{\sqrt\sM}\, e^{\lambda_0\ssp \a_0^\dagger\ssp \a_0^\dagger}\,\ketb{\vac}\\
%  &= \frac{1}{\sqrt \NFB}\, \sum_{n=0}^\infty \frac{\lambda_0^n}{n!}\, \big(\a_0^\dagger\big)^{2n}\,\ket{\vac}\\
 &= \frac{1}{\sqrt\sM}\, \sum_{n=0}^\infty \frac{\lambda_0^n}{n!}\,\sqrt{(2n)!}\;\ketb{2n}_0\otimes \ketb{\vac}_\perp\\%\label{eq:number_state_representation_c}
 &\simeq \frac{1}{\sqrt\sM}\,\sum_{n=0}^\infty
 \frac{\;(2 \lambda_0)^n}{\sqrt[4]{\pi n}}\:\ketb{2n}_0\otimes \ketb{\vac}_\perp\;,%\label{eq:number_state_representation_d}
\end{split}
\end{align}
where the subscript $\perp$ denotes the modes labeled by $k\ne0$, which are orthogonal to the zero mode, and we use Stirling's formula in the last step.  It turns out this exponential distribution is valid even when more than one $\lambda_k$ is nonzero.  Unlike the pairing wave function for fermions, which is always normalizable, the Valatin-Butler wave function~(\ref{eq:valatin_butler_wave function}) can only be normalized when $\norm{\lambda_k}< 1/2$ for all $k\in \{0,1,\ldots\}$. The Valatin-Butler wave function has been used to investigate the transition from a single condensate to a multimode condensate~\cite{coniglio_condensation_1967, evans_pairing_1969, nozieres_particle_1982}.

A number-conserving version of the Valatin-Butler wave function,
\begin{align}\label{eq:leggett_wave function}
\ketb{\varPsi_\mathrm{Legg}} \propto \Big(\,\a_0^\dagger\ssp \a_0^\dagger
+ 2\sum_k \lambda_k \b^\dagger_k \b^\dagger_{-k}\Big)^{\!N/2}\,\ketb{\vac}\;,
\end{align}
was introduced by Leggett~\cite{leggett_bose-einstein_2001, leggett_relation_2003}.  Leggett's state is the $n$th term, where $n=N/2$, in the expansion of the exponential in the Valatin-Butler state.  It is a special case of the BPCS with $\sigma$ pure and $l=2$; i.e., it is a PCS.  Indeed, by introducing modes with creation operators,
\begin{align}\label{eq:aksigma}
\a_{k0}^\dagger=\frac{1}{\sqrt2}(\b_k^\dagger+\b_{-k}^\dagger)\;,\qquad
\a_{k1}^\dagger=-\frac{i}{\sqrt2}(\b_k^\dagger-\b_{-k}^\dagger)\;,
\end{align}
the Leggett state takes on the standard PCS form, but with each pair of modes for $k\ge1$ sharing a Schmidt coefficient~$\lambda_k$.  It is worth mentioning that the ground state of a spin-1 Bose gas with an antiferromagnetic interaction takes a form similar to Eq.~(\ref{eq:leggett_wave function})~\cite{ho_fragmented_2000}; the fermion version of this state was used by Leggett \emph{et al.} to treat the BEC-BCS crossover problem~\cite{leggett_diatomic_1980,leggett_becbcs_2012} and by others to discuss the composite boson problem~\cite{combescot_n-exciton_2003, law_quantum_2005, combescot_many-body_2008, tichy_how_2014}.

Dziarmaga and Sacha~\cite{dziarmaga_bogoliubov_2003} generalized Leggett's wavefunction to the inhomogeneous case  (i.e., translational symmetry is broken) while retaining a similar pair-correlated form,
\begin{align}\label{eq:pcs_Sacha}
\ket{\varPsi_\mathrm{pcs}}
\propto \Big(\,\a_0^\dagger \a_0^\dagger
+\sum_{m>0}^{\rank-1} \lambda_m \a^{\dagger}_m\a^{\dagger}_m\Big)^{N/2}\,\ket{\vac}\;,
\end{align}
where $\a^\dagger_0$ and $\a^{\dagger}_m$ are the creation operators for the condensate mode and the modes orthogonal to the condensate wavefunction; the orthogonal modes and the corresponding real numbers $\lambda_m$ are derived from the Bogoliubov Hamiltonian by using a singular-value decomposition.  Later, Dziarmaga and Sacha generalized their results to include time-dependent evolutions~\cite{dziarmaga_images_2006}, where they showed that the state of the system retains the same structure if it starts in a Bogoliubov vacuum state.   Although derived from a different perspective, the wavefunction~(\ref{eq:pcs_Sacha}) is essentially identical to the \PCS\ introduced in Eq.~(\ref{eq:pcs_rank2_pure_second_quantization}).  We derive analytical expressions for the physical quantities (particularly, the 1RDM and 2RDM) associated with the \PCS\ in the large-$N$ limit when more than one mode is macroscopically occupied, i.e., $\lambda_m\sim1+O(1/N)$ for several modes, so the Bogoliubov approximation fails.  These results enable us to analyze a \PCS\ with constant computational cost (independent of $N$), whereas a na{\"\i}ve full numerical simulation of the same quantum state requires a computational resource of order $N^{2\rank-2}$.  Moreover, in App.~\ref{app:Iterative_Relations}, we present an iterative algorithm to calculate the RDMs of \PCS\ exactly with computational cost $O(N)$.

We discuss the single-particle, two-particle, and higher-order RDMs of the $N$-particle PCS in Secs.~\ref{sec:normalization_factor}--\ref{sec:allrdms} and present PCS results for the two-site Bose-Hubbard model in Sec.~\ref{sec:two_site_bose_hubbard_model}.  We plan to present a set of coupled equations to determine the Schmidt coefficients and orbitals for the \PCS\ ground state, as well as to formulate a generalized GP-type equation for the dynamics of the two-particle \PCS\ wave function, in a subsequent paper~\cite{pcs2}.

\section{PCS and the Bogoliubov Approximation} % (fold)
\label{sec:pcs_and_the_bogoliubov_approximation}

When the Schmidt rank of the two-particle state~(\ref{eq:schmidt_second_quantization}) is 1, the PCS~(\ref{eq:pcs_rank2_pure_second_quantization}) is a product of single-particle states and reduces to the product-state \emph{Ansatz\/} that is the basis of the mean-field GP equation.  In the case that the Schmidt rank is greater than 1, but there is a single dominant Schmidt coefficient, Dziarmaga and Sacha~\cite{dziarmaga_bogoliubov_2003} showed that the PCS reproduces the particle-number-conserving ($N$-conserving) Bogoliubov approximation~\cite{girardeau_theory_1959, gardiner_particle-number-conserving_1997, castin_low-temperature_1998}.  This result validates using PCS for BECs with small depletion, and it provides a different way to formulate the $N$-conserving Bogoliubov approximation.  We give an alternative demonstration of this equivalence in this section.

The Bogoliubov approximation fails when depletion from a single condensate mode is large and, hence, the picture of quasi-particle excitations on top of a condensate becomes invalid.  The \PCS\ with more than one dominant Schmidt coefficient (a multimode condensate) allows for correlations beyond the weak correlations present in the Bogoliubov approximation.  In Sec.~\ref{sec:two_site_bose_hubbard_model}, we discuss the superfluid-Mott insulator transition in the two-site Bose-Hubbard model using a large-$N$ expansion of the \PCS\ states.  In this expansion about a multiple-mode \PCS, we find that terms beyond Bogoliubov order must be retained to get an accurate description of the transition.

Consider a Bose-Einstein condensate of $N$ particles in a trapping potential.  We restrict our considerations here to the ground state of the condensate, but similar considerations apply to mean-field dynamics and accompanying Bogoliubov excitations.  The Gross-Pitaevskii (GP) ground state takes the form
\begin{align}\label{eq:gp_ground}
 \ket{\varPsi_\mathrm{gpgs}} &=  \ket{N}_0\otimes\ket{\vac}_\perp\;,
\end{align}
where the subscript $\perp$ denotes all the modes that are orthogonal to the condensate mode.  The essence of the $N$-conserving Bogoliubov approximation is to perturb about the state~(\ref{eq:gp_ground}) by introducing the operators \begin{equation}
\tilde\b_j^\dagger = \b_j^\dagger\a_0/\sqrt{N}
\quad\mbox{for $j=1,2,\ldots, \rank-1$,}
\end{equation}
where $\a_0$ is the annihilation operator of the condensate mode and $\b_j^\dagger$ is the creation operator of the $j$th orthogonal mode.  Even though the operators $\tilde\b_j^\dagger$ and $\tilde\b_j$ are not exactly creation and annihilation operators, they satisfy the canonical commutation relations approximately in the limit of large $N$ and small depletion (small number of particles not in the condensate mode).  Indeed, for a condensate with small depletion, $\a_0$ is of order $\sqrt N$ and the $\tilde\b_j$s are of order~$1$; the modification of the ground-state energy in the Bogoliubov approximation is also of order~$1$.  When we do operator algebra at Bogoliubov order, we can regard $\a_0$ and $\a_0^\dagger$ and the $\tilde\b_j$s and $\tilde\b_j^\dagger$s as making up a canonical set of creation and annihilation operators and neglect the small corrections to the canonical commutation relations, all of which are order $1/\sqrt N$ or smaller.

The $N$-conserving Bogoliubov Hamiltonian, with the mean field removed, is a quadratic function of $\tilde\b_j^\dagger$ and $\tilde\b_j$,
\begin{align}\label{eq:ncb_hamiltonian}
 \sH_\mathrm{ncb}= \sum_{j,k=1}^{\rank-1} H_{jk}\tilde\b_j^\dagger \tilde\b_k+\half\Big(H_{jk}' \tilde\b_j^\dagger \tilde\b_k^\dagger+\mathrm{H.c.}\Big)\;,
\end{align}
where $H^\dagger = H$ and $(H')^T=H'$.  This Hamiltonian can be diagonalized by a Bogoliubov transformation,
\begin{align}
 \sB^\dagger\, \sH_\mathrm{ncb}\, \sB
 = \sum_{k=1}^{\rank-1} \epsilon_k\, \tilde\b_k^\dagger \tilde\b_k\;,
\end{align}
where $\sB$ is the Gaussian unitary operator corresponding to a symplectic transformation of the operators $\tilde\b_j^\dagger$ and $\tilde\b_j$ for $j=1,2,\ldots,\rank-1$.  We can use the Bloch-Messiah reduction theorem~\cite{bloch_canonical_1962} to decompose the Gaussian unitary $\sB$ as
\begin{align}
\sB=\sU\,\bigg(\prod_{k=1}^{\rank-1}\,\sS(\gamma_k,\tilde\b_k)\bigg)\, \sV^\dagger\;,
\end{align}
where $\sV^\dagger$ and $\sU$ are multiport beamsplitters (which preserve the number of particles and thus have no effect on the vacuum), and the operators in the product are single-mode squeeze operators,
\begin{align}\label{eq:smsqueeze}
\sS(\gamma,\b)
=\exp\!\left(\half\big(\gamma\b^2-\gamma\b^{\dagger\,2}\big)\right)\;,
\quad\mbox{$\gamma$ real.}
\end{align}
The ground state of the Bogoliubov Hamiltonian~(\ref{eq:ncb_hamiltonian}) thus takes the form
\begin{align}\label{eq:bgs_a}
 \ketb{\varPsi_\mathrm{bgs}} &= \sB\:\ketb{N}_0\otimes\ketb{\vac}_\perp
 =\sU\,\bigg(\prod_{k=1}^{\rank-1}\,\sS(\gamma_k,\tilde\b_k)\bigg)\, \sU^\dagger\, \ketb{N}_0\otimes\ketb{\vac}_\perp\;,
\end{align}
Note that we have replaced $\sV^\dagger$ with $\sU^\dagger$ in the final form.  Neither $\sU^\dagger$ nor $\sV^\dagger$ has any effect on the GP ground state $\ket{N}_0\otimes\ket{\vac}_\perp$, so we can include either or neither of them on the right side of the product of the squeeze operators.

The action of the multiport beamsplitter $\sU$ is given by
\begin{align}
 \sU\, \tilde\b_j\, \sU^\dagger
 = \sum_{k=1}^{\rank-1} \tilde\b_k U_{kj}
 = \tilde\a_j = \a_0^\dagger\ssp \a_j/\sqrt N\;,
\end{align}
where $U$ is the unitary matrix that specifies the multiport beamsplitter.  The $\a_j$s are the annihilation operators for a new set of modes orthogonal to the condensate mode, in terms of which we can write the Bogoliubov ground state~(\ref{eq:bgs_a})~as
\begin{align}\label{eq:bgs_a2}
\ketb{\varPsi_\mathrm{bgs}}
=\prod_{k=1}^{\rank-1}\,\sS(\gamma_k,\tilde\a_k)\, \ketb{N}_0\otimes\ketb{\vac}_\perp\,.
\end{align}
Now we apply the ``quasi-normal-ordered'' factored form of the squeeze operator~\cite{perelomov_generalized_1977, hollenhorst_quantum_1979, Schumaker1986a},
\begin{equation}\label{eq:quasi_normal_ordered}
\sS(\gamma, \a)
=\frac{1}{\sqrt{\cosh \gamma}}\,
\exp\!\bigg(\mathord{-}\half \a^{\dagger\,2}\tanh\gamma\bigg)\,
\bigl(\cosh \gamma\bigr)^{- \a^{\dagger} \a}\,
\exp\!\bigg(\half \a^2 \tanh \gamma\bigg)\;,
\end{equation}
to the squeeze operators in Eq.~(\ref{eq:bgs_a2}), with the result
\begin{align}
\ketb{\varPsi_\mathrm{bgs}}
=\bigg(\prod_{k=1}^{\rank-1}\cosh\gamma_k\bigg)^{-1/2}
\exp\!\Bigg(\mathord{-}\half\sum_{k=1}^{\rank-1}\tilde\a_k^{\dagger\,2}\tanh\gamma_k\Bigg)\; \ketb{N}_0\otimes\ketb{\vac}_\perp\;.\label{eq:nc_bog_a}
\end{align}
Note that we can always make the coefficients $\mathord{-}\tanh\gamma_k$ in Eq.~(\ref{eq:nc_bog_a}) positive by redefining the phase of $\tilde\b_k$ to make $\gamma_k$ negative.

On the other hand, if we separate out a single, dominant Schmidt coefficient $\lambda_0$, the PCS~(\ref{eq:pcs_rank2_pure_second_quantization}) of $N=2n$ particles takes the form
\begin{align}\label{eq:pcstwo}
\begin{split}
 \ketb{\varPsi_\mathrm{pcs}}
 &= \frac{1}{\sqrt \NFB}\,
 \bigg(\lambda_0\big(\a_0^\dagger\big)^2
 +\sum_{k=1}^{\rank-1}\lambda_k\big(\a_k^\dagger\big)^2\bigg)^n\,\ketb{\vac}\\
 &= \frac{1}{\sqrt \NFB}\,
 \sum_{m=0}^n\binom{n}{m} \bigg(\sum_{k=1}^{\rank-1}\lambda_k\big(\a_k^\dagger\big)^2\bigg)^{m}
 \lambda_0^{n-m}(\a_0^\dagger)^{2(n-m)}
 \,\ketb{\vac}\;.
\end{split}
\end{align}
We have made no approximations as yet, but we now use $N!/(N-M)!\simeq N^M$, good when $M\ll N$, which holds in our case when $N=2n$ is large and the depletion $M=2m$ is small, to write
\begin{align}
(\a_0^\dagger)^{N-M}\ketb{\vac}
=\frac{(N-M)!}{N!}\a_0^M(\a_0^\dagger)^N\,\ketb{\vac}
\simeq\frac{\sqrt{N!}}{N^{M/2}}
\bigg(\frac{\a_0}{\sqrt N}\bigg)^M\,\ketb{N}_0\otimes\ketb{\vac}_\perp\;.
\end{align}
Plugging this into Eq.~(\ref{eq:pcstwo}) yields
\begin{align}
\begin{split}
 \ketb{\varPsi_\mathrm{pcs}}
 &\simeq\frac{\lambda_0^n\sqrt{(2n)!}}{\sqrt{\NFB}}\,
 \bigg(1+\frac{1}{2n}\sum_{k=1}^{\rank-1}\frac{\lambda_k}{\lambda_0}
 \big(\tilde\a_k^\dagger\big)^2\bigg)^n \,\ketb{2n}_0\otimes\ketb{\vac}_\perp\\
 &\simeq\sqrt{\frac{N!\,\lambda_0^N}{ \NFB}}\,
 \exp\!\bigg(\half\sum_{k=1}^{\rank-1}\frac{\lambda_k}{\lambda_0}\big(\tilde\a_k^\dagger \big)^2\bigg)\, \ketb{N}_0\otimes\ketb{\vac}_\perp\;,
 \end{split}\label{eq:nc_bog_b}
\end{align}
where the final approximation is good when $N$ is large.  Equations~(\ref{eq:nc_bog_a}) and~(\ref{eq:nc_bog_b}) can be made the same by choosing
\begin{gather}
  \NFB = N!\,\lambda_0^N \prod_{k=1}^{\rank-1} \cosh \gamma_k\;,\\
  \frac{\lambda_j}{\lambda_0} = -\tanh \gamma_j\,,\;\;\mbox{for $j=1,2,\ldots,\rank-1$.}
\end{gather}
Hence, as promised, the PCS~(\ref{eq:pcs_rank2_pure_second_quantization}) encompasses the Bogoliubov approximation to the BEC ground state.

Another way to display the same result is to notice that if $\sP_N$ is the projection operator onto the $N$-particle sector and $\ket{\alpha}_0$ is a coherent state for the condensate mode, with $\alpha=\sqrt N$,
then $\ket{N}_0\otimes\ket{\vac}_\perp\propto\sP_N \ket{\alpha}_0\otimes\ket{\vac}_\perp$.  Since $\tilde\a_k^\dagger$ is number conserving, we can write the approximate PCS of Eq.~(\ref{eq:nc_bog_b}) as
\begin{align}
 \ketb{\varPsi_\mathrm{pcs}}
 &\propto
 \sP_N\,\exp\!\bigg(\frac{1}{2}\sum_{k=1}^{\rank-1}
 \frac{\lambda_k}{\lambda_0}\big(\tilde\a_k^\dagger \big)^2\bigg)\,
 \ketb{\alpha}_0\otimes\ketb{\vac}_\perp\;.
\end{align}
The state before application of the projection into the $N$-particle sector coincides with the ``extended catalytic state'' that we have introduced for a treatment of the number-conserving Bogoliubov approximation~\cite{Nconserv}.

Now that we have the squeeze operator and its quasi-normal-ordered form available, we can display some general relations among the states we have introduced.  In particular, a tensor product of single-mode squeezed states can be written as
\begin{align}\label{eq:squeezedvacua}
\begin{split}
\prod_{j=1}^{\rank}\,\sS(\gamma_j,\a_j)\,\ketb{\vac}
&=\bigg(\prod_{k=1}^{\rank}(1-\lambda_j^2)\bigg)^{1/4}
\exp\!\bigg(\half\sum_{j=1}^{\rank}\lambda_j\,\a_j^{\dagger\,2}\bigg)\,\ketb{\vac}\\
&=\bigg(\prod_{k=1}^{\rank}(1-\lambda_j^2)\bigg)^{1/4}
e^{\sA^\dagger/2}\,\ketb{\vac}\\
&=\bigg(\prod_{k=1}^{\rank}(1-\lambda_j^2)\bigg)^{1/4}
\sum_{n=0}^\infty\frac{\sqrt{\NFB_{\smash{\vec\lambda},n}}}{2^n n!}\,\ketb{\varPsi_\mathrm{pcs}^{(n)}}\;,
\end{split}
\end{align}
where $\lambda_j=-\tanh\gamma_j$.  A product of single-mode squeezed vacua generates \PCS\ as the terms in its expansion into number sectors, thus, as noted above, making a product of single-mode squeezed vacua an extended catalytic state for a \PCS.  The Valatin-Butler state~(\ref{eq:valatin_butler_wave function}) is seen to be a product of single-mode squeezed vacua once it is written in terms of the decoupled modes~(\ref{eq:aksigma}), with the modes except for $k=0$ coming in pairs that share the same amount of squeezing.

\section{Normalization Factor}
\label{sec:normalization_factor}

The importance of the normalization factor $\NFB$ in the manipulation and analysis of PCS is analogous to the utility of the partition function in statistical physics.  By taking derivatives of the normalization factor with respect to the Schmidt coefficients $\lambda_j$, one can calculate the reduced (marginal) density matrices (RDMs) in the Schmidt basis, and these, given the Schmidt-basis wave functions (orbitals), give all the physical observables.

Generally, for any density operator $\rho$ in the symmetric subspace, the $q$-particle RDM $\rho^{(q)}$ has the following matrix elements:
\begin{align}\label{eq:qRDM}
\bra{k_1,\ldots,k_q}\,\rho^{(q)}\ket{j_1,\ldots,j_q}
=\rho^{(q)}_{k_1\cdots k_q,\,j_1\cdots j_q}
=\tr\!\big(\rho\,\a_{j_1}^\dagger\cdots\a_{j_q}^\dagger\a_{k_q}\cdots\a_{k_1}\big)\;.
\end{align}
We normalize the $q$RDM to $N!/(N-q)!=N(N-1)\cdots (N-q+1)$ unless stated otherwise; normalized in this way, the RDMs are clearly identically equal to normally ordered correlation functions.  For a PCS, we default to the Schmidt basis for the matrix elements of the $q$RDM:
\begin{align}\label{eq:qRDM_PCS}
\rho^{(q)}_{k_1\cdots k_q,\,j_1\cdots j_q}
=\frac{1}{\NFB}\bra{\vac}\sA^n\a_{j_1}^\dagger\cdots\a_{j_q}^\dagger
\a_{k_q}\cdots\a_{k_1}(\sA^\dagger)^n\ket{\vac}\;.
\end{align}

In the second-quantized picture, the normalization factor $\NFB$ introduced in Eq.~(\ref{eq:pcs_rank2_pure_second_quantization}) takes the form
\begin{align}
\begin{split}
 \NFB_{\vec\lambda,n}
 &=\brab{\vac} \sA^{n} \big(\sA^\dagger\big)^n  \ketb{\vac}\\
 &=\frac{1}{\pi^\rank}\int  \brab{\vac} \sA^{n} \projb{\vec{\alpha}} \big(\sA^\dagger\big)^n  \ketb{\vac}\; \dif^2 \alpha_1\cdots\dif^2 \alpha_\rank\\
 &=\frac{1}{\pi^\rank}\int  e^{-\norm{\vec{\alpha}}^2}\: \biggl|\,\sum_{j=1}^\rank \lambda_j\,\alpha_j^2\,\biggr|^{2n}\,\dif^2 \alpha_1\cdots\dif^2 \alpha_\rank\;,\label{eq:nfb_integral}
\end{split}
\end{align}
where we insert a basis of coherent states $\ket{\vec\alpha}=\ket{\alpha_1}\otimes\cdots\otimes\ket{\alpha_\rank}$ for the Schmidt modes.  Expanding the monomial in Eq.~(\ref{eq:nfb_integral}), we have
\begin{align}
 \NFB  &=\sum_{\{\vec{\idu},\, \vec{\idv}\}}\frac{n!}{\idu_1!\, \idu_2!\cdots\idu_\rank!}\: \frac{n!}{\idv_1!\, \idv_2!\cdots\idv_\rank!}\,\biggl(\prod_{j=1}^\rank \lambda_j^{\idu_j+\idv_j}\biggr)
 \frac{1}{\pi^\rank} \int e^{-\norm{\vec{\alpha}}^2} \bigg(\prod_{j=1}^\rank \alpha_j^{2\idu_j} \big(\alpha_j^*\big)^{2\idv_{j}}\bigg)\,\dif^2 \alpha_1\cdots\dif^2 \alpha_\rank\nonumber\\[2pt]
 &=\sum_{\{\vec{\idu}\}}\frac{(n!\ssp)^2}{\idu_1!\, \idu_2!\cdots\idu_\rank!}\:
 \prod_{j=1}^\rank\frac{(2\idu_j)!\,\lambda_j^{2\idu_j}}{\idu_j!}\;,\label{eq:nfb_a}
\end{align}
where $\idu_j$($\idv_j$) are nonnegative integers satisfying $\sum_{j=1}^\rank \idu_j=\sum_{j=1}^\rank \idv_j=n$.  Although the sum in Eq.~(\ref{eq:nfb_a}) appears to require exponential time to evaluate, it can be evaluated in polynomial time by using an iterative algorithm.  It is still, however, computationally demanding for large $N$, in addition to being unintuitive.

It is useful to note here that in the case of two Schmidt modes ($\rank=2$), Eq.~(\ref{eq:nfb_a}) reduces to
\begin{align}
 \NFB
 = 2^n n!\sum_{k=0}^n \binom nk (2k-1)!!\,[2(n-k)-1]!!\,\lambda_1^{2 (n-k)}\lambda_2^{2 k}
 =\Gamma(2 n+1)\lambda_1^{2 n}\, F\bigg(\frac{1}{2},-n;\frac{1}{2}-n;\,\frac{\lambda_2^2}{\lambda_1^2}\,\bigg)\;.\label{NFB_hypergeo}
\end{align}
Here
\begin{align}
F(a,b;c;z)
= \frac{\Gamma(c)}{\Gamma(a)\Gamma(b)}\sum_{k=0}^{\infty} \frac{\Gamma(a+k)\Gamma(b+k)}{\Gamma(c+k)}\frac{z^{k}}{k!}\label{GaussHypergeoSeries}
\end{align}
is a Gauss hypergeometric function [often denoted ${}_2 F_1(a,b;c;z)$, but missing the two subscripts here for brevity and clarity], which for the particular parameters in Eq.~(\ref{NFB_hypergeo}) has a series that terminates and is thus a polynomial.  Notice that $\NFB/\Gamma(2n+1)=\lambda_1^{2n}F\big(\frac12,-n;\frac12-n;\lambda_2^2/\lambda_1^2\big)=\lambda_2^{2n}F\big(\frac12,-n;\frac12-n;\lambda_1^2/\lambda_2^2\big)$, reflecting the symmetry under exchange of labels of the two Schmidt modes.  We apply the results of this section to this expression in App.~\ref{sec:pcs_2rdm_for_schmidt_rank_two} to find the exact 1RDM and 2RDM for Schmidt rank $\rank=2$.

To make progress in interpreting and evaluating $\NFB$, we now take what might be construed as a backward step by writing $\NFB$ in a different integral form.  To do so, we use $(2\idu_j)!/\idu_j!=2^{\idu_j}(2\idu_j-1)!!$ to write
\begin{align}
\prod_{j=1}^\rank\frac{(2\idu_j)!\,\lambda_j^{2\idu_j}}{\idu_j!}
=\frac{2^{n}}{(2\pi)^{\rank/2}}
\int_{-\infty}^\infty e^{-\norm{\vec{y}\ssp}^2/2}\bigg(
\prod_{j=1}^\rank \big(\lambda_j^2\,y_j^2\big)^{\idu_j}\bigg)
\,\dif y_1\cdots \dif y_\rank\;,
\end{align}
and we then put this into Eq.~(\ref{eq:nfb_a}) to give
\begin{equation}
 \NFB
 =\frac{2^{n}n!}{(2\pi)^{\rank/2}}
 \int_{-\infty}^\infty  e^{-\norm{\vec{y}\ssp}^2/2}\biggl( \sum_{j=1}^\rank \lambda_j^2\, y_j^2\biggr)^{\! n}\,\dif y_1\cdots \dif y_\rank\;. \label{eq:nfb_b}
\end{equation}
We use the above expression to evaluate the normalization factor in the large-$N$ limit in the next section.  For now, however, we employ Eq.~(\ref{eq:nfb_b}) to derive a generating function,
\begin{align}
\begin{split}
 \NFB
 &=\frac{2^{{2n}}n!}{(\sqrt{2\pi}\,)^{\rank}  }\:
 \frac{\partial^{n}}{\partial \tau^{n}}
 \bigg[\!\int%_{-\infty}^\infty
 \exp\!\bigg(\!-\half\Big(\norm{\vec{y}\ssp}^2+\tau\!\sum\limits_{j=1}^\rank\!\lambda_j^2 y_j^2\Big)\bigg)
 \dif y_1\cdots \dif y_\rank\bigg]\bigg|_{\tau=0}\\[3pt]
 &=2^{{2n}}n!\, \frac{\partial^{n}}{\partial \tau^{n}} \bigg(\prod_{j=1}^\rank\frac{1}{\sqrt{1-\tau\lambda_{\smash j}^2}}\bigg)\bigg|_{\tau=0}
 \label{eq:nfb_c}\;.
\end{split}
\end{align}
An alternate way to obtain the generating function~(\ref{eq:nfb_c}) is to evaluate the quantity $\bra{\vac}e^{\sqrt{\tau}\sA/2}e^{\sqrt{\tau}\sA^\dagger/2}\ket{\vac}$ using the ``quasi-normal-ordered'' factored form of the squeeze operator found in Eq.~(\ref{eq:quasi_normal_ordered}).  Indeed, we can read off the result directly from Eq.~(\ref{eq:squeezedvacua}),
\begin{align}
\bra{\vac}e^{\sqrt{\tau}\sA/2}e^{\sqrt{\tau}\sA^\dagger/2}\ket{\vac}
=\prod_{j=1}^\rank\frac{1}{\sqrt{1-\tau\lambda_{\smash j}^2}}
=\sum_{n=0}^\infty\frac{\NFB_{\smash{\vec\lambda}, n}}{2^{2n}(n!)^2}\,\tau^n\;,
\end{align}
which is equivalent to Eq.~(\ref{eq:nfb_c}).

The normalization factor can be used to find expressions for the diagonal elements of RDMs.  As a first example, we are able to represent, using Wick's theorem, the $j$th diagonal element of the 1RDM $\rho^{(1)}$ in terms of the normalization factor:
\begin{align}
\begin{split}
 \varrho_j=\rho_{j j}^{(1)} &=
 \frac{1}{\NFB}\: \brab{\vac} \sA^{n}\, \a_j^\dagger \a_j\, \big(\sA^\dagger\big)^n  \ketb{\vac}\\[3pt]
 &= \frac{n\,\lambda_j}{\NFB}\,\brab{\vac} \sA^{n-1}\, \a_j^2\, \big(\sA^\dagger\big)^n %+ \sA^n \, \big(\a_j^\dagger\ssp\big)^2\, \big(\sA^\dagger\big)^{n-1}\Big)
 \ketb{\vac} +\mathrm{c.c.}\\[3pt]
 &=\frac{\lambda_j}{\NFB}\, \brab{\vac} \bigg( \pa{\sA^{n}}{ \lambda_j}\,\big(\sA^\dagger\big)^n  +\sA^{n}\, \pa{\big(\sA^\dagger\big)^n }{ \lambda_j}\bigg)\ketb{\vac}\\[3pt]
 &= \frac{\lambda_j}{\NFB}\, \pa{\, \NFB}{\lambda_j}\;.\label{eq:diagonal_elements_rdm_single_particle}
\end{split}
\end{align}
In the second line of Eq.~(\ref{eq:diagonal_elements_rdm_single_particle}), the term and its complex conjugate, which correspond to contracting $\a^\dagger_j$ and $\a_j$ with the pair annihilation and creation operators, are equal.  In addition, by using Wick's theorem, it is not hard to prove that all the off-diagonal elements of $\rho^{(1)}$ in the Schmidt basis are zero; therefore, the normalization factor and its first derivative determine the \hbox{1RDM}.

We can apply the same thinking more generally to the $q$RDM matrix elements~(\ref{eq:qRDM_PCS}).  Indeed, Wick's theorem [or just thinking hard about the form of Eq.~(\ref{eq:qRDM_PCS})] shows that $\rho_{k_1\cdots k_q,\,j_1 \cdots j_q}^{(q)}$ is nonzero only when for all $j=1,2,\ldots,\rank$, the number of times index $j$ occurs in the matrix element,
\begin{align}\label{eq:qj}
q_j = \sum_{m=1}^q \delta_{j,j_m} + \delta_{j,k_m}\;,
\end{align}
is even.  We use this result in Sec.~\ref{sec:higher_rdms} to help find the off-diagonal matrix elements of $\rho^{(q)}$ in terms of the diagonal elements.

Returning to diagonal matrix elements, we have the following result for the diagonal elements of the $q$RDM $\rho^{(q)}$:
\begin{align}
\rho_{j_1\cdots j_q,\, j_1 \cdots j_q}^{(q)}
=\frac{1}{\NFB}\:
\brab{\vac}\sA^{n}\,
\a_{j_1}^\dagger\cdots\a_{j_q}^\dagger\ssp\a_{j_q}\cdots\a_{j_1}
\,\big(\sA^\dagger\big)^n  \ketb{\vac}
=\frac{\lambda_{j_1}\cdots\lambda_{j_q}}{\NFB}\:
\frac{\partial^q \NFB}{\partial\lambda_{j_1} \cdots \partial\lambda_{j_q}}\;,\label{eq:diagonal_elements_rdm_a}
\end{align}
This result is most easily proved by mathematical induction.  We already have that it holds for $q=1$, so to show that it holds for all positive integers $q$ is that if it holds for $q$, it is satisfied for $q+1$.  The inductive hypothesis is thus that
\begin{align}
\frac{\partial^q \NFB}{\partial\lambda_{j_1} \cdots\partial\lambda_{j_q}}
=\frac{1}{\lambda_{j_1}\cdots\lambda_{j_q}}\, \brab{\vac} \sA^{n}\,
\a_{j_1}^\dagger\cdots\a_{j_q}^\dagger\ssp\a_{j_q}\cdots\a_{j_1}
\,\big(\sA^\dagger\big)^n  \ketb{\vac}\;.\label{eq:diagonal_elements_rdm_b}
\end{align}
By taking derivatives with respect to $\lambda_{j_{q+1}}$ of both sides of Eq.~(\ref{eq:diagonal_elements_rdm_b}), we have
\begin{align}
\begin{split}
\frac{\partial^{q+1}\NFB}{\partial\lambda_{j_1} \cdots \partial\lambda_{j_{q+1}}}
&=\frac{1}{\lambda_{j_1}\cdots\lambda_{j_q}}\,\bigg(n\,
\brab{\vac}\sA^{n-1}\,\a_{j_{q+1}}^2\,
\a_{j_1}^\dagger\cdots\a_{j_q}^\dagger\ssp\a_{j_q}\cdots\a_{j_1}
\,\big(\sA^\dagger\big)^n \ketb{\vac}+\mathrm{c.c.}\\[3pt]
&\mskip125mu-\frac{1}{\lambda_{j_{q+1}}}\,
\brab{\vac} \sA^{n}\,
\a_{j_1}^\dagger\cdots\a_{j_q}^\dagger\ssp\a_{j_q}\cdots\a_{j_1}\,\big(\sA^\dagger\big)^n\ketb{\vac}
\sum_{k=1}^q\delta_{j_k,j_{q+1}}\bigg)\\[3pt]
&=\frac{n}{\lambda_{j_1}\cdots\lambda_{j_q}}\,\bigg(
\brab{\vac} \sA^{n-1}\,
\a_{j_{q+1}}\,\a_{j_1}^\dagger\cdots\a_{j_q}^\dagger\ssp\a_{j_q}\cdots\a_{j_1}\,\a_{j_{q+1}}
\,\big(\sA^\dagger\big)^n\ketb{\vac}+\mbox{c.c.}\bigg)\\[3pt]
&=\frac{1}{\lambda_{j_1}\cdots\lambda_{j_{q+1}}}\,
\brab{\vac}\sA^{n}\,
\a_{j_1}^\dagger\cdots\a_{j_{q+1}}^\dagger\ssp\a_{j_{q+1}}\cdots\a_{j_1}
\,\big(\sA^\dagger\big)^n \ketb{\vac} \;,
\end{split}\label{eq:diagonal_elements_rdm_c}
\end{align}
which is the required result.  We note that for $q>1$, the $q$RDM is generally not diagonalized in the Schmidt basis.  In Sec.~\ref{sec:higher_rdms}, we show how to construct the entire $q$RDM using only the diagonal elements calculated by the above method.

Although we can avail ourselves of the power of Wick's theorem to derive the results of this section, the content of Wick's theorem here and elsewhere in this paper is contained in two commutators, $[a_j,(\sA^\dagger)^n]=2n\lambda_j\a_j^\dagger(\sA^\dagger)^{n-1}$ and $[\sA^n,\a_j^\dagger]=2n\lambda_j\sA^{n-1}\a_j$, and the following relations, which follow immediately from the commutators:
\begin{align}
\a_j(\sA^\dagger)^n\ket\vac&=2n\lambda_j\a_j^\dagger(\sA^\dagger)^{n-1}\ket\vac\;,\\
\bra\vac\sA^n\a_j^\dagger&=2n\lambda_j\bra\vac\sA^{n-1}\a_j\;.
\end{align}

\section{Limit of Large Particle Number}
\label{sec:large_N}

\subsection{Diagonal matrix elements of \textit{q}RDMs in the large-\textit{N} limit}
\label{sec:diagonal_large_N}

Often there are thousands to millions of atoms in a BEC, and it is informative to have available results that
are only valid in the large-$N$ limit.  In this section, we discuss how to derive an asymptotic form of the normalization factor for large $N$.  To get the desired analytical results, terms that are of order $1/N$ smaller than the leading terms are neglected.  The depletion predicted by the Bogoliubov approximation is of order~$1$; i.e., it modifies the 1RDM by order~$1$, which is $1/N$ times smaller than the 1RDM itself.  This means that the large-$N$ results derived in this section do not encompass the Bogoliubov approximation; instead we should think of them as a generalization of the mean-field GP description to a multimode condensate.

\begin{figure}[htb]
   \includegraphics[width=0.5\textwidth]{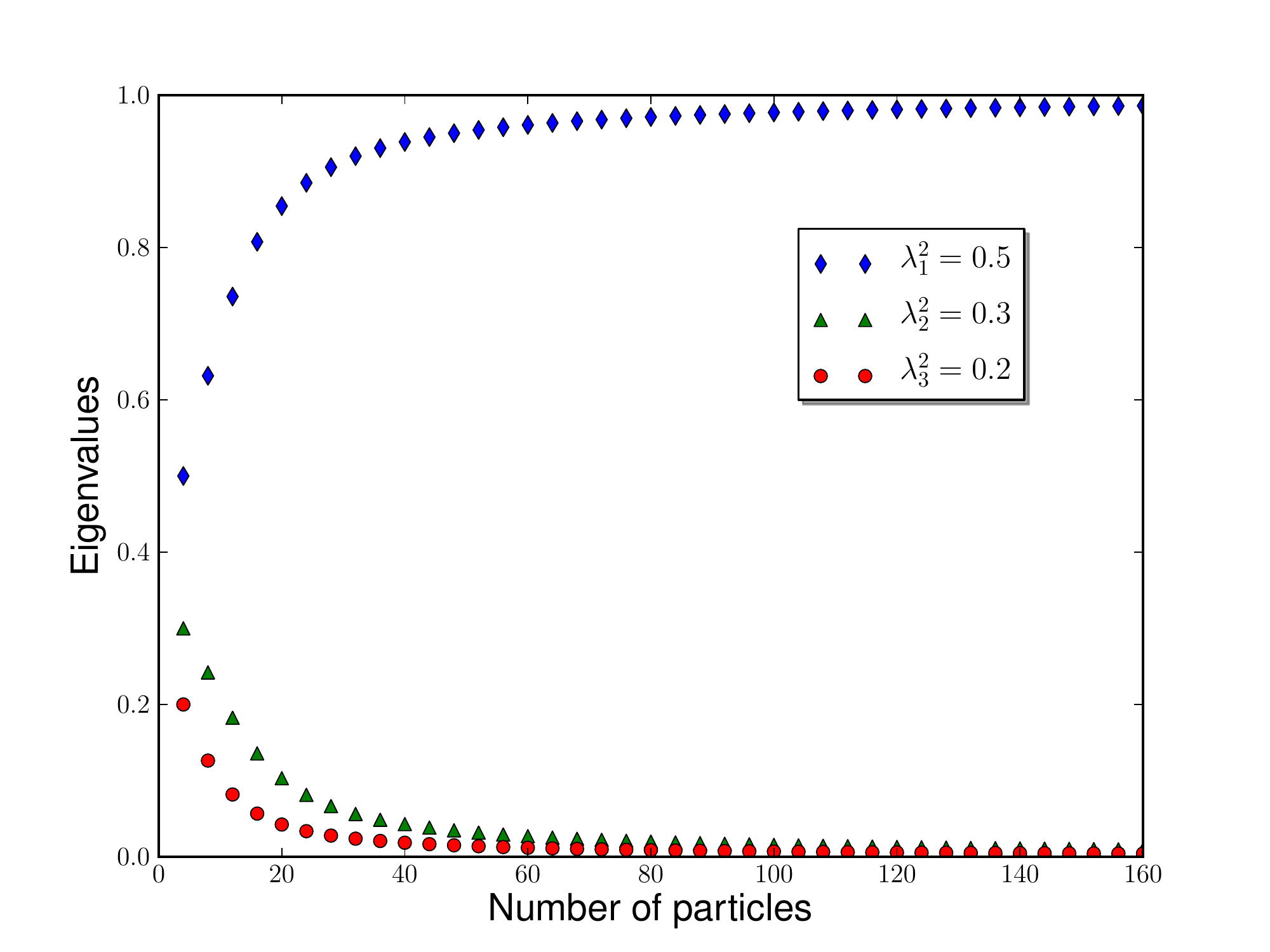}
   \caption[Eigenvalues of 1RDMs as Functions of $N$ I]{The three eigenvalues of the single-particle RDM (normalized to unity) plotted as a function of the number of particles~$N$ for the case $\lambda_1^2=0.5$, $\lambda_2^2=0.3$, $\lambda_3^2=0.2$.   As $N$ becomes large, the eigenvalue corresponding to the biggest $\lambda$ approaches 1 while the other eigenvalues become negligible.}
   \label{fig:eigenv_a}
\end{figure}

Before the analytical calculation, let us look first at some numerical examples to get some intuition. In Fig.~\ref{fig:eigenv_a}, the eigenvalues of the 1RDM (normalized to unity instead of to $N$) are plotted as a function of the number~$N$ of particles for $\lambda_1^2=0.5$, $\lambda_2^2=0.3$, and $\lambda_3^2=0.2$.  For $N=2$, the eigenvalues of the 1RDM are, of course, equal to the square of the $\lambda_j$s, but as $N$ gets larger, the eigenvalues become farther apart.  Eventually, the biggest eigenvalue approaches 1, leaving the other two eigenvalues negligible; thus, as far as the 1RDM can tell, the PCS becomes an uninteresting product state for large $N$.

\begin{figure}
%   \centering
   \includegraphics[width=0.5\textwidth]{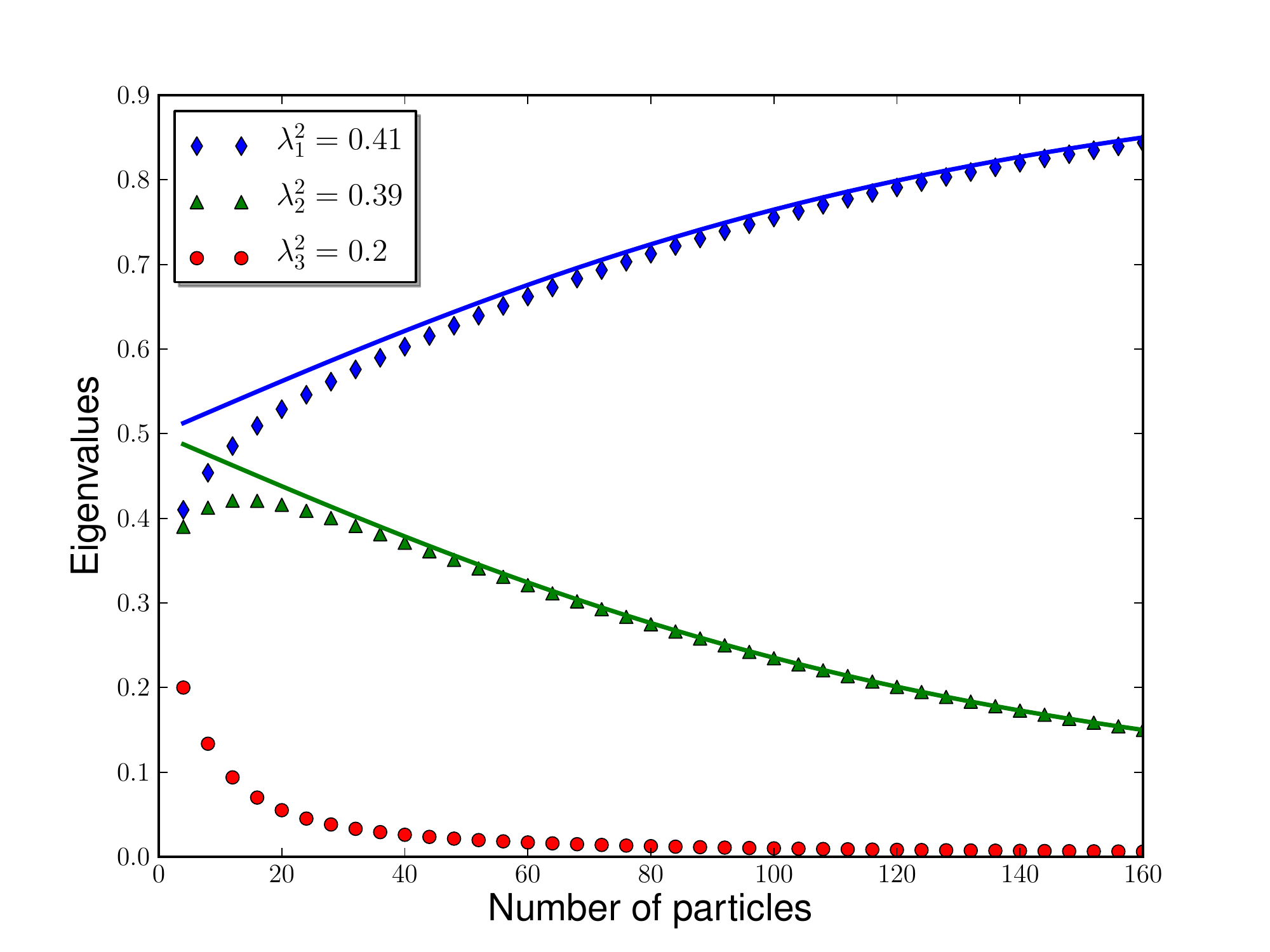}
   \caption[Eigenvalues of 1RDMs as Functions of $N$ II]{The three eigenvalues of the 1RDM (normalized to unity) plotted as a function of the number of particles~$N$ for the case $\lambda_1^2=0.41$, $\lambda_2^2=0.39$, $\lambda_3^2=0.2$.  The solid lines are the eigenvalues calculated by setting $\lambda_3=0$ and keeping $\lambda_{1}$ and $\lambda_2$ unchanged except for renormalizing; they conform pretty well to the other results for large $N$.}
   \label{fig:eigenv_c}
\end{figure}

In the second numerical example, plotted in Fig.~\ref{fig:eigenv_c}, we consider the situation where there are two $\lambda_j$s that are nearly degenerate, $\lambda_1^2=0.41$ and $\lambda_2^2=0.39$, and a third smaller value, $\lambda_3^2=0.2$.  When $N$ is of the order $1/(\lambda_1-\lambda_2)=40$ or larger, the third eigenvalue has died out, and only the two biggest eigenvalues play much of a role in determining the \hbox{1RDM}.  Our numerics suggest that in the large-$N$ limit, only those $\lambda_j$s that are within order $1/N$ of $\lambda_1$ survive.

Generally, we speculate that one only needs to keep those $\lambda_j$s that are within $1/N$ of $\lambda_1$, the largest eigenvalue given our ordering convention~(\ref{eq:lambda_convention}).  The other $\lambda_j$s can be omitted without affecting the PCS; i.e., the relevant low-order RDMs are not affected.  This speculation is supported not only by the above numerical results, but also by the following analytical results for the large-$N$ limit.   The dominant $\lambda_j$s, being very close to each other, can be rescaled (i.e., they are no longer normalized to 1) and parametrized as
\begin{align}\label{eq:lambda_zeta}
\lambda_j^2\equiv 1+\frac{\pcss_j}{n}\quad \mathrm{or}\quad\lambda_j=1+\frac{\pcss_j}{2n}+\bigO\!\left(\frac{1}{n^2}\right)\simeq1+\frac{\pcss_j}{2n}\;,
\end{align}
where the $\pcss_j$s are real parameters of order unity.  Notice that
\begin{align}
\lambda_j\frac{\partial}{\partial\lambda_j}=2(n+\pcss_j)\frac{\partial}{\partial\pcss_j}\;.
\end{align}
Because of the rescaling, all the $\lambda_j$s are very close to $1$, and their differences are of order $1/N$.  Putting Eq.~(\ref{eq:lambda_zeta}) into Eq.~(\ref{eq:nfb_b}), we manipulate the normalization factor $\NFB$ through the following sequence of steps:
\begin{align}
\begin{split}
\NFB_{\smash{\vec\pcss},\ssp n}
%&=\frac{2^{n}n!}{(\sqrt{2\pi}\,)^{\rank}  } \int_{-\infty}^\infty  e^{-\norm{\vec{y}\ssp}^2/2}\biggl( \sum_{j=1}^\rank \lambda_j^2\ssp y_j^2\biggr)^{\! n}\,\dif y_1\cdots \dif y_\rank\\
&=\frac{2^{n}n!}{(\sqrt{2\pi}\,)^{\rank}  } \int_{-\infty}^\infty  e^{-\norm{\vec{y}\ssp}^2/2}\biggl( \sum_{j=1}^\rank \Big(1+\frac{\pcss_j}{n}\Big) y_j^2\biggr)^{\! n}\,\dif y_1\cdots \dif y_\rank\\
&=\frac{2^{n}n!}{(\sqrt{2\pi}\,)^{\rank}  } \int_{-\infty}^\infty  e^{-\norm{\vec{y}\ssp}^2/2} \norm{\vec{y}\ssp}^{2n} \biggl(1+\frac{1}{n\norm{\vec{y}\ssp}^{2}} \sum_{j=1}^\rank \pcss_j\ssp y_j^2\biggr)^{\! n}\,\dif y_1\cdots \dif y_\rank\\[3pt]
&\simeq \frac{2^{n}n!}{(\sqrt{2\pi}\,)^{\rank}}
\int_{-\infty}^\infty e^{-\norm{\vec{y}\ssp}^2/2} \norm{\vec{y}\ssp}^{2n} \exp\!\bigg(\frac{1}{\norm{\vec{y}\ssp}^2}\sum_{j=1}^\rank \pcss_j\ssp y_j^2\bigg)\,\dif y_1\cdots \dif y_\rank\\
&= \frac{2^{n}n!}{(\sqrt{2\pi}\,)^{\rank}}
\int_{0}^\infty e^{\mathord{-}r^2/2} r^{2n+\rank-1}\,\dif r\int_{\norm{\vec y}=1}  \exp\!\Big(\sum_{j=1}^\rank \pcss_j y_j^2\Big)\:\dif \Omega\\
&=\frac{4^{n}n!}{2 \pi^{\rank/2}} \; \Gamma\Big(n+\frac{\rank}{2}\Big)\int_{\norm{\vec y}=1}  \exp\!\Big(\sum_{j=1}^\rank \pcss_j y_j^2\Big)\:\dif \Omega\;,
\end{split}
\label{eq:nfb_d}
\end{align}
where $\dif \Omega$ denotes the area element on the unit $(\rank-1)$-dimensional sphere defined by $\norm{\vec{y}\ssp}=1$.  The only approximation here is to replace, in the fourth line of Eq.~(\ref{eq:nfb_d}), the power function by the exponential function.  For each low-degree monomial of $\pcss_j$s, the error in its expansion coefficient as a result of this replacement is of order $1/n$; such error only becomes substantial when the degree of the monomial approaches $n$.  This is an excellent approximation for our purpose of calculating low-order RDMs, because the high-degree monomials only affect high-order RDMs.

Now denote the Gaussian integral in Eq.~(\ref{eq:nfb_d}) by
\begin{align}\label{eq:upsilon_definition}
\Upsilon\big(\vec{\pcss}\,\big)
\equiv\frac{1}{2 \pi^{\rank/2}}
\int_{\norm{\vec y}=1}\exp\!\bigg(\sum_{j=1}^\rank \pcss_j\ssp y_j^2\bigg)\:\dif\Omega
=\frac{1}{2 \pi^{\rank/2}}
\int_{-\infty}^\infty \delta\big(\norm{\vec y}-1\big)
\exp\!\bigg(\sum_{j=1}^\rank \pcss_j\ssp y_j^2\bigg)\,\dif y_1\cdots \dif y_\rank\;.
\end{align}
Since the area of the unit $(\rank-1)$-dimensional sphere is $2\pi^{\rank/2}/\Gamma(\rank/2)$, we have $\Upsilon(0)=1/\Gamma(\rank/2)$.  In terms of $\Upsilon(\vec s)$, the normalization factor takes the form
\begin{align}\label{eq:nfb_upsilon}
\NFB_{\smash{\vec\pcss},\ssp n}&\simeq 4^{n}n!\: \Gamma\Big(n+\frac{\rank}{2}\Big)\,\Upsilon\big(\vec{\pcss}\,\big)\;.
\end{align}
The significance of this expression is that the dependencies on $n$ and on $\vec\pcss$ (or~$\vec\lambda$) factorize.  According to Eq.~(\ref{eq:diagonal_elements_rdm_a}), the diagonal elements of the RDMs can now be expressed approximately, with errors of order $1/N$, as
\begin{align}
\begin{split}
\rho_{j_1\cdots j_q,\, j_1 \cdots j_q}^{(q)}
&\simeq\frac{\lambda_{j_1}\cdots\lambda_{j_q}}{\Upsilon\big(\vec{\lambda}\,\big)}
\:\frac{\partial^q \Upsilon\big(\vec{\lambda}\,\big)}{\partial\lambda_{j_1} \cdots \partial\lambda_{j_q}}\\[3pt]
&=\frac{(2n)^q\lambda^2_{j_1}\cdots\lambda^2_{j_q}}{\Upsilon\big(\vec{\lambda}\,\big)}
\:\frac{\partial^q \Upsilon\big(\vec{\lambda}\,\big)}{\partial\pcss_{j_1} \cdots \partial\pcss_{j_q}}\\[3pt]
&\simeq \frac{(2n)^q}{\Upsilon\big(\vec{\pcss}\,\big)}\: \frac{\partial^q \Upsilon\big(\vec{\pcss}\,\big)}{\partial\pcss_{j_1} \cdots \partial\pcss_{j_q}}\;. \label{eq:upsilon_diagonal_elements_rdm}
\end{split}
\end{align}
The function $\Upsilon\big(\vec{\pcss}\,\big)$ determines the diagonal elements of the $q$RDMs in the large-$N$ limit, with all the $n$ dependence removed to the factor $(2n)^q$.  Relative to exact expressions like Eq.~(\ref{eq:nfb_a}), the complexity of evaluating $\Upsilon\big(\vec{\pcss}\,\big)$ is dramatically reduced because of the removal of the $n$ dependence.

As an example, the $k$th eigenvalue of the 1RDM in the large-$N$ limit takes the form
\begin{align}\label{eq:single_particle_upsilon}
 \rho_{kk}^{(1)}= \frac{2n}{2 \pi^{\rank/2} \Upsilon(\vec\pcss)}
\int_{\norm{\vec y}=1}y_k^2 \exp\!\bigg(\sum_{j=1}^\rank \pcss_j\ssp y_j^2\bigg)\:\dif\Omega\;.
\end{align}
Consider the case where $\lambda_\rank$, the smallest eigenvalue given our ordering convention~(\ref{eq:lambda_convention}), satisfies $1-\lambda_\rank\gg1/2n$ and all the other $\lambda_j$s are within roughly $1/2n$ of $1$.  In this situation $\pcss_\rank$ is a negative number with large magnitude, and all the other $s_j$s are of order unity.  Because of the exponential function in Eq.~(\ref{eq:single_particle_upsilon}), the magnitude of $y_\rank$ must be very small to contribute to the integral; this tells us that $\rho_{\rank\rank}^{(1)} \ll 2n$.  On the other hand, we have $\sum_{j=1}^{\rank-1} y_j^2 \simeq 1$ for the other dimensions, so $\rho_{jj}^{(1)}$ for $j=1,2,\ldots,\rank-1$ can be calculated by neglecting the last dimension $y_\rank$; in effect, the integral in Eq.~(\ref{eq:single_particle_upsilon}) is reduced to an integral over a hypersphere of one less dimension.  This argument can be easily generalized to higher-order RDMs, and it confirms our speculation that we need only keep those $\lambda_j$s that are within $1/N$ of $\lambda_1$.

Although we have already made life easier by introducing $\Upsilon\big(\vec{\pcss}\,\big)$, it is still a difficult task to evaluate the Gaussian integral~(\ref{eq:upsilon_definition}) over the hypersphere.  Fortunately, we can reduce the expression for $\Upsilon\big(\vec\pcss\,)$ to a single-variable integral.  To do so, notice that
\begin{align}\label{eq:upsilon_b}
\begin{split}
\prod_{j=1}^\rank \frac{1}{\sqrt{\tau-\pcss_j}}
&=\frac{1}{(\sqrt{2\pi}\,)^\rank}\int_{-\infty}^\infty
\exp\!\bigg(\mathord{-}\half\! \sum\limits_{j=1}^\rank (\tau-\pcss_j) y_j^2\bigg)
\,\dif y_1\cdots \dif y_\rank\\
&=\frac{2}{2^{\rank/2}}\int_0^{\infty} r^{\rank-1}\, e^{-\tau r^2/2}\, \Upsilon\Big(\,\frac{r^2\ssp \vec{\pcss}}{2}\,\Big)\, \dif r\\
 &=\int_0^{\infty}\nsp \chi^{\rank/2-1}\, e^{-\tau \chi}\, \Upsilon\big(\chi\ssp \vec{\pcss}\,\big)\,\dif \chi\;,
\end{split}
\end{align}
where we do the substitution $\chi=r^2/2$ and where $\tau > \pcss_1$ for convergence ($\pcss_1$ is the largest of the $\pcss_j$s). Because Eq.~(\ref{eq:upsilon_b}) is the Laplace transform of the function  $\chi^{\rank/2-1}\, \Upsilon\big(\chi\ssp \vec{\pcss}\,\big)$, we have
\begin{align}
\Upsilon\big(\chi\ssp \vec{\pcss}\,\big)
= \chi^{1-\rank/2} \sL^{-1}\bigg(\prod_{j=1}^\rank \frac{1}{\sqrt{\tau-\pcss_j}}\bigg)
= \frac{\chi^{1-\rank/2}}{2\pi i}\, \int_{\Delta-i\infty}^{\Delta+i\infty} e^{\tau \chi}\bigg(\prod_{j=1}^\rank \frac{1}{\sqrt{\tau-\pcss_j}}\bigg)\,\dif \tau\;,\label{eq:upsilon_laplace}
\end{align}
where $\sL^{-1}$ stands for the inverse Laplace transformation and the real parameter $\Delta>\pcss_1$ for convergence.  We have thus succeeded in reducing the high-dimensional integral~(\ref{eq:upsilon_definition}) to the one-dimensional integral~(\ref{eq:upsilon_laplace}).

For numerical calculations, one might find a straightforward series expansion of the function $\Upsilon(\vec\pcss\,)$ to be useful:
\begin{align}
 \Upsilon\big(\vec{\pcss}\,\big)
 &= \frac{1}{2\pi^{\rank/2}}\int_{\norm{\vec y}=1} \prod_{j=1}^\rank e^{\pcss_j\ssp y_j^2}\:\dif \Omega= \frac{1}{2\pi^{\rank/2}} \sum_{m_1,\ldots, m_\rank=0}^\infty\, \frac{\pcss_1^{m_1}\pcss_2^{m_2}\cdots\pcss_\rank^{m_\rank}}{m_1!\, m_2!\cdots m_\rank!}\,\int_{\norm{\vec y}=1} y_1^{2 m_1} y_2^{2 m_2}\cdots y_\rank^{2 m_\rank}\:\dif \Omega \label{eq:series_expression_b}\;.
\end{align}
The integral in Eq.~(\ref{eq:series_expression_b}) can be manipulated by a change of variables into the form
\begin{align}
\begin{split}
\int_{\norm{\vec y}=1} y_1^{2 m_1}\cdots y_\rank^{2 m_\rank}\:\dif \Omega
&=\int_{-\infty}^\infty\,\delta\big(\norm{\vec y}-1\big)
y_1^{2 m_1}\cdots y_\rank^{2 m_\rank}\,\dif y_1\cdots \dif y_\rank\\
&=2\int_0^\infty\,\delta\bigg(\sum_{j=1}^\rank z_j-1\bigg)
z_1^{m_1-1/2}\cdots z_\rank^{m_\rank-1/2}\,\dif z_1\cdots \dif z_\rank\\
&=2B\big(m_1+1/2,\ldots,m_\rank+1/2\big)\;,
\end{split}
\end{align}
where
\begin{align}
B\big(m_1+1/2,\ldots,m_\rank+1/2\big)
&=\frac{\prod_{j=1}^\rank\Gamma(m_j+1/2)}{\Gamma(m+\rank/2)}
=\frac{\pi^{\rank/2}}{2^m\Gamma(m+\rank/2)}\prod_{j=1}^\rank(2m_j-1)!!\;,
\qquad m=\sum_{j=1}^\rank m_j\;,
\end{align}
is the multivariable Beta function [notice that $(-1)!!=1$].  Putting this back into Eq.~(\ref{eq:series_expression_b}) gives
\begin{align}\label{eq:UpsilonBetaFunctionExpansion}
\Upsilon\big(\vec{\pcss}\,\big)
=\frac{1}{\pi^{\rank/2}} \sum_{m_1,\ldots, m_\rank=0}^\infty\, \frac{\pcss_1^{m_1}\pcss_2^{m_2}\cdots\pcss_\rank^{m_\rank}}{m_1!\, m_2!\cdots m_\rank!}\,
B\big(m_1+1/2,\ldots,m_\rank+1/2\big)\;.
\end{align}

It is worth pointing out that we can also represent the function $\Upsilon\big(\chi\ssp \vec{\pcss}\,\big)$ as a convolution.  This representation is particularly useful in Sec.~\ref{sec:examples}, when we discuss exactly solvable examples.  Note that
\begin{align}
 \frac{1}{\sqrt{\tau-\pcss_j}}
 = \frac{1}{\sqrt{\pi}}\int_0^{\infty} \chi^{-1/2} \, e^{-(\tau-\pcss_j) \chi}  \,\dif \chi
 =\frac{1}{\sqrt{\pi}}\int_0^{\infty} e^{-\tau \chi}\, \idG_{\pcss_j}(\chi) \,\dif \chi\;, \label{eq:upsilon_c}
\end{align}
where
\begin{align}
 \idG_{\pcss_j}(\chi)=\left\{
 \begin{aligned}
  &\chi^{-1/2}\, e^{\pcss_j \chi},\; &\mbox{for $\chi>0$}\;,\\
  &\quad 0, \; &\mbox{for $\chi\leq 0$}\;.
 \end{aligned}
 \right.
\end{align}
Putting Eq.~(\ref{eq:upsilon_c}) into Eq.~(\ref{eq:upsilon_b}), we have
\begin{align}% \label{eq:upsilon_d}
\begin{split}
\int_0^{\infty} \chi^{\rank/2-1}\,
e^{-\tau \chi}\, \Upsilon\big(\chi\ssp \vec{\pcss}\,\big)\:\dif \chi
& = \prod_{j=1}^{\rank} \bigg(\frac{1}{\sqrt{\pi}}\int_0^{\infty}\,e^{-\tau \chi} G_{s_j}(\chi) \:\dif \chi\bigg)\\
&=\frac{1}{\pi^{\rank/2}} \int_0^{\infty} e^{-\tau \chi} \big(\idG_{\pcss_1}*\idG_{\pcss_2}*\cdots * \idG_{\pcss_{\rank}}\big)(\chi)\:\dif \chi\;,\label{eq:upsilon_e}
\end{split}
\end{align}
where $*$ stands for the convolution,
\begin{align}
\big(\idG_{\pcss_j}*\idG_{\pcss_k}\big)(\chi)&=\int_{0}^{\chi} \idG_{\pcss_j}(\chi-\chi')\, \idG_{\pcss_k}(\chi') \,\dif \chi'\;.
\end{align}
Doing the inverse Laplace transformation, we have
\begin{align}\label{eq:upsilon_f}
\Upsilon\big(\chi\ssp \vec{\pcss}\,\big)
=\frac{\chi^{1-\rank/2}}{\pi^{\rank/2}}\, \big(\idG_{\pcss_1}*\idG_{\pcss_2}*\cdots *\idG_{\pcss_{\rank}}\big)(\chi)\;.
\end{align}

We now have four representations of $\Upsilon\big(\ssp \vec{\pcss}\,\big)$, expressed in Eqs.~(\ref{eq:upsilon_definition}), (\ref{eq:upsilon_laplace}), (\ref{eq:UpsilonBetaFunctionExpansion}), and~(\ref{eq:upsilon_f}).  All of these turn out to be useful, and we use whichever is most convenient.

We turn now to an exploration of relations among the RDMs (in the large-$N$ limit) that can be derived and expressed through the function $\Upsilon\big( \vec{\pcss}\,\big)$.  First, from the definition~(\ref{eq:upsilon_definition}) or from the Laplace transform~(\ref{eq:upsilon_laplace}), we have
\begin{align}
\Upsilon\big(\vec{\pcss}+\delta\ssp\vec{1}\big)
&=e^{\delta}\,\Upsilon\big(\vec{\pcss}\,\big)\;,\label{eq:upsilon_scale}
\end{align}
where $\vec{1}=(1,1,\ldots,1)^T$ and $\delta$ is a $c$-number.  The only effect of adding a constant $\delta$ to all the $\pcss_j$s, which changes the normalization of the $\lambda_j^2$s by $\rank\delta/n$, is to change $\Upsilon\big(\vec{\pcss}\,\big)$ and, hence, $\NFB_{\vec s,n}$ by multiplying by a factor $e^\delta$.  This trivial fact implies that
\begin{align}\label{eq:upsilon_normalization_single_rdm}
 \sum_{k=1}^{\rank} \frac{\partial \Upsilon\big(\vec{\pcss}\,\big)}{\partial \pcss_k}=\frac{\partial \Upsilon\big(\vec{\pcss}+\delta\ssp\vec{1}\,\big)}{\partial \delta}\bigg|_{\delta=0}=\Upsilon\big(\vec{\pcss}\,\big)\;,
\end{align}
which applied to Eq.~(\ref{eq:upsilon_diagonal_elements_rdm}), confirms the normalization condition for the 1RDM:
\begin{align}
\sum_{k=1}^\rank \rho_{k\ssp k}^{(1)}
&\simeq \frac{2n}{\Upsilon\big(\vec{\pcss}\,\big)}\: \sum_{k=1}^\rank \frac{\partial \Upsilon\big(\vec{\pcss}\,\big)}{\partial\pcss_k}=2n\;.
\end{align}
Equation~(\ref{eq:upsilon_normalization_single_rdm}) can be generalized to
\begin{align}
 \sum_{k=1}^{\rank} \pa{}{\pcss_k}\,\frac{\partial^{q} \Upsilon\big(\vec{\pcss}\,\big)}{\partial \pcss_{j_1}\cdots \partial \pcss_{j_q}}=\frac{\partial^{q} \Upsilon\big(\vec{\pcss}\,\big)}{\partial \pcss_{j_1}\cdots \partial \pcss_{j_q}}\;,
\end{align}
which corresponds to the partial-trace condition on the higher RDMs:
\begin{align}
\begin{split}
\sum_{k=1}^\rank \rho_{j_1\cdots j_q\ssp k\ssp ,\, j_1 \cdots j_q\ssp k}^{(q+1)}
&\simeq \frac{(2n)^{q+1}}{\Upsilon\big(\vec{\pcss}\,\big)}\, \sum_{k=1}^\rank\, \pa{}{\pcss_k}\, \frac{\partial^q \Upsilon\big(\vec{\pcss}\,\big)}{\partial\pcss_{j_1} \cdots \partial\pcss_{j_q}}\\
&= \frac{(2n)^{q+1}}{\Upsilon\big(\vec{\pcss}\,\big)}\,  \frac{\partial^q \Upsilon\big(\vec{\pcss}\,\big)}{\partial\pcss_{j_1} \cdots \partial\pcss_{j_q}}\\[2pt]
&= 2n\,  \rho_{j_1\cdots j_q,\, j_1 \cdots j_q}^{(q)}\;. \label{eq:upsilon_normalization_q_rdm}
\end{split}
\end{align}
Notice that we have the factor $2n$, instead of $2n-q$, in Eq~(\ref{eq:upsilon_normalization_q_rdm}); this is a consequence of and illustrates the approximations we have used, which are fine for large $n$ and $q$RDMs such that $q\ll N$.

A summary of these considerations is that $\sum_{j=1}^\rank\partial/\partial\pcss_j$ is the unit operator on $\Upsilon\big(\vec{\pcss}\,\big)$ and its derivatives.  Thus, letting
\begin{align}
\sum_{j=1}^\rank \pcss_j=\rank \pcss_+\;,
\end{align}
we can say that $\Upsilon\big(\vec{\pcss}\,\big)$ has the dependence $e^{\pcss_+}$ on the sum of the PCS parameters;
as a consequence, the RDMs do not depend on $s_+$, and the differential operator $\sum_{j=1}^\rank\partial/\partial\pcss_j$ gives zero when acting on RDMs.

More powerful relations for the RDMs can be derived using the Laplace form~(\ref{eq:upsilon_laplace}),
\begin{align}\label{eq:upsilon_laplace_s=1}
 \Upsilon\big(\vec{\pcss}\,\big)
   &= \frac{1}{2\pi i}\,  \int_{\Delta-i\infty}^{\Delta+i\infty} e^{\tau }\bigg(\prod_{m=1}^\rank \frac{1}{\sqrt{\tau-\pcss_m}}\bigg)\:\dif \tau\;.
\end{align}
For $\pcss_j\neq \pcss_k$, we find
\begin{align}
\begin{split}
 \frac{\partial^2 \Upsilon\big(\vec{\pcss}\,\big)}{\partial \pcss_j\, \partial \pcss_k}
   &= \frac{1}{2\pi i}  \int_{\Delta-i\infty}^{\Delta+i\infty}\frac{e^{\tau }}{4\,(\tau-\pcss_j)(\tau-\pcss_k)}\bigg(\prod_{m=1}^\rank \frac{1}{\sqrt{\tau-\pcss_m}}\bigg)\:\dif \tau
   %&= \frac{1}{2\pi i}  \int_{\Delta-i\infty}^{\Delta+i\infty}\frac{e^\tau }{4\,(\pcss_j-\pcss_k)}\,\bigg(\frac{1}{\tau-\pcss_j}-\frac{1}{\tau-\pcss_k}\bigg)\bigg(\prod_{m=1}^\rank \frac{1}{\sqrt{\tau-\pcss_m}}\bigg)\:\dif \tau\\[3pt]
   =\frac{1}{2\,(\pcss_j-\pcss_k)}\, \bigg(\frac{\partial \Upsilon\big(\vec{\pcss}\,\big)}{\partial \pcss_j}-\frac{\partial\Upsilon\big(\vec{\pcss}\,\big)}{\partial \pcss_k}\bigg)\;.\label{eq:upsilon_derivative_a}
\end{split}
\end{align}
Equations~(\ref{eq:upsilon_diagonal_elements_rdm}) and (\ref{eq:upsilon_derivative_a}) together give the following relation between the single and two-particle RDMs,
\begin{align}\label{upsilon_relations_rdms}
 \rho^{(2)}_{j\ssp k,\,j\ssp k}\simeq n\,\frac{\rho^{(1)}_{jj}-\rho^{(1)}_{kk}}{\pcss_j-\pcss_k}\;,
 \quad\pcss_j\ne\pcss_k,
\end{align}
which we rederive in its exact form, using Wick's theorem, in Sec.~\ref{sec:2rdms} [see Eq.~(\ref{eq:rho2jkjkexact})].  We can also write $\rho^{(2)}_{j\ssp k,\,j\ssp k}$ in terms of derivatives of the 1RDM,
\begin{align}\label{upsilon_relations_rdms2}
\rho^{(2)}_{j\ssp k,\,j\ssp k}
=\frac{(2n)^2}{\Upsilon}
\frac{\partial^2 \Upsilon}{\partial \pcss_j\, \partial \pcss_k}
=\frac{2n}{\Upsilon}\frac{\partial}{\partial\pcss_j}\big(\Upsilon\rho^{(1)}_{kk}\big)
=\rho^{(1)}_{jj}\rho^{(1)}_{kk}+2n\frac{\partial\rho^{(1)}_{kk}}{\partial\pcss_j}\;,
\end{align}
which can be used to evaluate $\rho^{(2)}_{j\ssp k,\,j\ssp k}$ when $s_j=s_k$ or even when $j=k$.  Once we know $\rho^{(2)}_{j\ssp k,\,j\ssp k}$ for all $k\neq j$, an alternative way to find $\rho^{(2)}_{j\ssp j,\,j\ssp j}$ is by using the marginalization condition,
\begin{equation}\label{eq:2marginalize1}
\sum_k\rho^{(2)}_{j\ssp k,\,j\ssp k}=(2n-1)\rho_{jj}^{(1)}\;.
\end{equation}
Our conclusion is that the diagonal elements of the 2RDM can be calculated from the diagonal elements of the \hbox{1RDM}.  We consider the off-diagonal elements of the 2RDM in Sec.~\ref{sec:2rdms}.

In view of the results in this section, we should squarely address the question of whether the  1RDM encodes all the information about a PCS.  Clear at the start of this discussion is that a PCS is determined by the single-particle Schmidt states $\ket{\psi_j}$, the Schmidt coefficients $\lambda_j$, and the number of particles, $N=2n$.  Yet the entire structure of RDMs, \emph{when written in the Schmidt basis}, has exactly the same form, regardless of the Schmidt basis; the relations among the Schmidt-basis RDMs depend only on $n$ and the Schmidt coefficients.  The 1RDM is diagonal in the Schmidt basis, and its eigenvalues encode the Schmidt coefficients, so we can determine all the other RDMs, \emph{in the Schmidt basis}, in terms of the \hbox{1RDM}.

Does this mean that the 1RDM determines all the physical properties of a \PCS?  The answer to this is a resounding no.  Given a 1RDM in some arbitrary basis, one can diagonalize it, extract the Schmidt coefficients (knowing $n$), and determine the Schmidt orbitals \emph{up to a phase for each orbital\/} (equivalently, the phase of the corresponding creation operator).  The phase of one orbital (or the associated creation operator) can be fixed by using the overall phase freedom of the \PCS, but the PCS depends on the remaining phases, so the 1RDM is clearly not sufficient to determine the \PCS.  Nonetheless, the form of the RDMs, in the Schmidt basis, is independent of these phases; the phases show up in the RDMs, or correlation coefficients, when they are written in other bases.  This is particularly clear when the $q$RDM is written in the position basis:
\begin{align}\label{eq:qRDMx}
\begin{split}
\rho^{(q)}(\ybf_1,\ldots,\ybf_q;\xbf_1,\ldots,\xbf_q)
&=\brab{\varPsi_\mathrm{pcs}}
\uppsi^\dagger(\xbf_1)\cdots\uppsi^\dagger(\xbf_q)\uppsi(\ybf_1)\cdots\uppsi(\ybf_q)
\ketb{\varPsi_\mathrm{pcs}}\\
&=\sum_{k_1,\ldots,k_q;j_1,\ldots,j_q}
\rho^{(q)}_{k_1\cdots k_q,\,j_1\cdots j_q}
\psi_{k_1}(\ybf_1)\cdots\psi_{k_q}(\ybf_q)
\psi^*_{j_1}(\xbf_1)\cdots\psi^*_{j_q}(\xbf_q)
\;.
\end{split}
\end{align}
Even when one takes into account that for PCS the off-diagonal elements vanish unless each index $j$ occurs an even number of times, this still allows the possibility that the phases of the orbitals appear in the position-basis $q$RDM.

\subsection{Examples}
\label{sec:examples}

In the following, we give some exactly solvable examples in the large-$N$ limit, which include the case of Schmidt rank $\rank=2$ and the cases where the $\pcss_j$ coefficients are either totally or pair-wise degenerate.

\subsubsection{PCS with Schmidt rank two}
\label{sub:pcs_with_schmidt_rank_rank_2}

For the $\rank=2$ case, we notice that
\begin{align}
 \big(\idG_{\pcss_1}*\idG_{\pcss_2}\big)(\chi)
 =\int_{0}^{\chi} \frac{e^{\pcss_1 (\chi-\chi')}\, e^{\pcss_2 \chi'}}{\sqrt{\chi'(\chi-\chi')}}\;
 \dif \chi'
 =\pi \exp\!\Big(\frac{\pcss_1 + \pcss_2}{2}\,\chi\Big)\, \BesselI_0\Big(\frac{\pcss_1 - \pcss_2}{2}\,\chi\Big)\;,\label{eq:convolution_rank_2}
\end{align}
where
\begin{equation}
\BesselI_0(x)=\frac{1}{\pi}\int_0^\pi e^{-x\cos\theta}\,d\theta
=\frac{e^x}{\pi}\int_0^1\frac{e^{-2xu}}{\sqrt{u(1-u)}}\,du
\end{equation}
is the zeroth-order modified Bessel function.  Putting Eq.~(\ref{eq:convolution_rank_2}) into Eq.~(\ref{eq:upsilon_f}), we have
\begin{align}\label{eq:upsilon_s=2}
\Upsilon\big(\pcss_1,\pcss_2\big)
&=e^{\pcss_+}\BesselI_0(\pcss_-)\;,
\end{align}
where
\begin{equation}\label{eq:spm}
\pcss_{\pm}=\frac{\pcss_1 \pm \pcss_2}{2}\;.
\end{equation}
Notice that if we add $\delta=-(\pcss_1+\pcss_2)/2=-\pcss_+$ to both $s_1$ and $s_2$, as in Eq.~(\ref{eq:upsilon_scale}), we would remove $\pcss_+$ from $\Upsilon\big(\pcss_1,\pcss_2\big)$.

\begin{figure}[ht]
   \centering
   \includegraphics[width=0.5\textwidth]{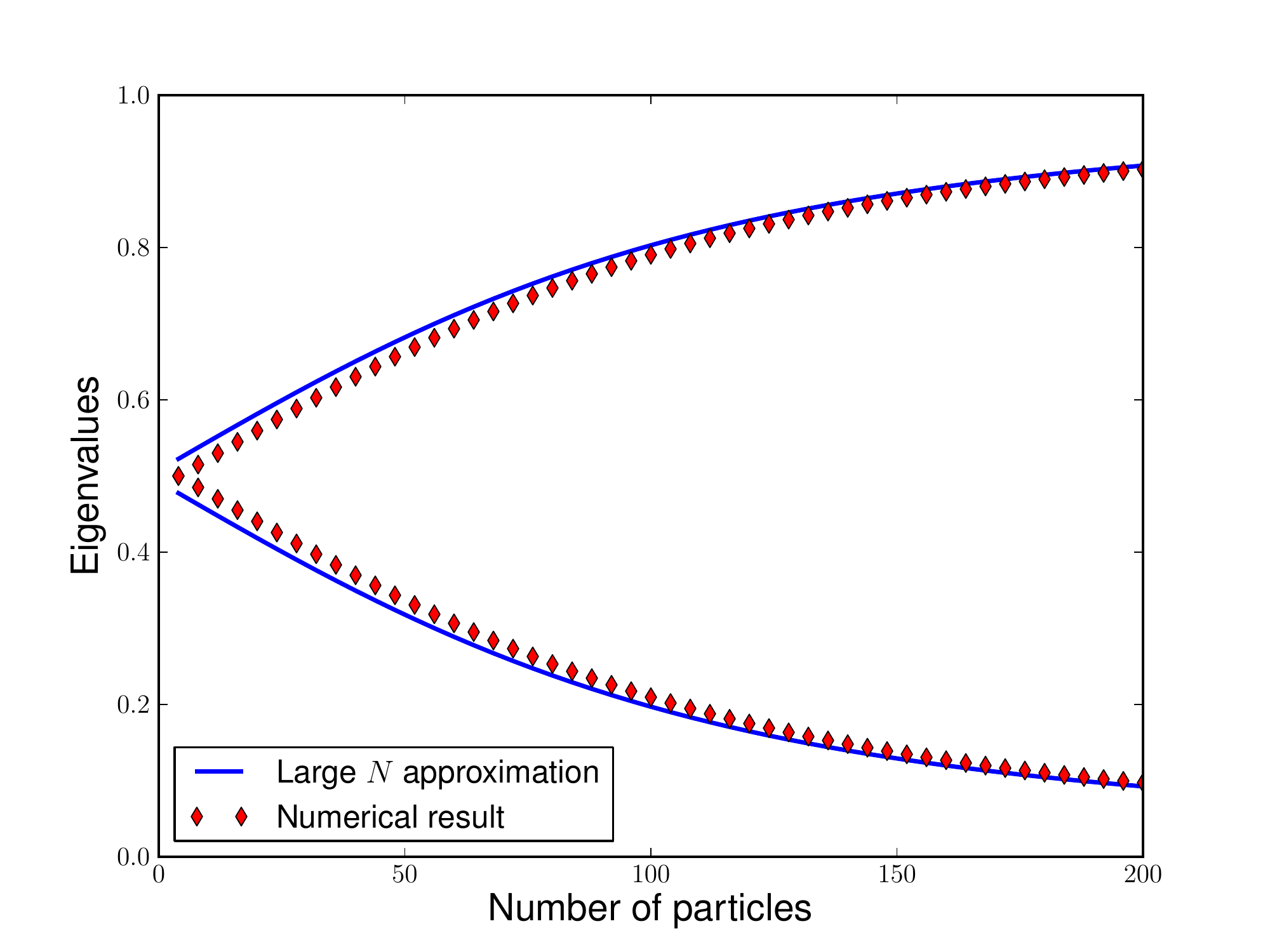}
   \caption[Large-$N$ Approximations vs Fully Numerical Results]{Eigenvalues of the 1RDM (here normalized to 1) as a function of the number of particles.  The coefficients $\lambda_1$ and $\lambda_2$ are fixed; i.e., the parameter $\pcss_-$ grows linearly in $N$.  The validity of our approximation in the large-$N$ limit is confirmed by the numerically determined eigenvalues.}
   \label{fig:eigenv_comparison}
\end{figure}

It is now straightforward to calculate the 1RDM using Eq.~(\ref{eq:upsilon_diagonal_elements_rdm}),
\begin{align}\label{eq:upsilon_s=2_1rdm_a}
%\begin{split}
 \rho_{11}^{(1)}
 \simeq \frac{2n}{\Upsilon\big(\vec{\pcss}\,\big)}\:
 \frac{\partial}{\partial\pcss_1}\,\big(e^{\pcss_+}I_0(\pcss_-)\big)
 =\frac{n}{\Upsilon\big(\vec{\pcss}\,\big)}\:e^{\pcss_+}\,\Big( \BesselI_0( \pcss_-)+ \BesselI_1(\pcss_-)\Big)
 =n\bigg( 1+ \frac{\BesselI_1( \pcss_-)}{\BesselI_0( \pcss_-)}\,\bigg)\;.
%\end{split}
\end{align}
Similarly, we have
\begin{align}\label{eq:upsilon_s=2_1rdm_b}
\rho_{22}^{(1)}
&\simeq n\bigg(1 - \frac{\BesselI_1( \pcss_-)}{\BesselI_0( \pcss_-)}\,\bigg)\;.
\end{align}
These equations can be compared with numerical evaluation of the eigenvalues of the \hbox{1RDM}.  As shown in Fig.~\ref{fig:eigenv_comparison}, the two conform quite well in the large-$N$ limit.

To examine particle-particle correlations in the $\rank=2$ PCS, we calculate the 2RDM.  By putting Eq.~(\ref{eq:upsilon_s=2}) into Eq.~(\ref{eq:upsilon_diagonal_elements_rdm}), we have
\begin{align}\label{eq:two_particle_rank2_a}
\begin{split}
 \rho_{11,11}^{(2)}
 &\simeq \frac{4 n^2}{\Upsilon\big(\vec{\pcss}\,\big)}\:
 \frac{\partial^2}{\partial \pcss_1^2}\big(e^{\pcss_+}\,I_0(\pcss_-)\big)\\[3pt]
 &=\frac{2 n^2}{\Upsilon\big(\vec{\pcss}\,\big)}\:
 \frac{\partial}{\partial \pcss_1}\bigg(e^{\pcss_+}\,
 \Big( \BesselI_0( \pcss_-)+ \BesselI_1( \pcss_- )\Big)\bigg)\\[3pt]
 &=n^2\,\bigg( \frac{3}{2}+ 2\,\frac{\BesselI_1(\pcss_-)}{\BesselI_0(\pcss_-)}+\half\, \frac{\BesselI_2(\pcss_-)}{\BesselI_0(\pcss_-)}\bigg)\;;
\end{split}
\end{align}
similarly, we have
\begin{gather}\label{eq:two_particle_rank2_b}
 \rho_{12,12}^{(2)} \simeq n^2\,\bigg( \half-\half\, \frac{\BesselI_2(\pcss_-)}{\BesselI_0(\pcss_-)}\bigg)\;,\\[6pt]
 \rho_{22,22}^{(2)} \simeq n^2\,\bigg( \frac{3}{2}- 2\,\frac{\BesselI_1(\pcss_-)}{\BesselI_0(\pcss_-)}+\half\, \frac{\BesselI_2(\pcss_-)}{\BesselI_0(\pcss_-)}\bigg)\;. \label{eq:two_particle_rank2_c}
\end{gather}
It is straightforward to check that these 2RDM elements marginalize correctly, i.e., $\rho_{11,11}^{(2)}+\rho_{12,12}^{(2)}=2n\,\rho_{11}^{(1)}$ and $\rho_{12,12}^{(2)}+\rho_{22,22}^{(2)}=2n\, \rho_{22}^{(1)}$, where we neglect the difference between $2n$ and $2n-1$ in the large-$N$ limit.  The relation~(\ref{upsilon_relations_rdms}) can be verified by using the recursion relations of the Bessel functions,
\begin{align}
\rho_{12,12}^{(2)}&\simeq \frac{n^2}{2}\,\frac{\BesselI_0(\pcss_-)-\BesselI_2(\pcss_-)}{\BesselI_0(\pcss_-)} =\frac{n^2}{\pcss_-}\,\frac{\BesselI_1(\pcss_-)}{\BesselI_0(\pcss_-)}=n\,\frac{\rho_{11}^{(1)}-\rho_{22}^{(1)}}{\pcss_1 - \pcss_2}\;.
\end{align}

\subsubsection{Totally degenerate coefficients}
\label{sub:totally_degenerate_case}

In the totally degenerate case ($\pcss_1=\pcss_2=\cdots =\pcss_{\rank}$), all of the eigenvalues of the 1RDM are the same and thus are determined by the normalization condition,
\begin{align}
  \rho_{jj}^{(1)}=\frac{2n}{\rank}\,,\;\; \mbox{for $j=1,2,\ldots,\rank$}\;.
\end{align}
Similarly, we have that the 2RDM matrix elements $\rho_{jj,\, jj}^{(2)}$ are the same for all $j$ and the matrix elements $\rho_{j\ssp k,\, j\ssp k}^{(2)}$ are the same for all $j\ne k$; moreover, putting
the Laplace form~(\ref{eq:upsilon_laplace_s=1}) into Eq.~(\ref{eq:upsilon_diagonal_elements_rdm}) implies that
\begin{align}\label{eq:diagonal_elements_relation_degenerated}
 \rho_{jj,\, jj}^{(2)}\simeq 3\rho_{j\ssp k,\, j\ssp k}^{(2)}\;,\quad \mbox{for $j\neq k$}\;.
\end{align}
We can now use the marginalization condition~(\ref{eq:upsilon_normalization_q_rdm}) to determine that
\begin{align}\label{eq:diagonal_elements_degenerated}
  \rho_{j\ssp k,\, j\ssp k}^{(2)}\simeq
  \frac{2n}{\rank+2}\:(2\delta_{jk}+1)\rho_{jj}^{(1)}
  =\frac{4n^2}{\rank(\rank+2)}\:(2\delta_{jk}+1)\;.
\end{align}

In this situation, plugging Eq.~(\ref{eq:UpsilonBetaFunctionExpansion}) into Eq.~(\ref{eq:upsilon_diagonal_elements_rdm}) gives us the diagonal matrix elements of the RDMs of all orders,
\begin{align}
\begin{split}
\rho_{j_1\cdots j_q,\, j_1 \cdots j_q}^{(q)}
&\simeq\frac{(2n)^q}{\Upsilon\big(0\big)}
\:\bigg.\frac{\partial^q \Upsilon\big(\vec{\pcss}\,\big)}{\partial\pcss_{j_1} \cdots \partial\pcss_{j_q}}\bigg|_{\vec s=0}\\[4pt]
&=\frac{(2n)^q\Gamma(\rank/2)}{\pi^{\rank/2}}B(q_1+1/2,\ldots,q_\rank+1/2)\\[3pt]
&=\frac{n^q\,\Gamma(\rank/2)}{\Gamma(q+\rank/2)}\,\prod_{j=1}^\rank(2q_j-1)!!\\[1pt]
&=\frac{(2n)^q(\rank-2)!!}{(\rank+2q-2)!!}\,\prod_{j=1}^\rank(2q_j-1)!!\;,
\end{split}
\end{align}
where $q_j=\sum_{k=1}^q\delta_{j,j_k}$ is the number of times $j$ appears in the list of single-particle states, $j_1,\ldots,j_q$.

The degenerate case illustrates that when the system possesses special symmetries, the expressions for matrix elements can sometimes be solved exactly.

\subsubsection{Pair-degenerate coefficients}
\label{sub:degenerate_pairs}

Another example occurs when the $\pcss_j$s come in degenerate pairs, i.e., $\pcss_j=\pcss_{j+1}$ for odd $j$.  Note that Eq.~(\ref{eq:convolution_rank_2}) gives the following for $\pcss_1=\pcss_2$:%\%{A more general expression is: $\Omega_{\rank}\, \chi^{\half \rank}\, \idG_\pcss(\chi) =\Omega_{\rank-1}\big[\chi^{\half(\rank-1)} \idG_\pcss(\chi) \big]*\idG_{\pcss}(\chi)\;.$}
\begin{align}\label{eq:degenerate_rank_2}
\big(\idG_{\pcss}*\idG_{\pcss}\big)(\chi)
 &=\pi e^{\chi\ssp\pcss}\;.
\end{align}
Using Eq.~(\ref{eq:upsilon_f}) and (\ref{eq:degenerate_rank_2}), we have
\begin{align}\label{eq:pair_degenerate}
%\begin{split}
\Upsilon\big(\chi\ssp \vec{\pcss}\,\big)
=\chi^{1-\rank/2} \Big(e^{\chi\ssp \pcss_1}*e^{\chi\ssp \pcss_3}*\cdots *e^{\chi\ssp \pcss_{\rank-1}}\Big)
=\chi^{1-\rank/2}\! \sum_{j\,\in\, \mathrm{odd}}\! \bigg(e^{\chi \ssp \pcss_j} \prod_{\substack{k\,\in\, \mathrm{odd}\\[2pt] k\neq j}}\frac{1}{\pcss_j-\pcss_k}\bigg)\;,
%\end{split}
\end{align}
where $\mathrm{odd}=\{1,3,\ldots, \rank-1\}$.  The convolutions in Eq.~(\ref{eq:pair_degenerate}) correspond to the sum of many exponential distributions (hypoexponential distribution); results from probability theory can then be used to derive the final form.   Another way of deriving Eq.~(\ref{eq:pair_degenerate}) is to use the residue theorem.

\section{Higher-Order Reduced Density Matrices}
\label{sec:allrdms}

\subsection{2RDMs}
\label{sec:2rdms}

In preceding sections, we focused on the diagonal elements of RDMs.  In this section we first show how to represent the off-diagonal elements of the 2RDM in terms of its diagonal elements and thus to relate the entire 2RDM to the \hbox{1RDM}.  We do this exactly, without making the large-$N$ approximation in this section, except where we explicitly introduce that approximation to illustrate how the exact results simplify when $N$ is large.  We then go on to showing how to find the off-diagonal elements of $q$RDMs from the diagonal elements in the large-$N$ limit.

In the Schmidt basis, the 2RDM reads
\begin{align}\label{eq:two_particle_rdm_a}
 \rho_{k_1 k_2,\,j_1 j_2}^{(2)}= \frac{1}{\NFB}\,\brab{\vac}\sA^n \, \a^\dagger_{j_1} \a^\dagger_{j_2} \a_{k_2} \a_{k_1}\ssp \big(\sA^\dagger\big)^n \ketb{\vac}\;.
\end{align}
All the matrix elements of $\rho^{(2)}$ are real, and $\rho^{(2)}$ satisfies the normalization condition
\begin{equation}
\tr\rho^{(2)}=\sum_{j,k}\rho_{j\ssp k,\,j\ssp k}^{(2)}=2n(2n-1)\;.
\end{equation}
Using Wick's theorem (or doing some hard thinking about the definition of the PCS), one finds that the 2RDM has the form
\begin{align}\label{eq:two_particle_rdm_b}
%\begin{split}
 \rho_{k_1 k_2,\,j_1 j_2}^{(2)}
 = \xi_{j_1 k_1}\delta_{j_1 j_2}\delta_{k_1 k_2}
 +\xi'_{j_1\ssp j_2} \delta_{j_1 k_1}\delta_{j_2 k_2}
 +\xi'_{j_2\ssp j_1}\delta_{j_1 k_2}\delta_{j_2\ssp k_1}\;.
%\end{split}
\end{align}
The matrices $\xi$ and $\xi'$ are real because none of the creation and annihilation operators introduces any complex numbers into Eq.~(\ref{eq:two_particle_rdm_a}); they are symmetric as a consequence of the Hermiticity of the \hbox{2RDM}.

There are two instructive ways to write the 2RDM, which avoid the blizzard of Kronecker deltas in Eq.~(\ref{eq:two_particle_rdm_b}).  The first is to write out $\rho^{(2)}$ as an operator in the Schmidt basis:
\begin{align}\label{eq:rho2explicit}
\begin{split}
\rho^{(2)}&=
\sum_{k_1,k_2,j_1,j_2}\rho_{k_1 k_2,\,j_1 j_2}^{(2)}
\ket{\psi_{k_1}}\bra{\psi_{j_1}}\otimes\ket{\psi_{k_2}}\bra{\psi_{j_2}}\\
&=\sum_{j,k}\Big(\xi_{jk}
\ket{\psi_k}\bra{\psi_j}\otimes\ket{\psi_k}\bra{\psi_j}
+\xi'_{jk}\ket{\psi_{j}}\bra{\psi_{j}}\otimes\ket{\psi_{k}}\bra{\psi_{k}}
+\xi'_{jk}\ket{\psi_{k}}\bra{\psi_{j}}\otimes\ket{\psi_{j}}\bra{\psi_{k}}
\Big)\;.
\end{split}
\end{align}
A second instructive way to state the content of Eq.~(\ref{eq:two_particle_rdm_b}) is that the only nonzero matrix elements of the 2RDM are the diagonal elements,
\begin{align}\label{eq:rho2jkjk}
\rho^{(2)}_{jk,jk}=\xi_{jj}\delta_{jk}+\xi'_{jk}(1+\delta_{jk})\;,
\end{align}
and the following specific off-diagonal elements,
\begin{align}\label{eq:rho2kkjj}
\rho^{(2)}_{kk,jj}&=\xi_{jk}+2\xi'_{jj}\delta_{jk}\;,\\
\rho^{(2)}_{kj,jk}&=\xi_{jj}\delta_{jk}+\xi'_{jk}(1+\delta_{jk})=\rho^{(2)}_{jk,jk}\;.
\label{eq:rho2kjjk}
\end{align}
These latter are only off-diagonal when $j\ne k$; all the formulas coincide for $\rho_{jj,jj}^{(2)}$.  Note that the off-diagonal elements in Eq.~(\ref{eq:rho2kjjk}) are equal to the diagonal elements~(\ref{eq:rho2jkjk}) because of the symmetry under the exchange of the two particles.  The relations~(\ref{eq:rho2kjjk}) can be inverted in the following way:
\begin{equation}\label{eq:xi}
\begin{aligned}
&\xi_{jk}=\rho^{(2)}_{kk,jj}\;,\quad k\ne j,\\
&\xi'_{jk}=\rho^{(2)}_{jk,jk}=\rho^{(2)}_{kj,jk}\;,\quad k\ne j,\\
&\xi_{jj}+2\xi'_{jj}=\rho^{(2)}_{jj,jj}\;.
\end{aligned}
\end{equation}
The diagonal elements of $\xi$ and $\xi'$ cannot be determined separately, but we can always choose, for example, $\xi_{jj}=\xi'_{jj}$.

To proceed, we derive an exact result for how the 2RDM changes under exchange of $j_1$ and $k_1$,
\begin{align}
\begin{split}
 \NFB\,\lambda_{j_1}\rho_{k_1 k_2,\,j_1 j_2}^{(2)}
  &=2n\lambda_{j_1}\lambda_{k_1}\,\brab{\vac}\sA^{n}\, \a^\dagger_{j_1} \a^\dagger_{j_2} \a_{k_2} \a_{k_1}^\dagger \ssp\big(\sA^\dagger\big)^{n-1}\ketb{\vac}\\
 &=2 n\lambda_{j_1}\lambda_{k_1}\,
 \brab{\vac}\sA^{n}\,
 \big(\a_{k_1}^\dagger\a^\dagger_{j_2}\a_{k_2}\a_{j_1}^\dagger
 +\delta_{k_1k_2}\a_{j_2}^\dagger\a_{j_1}^\dagger-\delta_{j_1k_2}\a_{j_2}^\dagger\a_{k_1}^\dagger\big)
 \ssp\big(\sA^\dagger\big)^{n-1}\ketb{\vac}\,\\
 &=\lambda_{k_1}\,
 \brab{\vac}\sA^{n}\,\a_{k_1}^\dagger\a^\dagger_{j_2}\a_{k_2}\a_{j_1}\ssp\big(\sA^\dagger\big)^{n}\ketb{\vac}\\
 &\qquad+\delta_{k_1k_2}\lambda_{k_1}\,
 \brab{\vac}\sA^{n}\,\a_{j_2}^\dagger\a_{j_1}\ssp\big(\sA^\dagger\big)^n\ketb{\vac}
 -\delta_{j_1k_2}\lambda_{j_1}\,
 \brab{\vac}\sA^{n}\,\a_{j_2}^\dagger\a_{k_1}\ssp\big(\sA^\dagger\big)^n\ketb{\vac}\\
 %&=\frac{1}{\NFB}\: \brab{\vac}\sA^{n}\, \a_{k_1}^\dagger  \a^\dagger_{j_2} \a_{k_2}  \a_{j_1}\ssp \big(\sA^\dagger\big)^{n}\ketb{\vac}+ \bigO(n)\nonumber\\[2pt]
 &=\NFB\Big(\lambda_{k_1}\rho_{j_1 k_2,\,k_1 j_2}^{(2)}
 +\delta_{k_1k_2}\lambda_{k_1}\,\rho_{j_1j_2}^{(1)}-\delta_{j_1k_2}\lambda_{j_1}\,\rho_{k_1j_2}^{(1)}\Big)\;.
\end{split}
\end{align}
If we make explicit that $\rho^{(1)}$ is diagonal, this exchange takes the form
\begin{align}\label{eq:jkexchange}
\lambda_{j_1}\rho_{k_1 k_2,\,j_1 j_2}^{(2)}-\lambda_{k_1}\rho_{j_1 k_2,\,k_1 j_2}^{(2)}
=\delta_{j_1j_2}\delta_{k_1k_2}\lambda_{k_1}\,\rho_{j_1j_1}^{(1)}
-\delta_{j_1k_2}\delta_{k_1j_2}\lambda_{j_1}\,\rho_{k_1k_1}^{(1)}\;.
\end{align}
This expression contains information about $\bigO(n)$ contributions to the 2RDM, whose size is generally $\bigO(n^2)$, whereas the 1RDM is $\bigO(n)$.  In the large-$N$ limit, where  $\lambda_j/\lambda_k =1+\bigO(1/n)$ [see Eq.~(\ref{eq:lambda_zeta})], the expression~(\ref{eq:jkexchange}) simply says that $\rho_{k_1 k_2,\,j_1 j_2}^{(2)}$ is symmetric at leading order under the exchange of $j_1$ and $k_1$.  The useful content of Eq.~(\ref{eq:jkexchange}) comes from specializing to $j_1=j_2=j$ and $k_1=k_2=k$:
\begin{align}\label{eq:jkexchange2}
\lambda_j\rho_{kk,jj}^{(2)}-\lambda_{k}\rho_{jk,kj}^{(2)}
=\lambda_{k}(1-\delta_{jk})\rho_{jj}^{(1)}\;.
\end{align}
This expression relates the two kinds of off-diagonal elements of the \hbox{2RDM}, and since $\rho_{jk,kj}^{(2)}=\rho_{jk,jk}^{(2)}$, it also allows us to relate all the off-diagonal elements of the 2RDM to the 1RDM and diagonal elements of the 2RDM:
\begin{align}\label{eq:kkjj}
\lambda_j\rho_{kk,jj}^{(2)}
=\lambda_{k}\big[\rho_{jk,jk}^{(2)}+(1-\delta_{jk})\rho_{jj}^{(1)}\big]\,.
\end{align}

To see what is going on more clearly, it is useful to manipulate Eq.~(\ref{eq:jkexchange2}) into an equivalent form in which the Schmidt coefficients no longer multiply the 2RDM matrix elements.  To do this, we use the condition $\rho^{(2)}_{kk,jj}= \rho^{(2)}_{jj,kk}$ to write
\begin{align}\label{eq:jkexchange3}
\lambda_k\rho^{(2)}_{kk, jj} = \lambda_k\rho^{(2)}_{jj,kk} = \lambda_j\big[\rho^{(2)}_{jk,jk}+(1-\delta_{jk})\rho^{(1)}_{kk}\big]\,.
\end{align}
Combining Eqs.~(\ref{eq:kkjj}) and~(\ref{eq:jkexchange3}), we have
\begin{align}
\rho^{(2)}_{kk,jj} -\rho^{(2)}_{jk,jk} &= \frac{\lambda_k\rho^{(1)}_{jj}+\lambda_j\rho^{(1)}_{kk}}{\lambda_j+\lambda_k}-\delta_{jk}\rho^{(1)}_{jj}\nonumber\\
&=
\frac{1}{2}(1-\delta_{jk})
\big(\rho^{(1)}_{jj}+\rho^{(1)}_{kk}\big)
-\half\frac{\lambda_j-\lambda_k}{\lambda_j+\lambda_k}\big(\rho^{(1)}_{jj}-\rho^{(1)}_{kk}\big)\,.\label{eq:difference_rho2}
\end{align}
In the large-$N$ limit, we have, from Eq.~(\ref{eq:lambda_zeta}), $(\lambda_j-\lambda_k)\big/(\lambda_j+\lambda_k)= (s_j-s_k)/4n + \bigO(1/n^2)$ and thus
\begin{align}\label{eq:difference_xi}
\rho^{(2)}_{kk,jj}-\rho^{(2)}_{jk,jk}= \frac{1}{2}(1-\delta_{jk})
\big(\rho^{(1)}_{jj}+\rho^{(1)}_{kk}\big)-\frac{s_j-s_k}{8n}
\big(\rho^{(1)}_{jj}-\rho^{(1)}_{kk}\big)+\bigO(1/n)\,,
\end{align}
The 2RDM matrix elements on the left are $\bigO(n^2)$, whereas the two terms on the right-hand side are $\bigO(n^1)$ and $\bigO(n^0)$.

We see that to leading order, $\bigO(n^2)$, in the large-$N$ limit, we have $\rho^{(2)}_{kk,jj}=\rho^{(2)}_{jk,jk}$, confirming our assertion above that to leading order, $\rho_{k_1 k_2,\,j_1 j_2}^{(2)}$ is symmetric under exchange of $j_1$ and $k_1$.  Indeed, using Eqs.~(\ref{eq:rho2kkjj}) and~(\ref{eq:rho2kjjk}), we have that $\xi_{jk}-\xi_{jk}' = \rho^{(2)}_{kk,jj} -\rho^{(2)}_{jk,jk}$, so to leading order in $n$, we can conclude that $\xi$ and $\xi'$ are the same, and a single matrix~$\xi$ determines the 2RDM.  At leading order, Eq.~(\ref{eq:two_particle_rdm_b}) simplifies to
\begin{align}\label{eq:two_particle_rdm_c}
\rho_{k_1 k_2,\,j_1 j_2}^{(2)} &\simeq \xi_{j_1 k_1} \delta_{j_1 j_2}\delta_{k_1 k_2}
+\xi_{j_1 j_2}\bigl( \delta_{j_1 k_1}\delta_{j_2\ssp k_2}+\delta_{j_1 k_2}\delta_{j_2\ssp k_1}\bigr)\;.
\end{align}
Notice that from Eq.~(\ref{eq:rho2jkjk}), to this level of approximation, $\xi$---and thus the entire 2RDM---is determined by the diagonal elements of $\rho^{(2)}$, and from Eqs.~(\ref{upsilon_relations_rdms}) and~(\ref{upsilon_relations_rdms2}), the diagonal components of the 2RDM can be related to the 1RDM,
\begin{align}\label{eq:two_particle_rdm_diagonal}
(1+2\delta_{jk})\,\xi_{jk}\simeq \rho_{j\ssp k,\,j\ssp k}^{(2)}
\simeq n\,\frac{\rho^{(1)}_{jj}-\rho^{(1)}_{kk}}{\pcss_j-\pcss_k}
\simeq \rho^{(1)}_{jj}\rho^{(1)}_{kk}+2n\frac{\partial\rho^{(1)}_{kk}}{\partial\pcss_j}\,.
\end{align}
Thus, to leading order in the large-$N$ limit, we have determined the 2RDM in terms of the \hbox{1RDM}.  Notice that Eq.~(\ref{eq:two_particle_rdm_diagonal}) implies that to leading order, all the matrix elements of $\xi$ are nonnegative.  Since from Eq.~(\ref{eq:rho2kkjj}) we also have $\rho^{(2)}_{kk,jj}=\xi_{jk}$ for $j\ne k$, we can conclude that the off-diagonal elements $\rho^{(2)}_{kk,jj}$ are nonnegative to leading order.  We can make this last conclusion secure to all orders by considering the exact relation~(\ref{eq:difference_rho2}), which shows that the off-diagonal elements $\rho^{(2)}_{kk,jj}$ are nonnegative to all orders.

We must use the leading-order, large-$N$ approximation with great care, however, because the small terms that are neglected in this approximation are crucial to determining the \PCS\ ground state precisely in those circumstances where the \PCS\ ground state is different from ground state obtained from the product-state (GP) \emph{Ansatz}.  The first term on the right-hand side of Eq.~(\ref{eq:difference_xi})---this is the first correction to the leading-order behavior---is already a Bogoliubov-order correction that is not captured by the large-$N$ limit.  The second term on the right-hand side of Eq.~(\ref{eq:difference_xi}) is thus a post-Bogoliubov correction.   We realize this is not a standard Bogoliubov expansion, because we are expanding about our large-$N$ \PCS\ approximation, not about a single-mode condensate.  Nonetheless, we believe that referring to terms in this expansion as Bogoliubov order and post-Bogoliubov order is both useful and informative, so we adopt this terminology in what follows.  As we show in Sec.~\ref{sec:two_site_bose_hubbard_model}, this second term, of order $n^0$, although two orders smaller than the leading term in the 2RDM, is crucial in capturing the superfluid-insulator transition in the Bose-Hubbard model.

To complete our program of finding the exact relation of the 2RDM to the 1RDM, we need to find the exact analog of the large-$N$ relation~(\ref{upsilon_relations_rdms}) between the diagonal elements of $\rho^{(2)}$ and those of $\rho^{(1)}$.   Thus we rederive this relation here using Wick's theorem, without making the large-$N$ approximation.  By doing contractions of the single annihilation and creation operators with the pair creation and annihilation operators, we have for $j\neq k$,
\begin{align}
\begin{split}
 \NFB\, \rho_{j\ssp k,\,j\ssp k}^{(2)}&= \brab{\vac}\sA^n \, \a_j^\dagger \a_k^\dagger \a_k \a_j\, \big(\sA^\dagger\big)^n \ketb{\vac}\\[3pt]
 &= 4n^2 \lambda_j^2\, \brab{\vac}\sA^{n-1}\, \a_j \a_k^\dagger \a_k \a_j^\dagger\, \big(\sA^\dagger\big)^{n-1}\ketb{\vac}\\
 &= 4n^2 \lambda_j^2\, \brab{\vac}\sA^{n-1}\, \big(\a_j^\dagger \a_k^\dagger \a_k \a_j+\a_k^\dagger \a_k\big)\, \big(\sA^\dagger\big)^{n-1}\ketb{\vac}\;.\label{eq:diagonal_element_contraction_a}
\end{split}
\end{align}
Switching the roles of $j$ and $k$, we have
\begin{align}
 \NFB\,\rho_{j\ssp k,\,j\ssp k}^{(2)}
 = 4n^2 \lambda_k^2\, \brab{\vac}\sA^{n-1}\, \big(\a_j^\dagger \a_k^\dagger \a_k \a_j+\a_j^\dagger \a_j\big)\, \big(\sA^\dagger\big)^{n-1}\ketb{\vac}\;.\label{eq:diagonal_element_contraction_b}
\end{align}
Multiplying Eqs.~(\ref{eq:diagonal_element_contraction_b}) and (\ref{eq:diagonal_element_contraction_a}) by $\lambda_j^2$ and $\lambda_k^2$, respectively, and subtracting the results gives
\begin{align}\label{eq:diagonal_element_contraction_c}
\begin{split}
 \big(\lambda_j^2-\lambda_k^2\big)\, \NFB\, \rho_{j\ssp k,\,j\ssp k}^{(2)}
 &= 4n^2 \lambda_j^2 \lambda_k^2\, \brab{\vac}\sA^{n-1}\, \big(\a_j^\dagger \a_j-\a_k^\dagger \a_k \big)\, \big(\sA^\dagger\big)^{n-1}\ketb{\vac}\\
 &= \brab{\vac}\sA^n \, \big(\lambda_k^2\, \a_j^\dagger \a_j-\lambda_j^2\, \a_k^\dagger \a_k \big)\, \big(\sA^\dagger\big)^n \ketb{\vac}\\
 &=\NFB\,\big(\lambda_k^2\,\rho_{jj}^{(1)}-\lambda_j^2\,\rho_{kk}^{(1)}\big)\;,
\end{split}
\end{align}
which leads to
\begin{align}\label{eq:rho2jkjkexact}
 \rho_{j\ssp k,\,j\ssp k}^{(2)}
 = \frac{\lambda_k^2\,\rho_{jj}^{(1)}-\lambda_j^2\,\rho_{kk}^{(1)}}{\lambda_j^2-\lambda_k^2}
 = n\frac{\rho_{jj}^{(1)}-\rho_{kk}^{(1)}}{s_j-s_k}+\frac{s_k\rho_{jj}^{(1)}-s_j\rho_{kk}^{(1)}}{s_j-s_k}\;,
 \quad\lambda_j\ne\lambda_k.
\end{align}
In the large-$N$ limit, the first term on the right-hand side of Eq.~(\ref{eq:rho2jkjkexact}) dominates the second term and reproduces Eq.~(\ref{upsilon_relations_rdms}).  If $\lambda_j=\lambda_k$, we can fall back on the analog of Eq.~(\ref{upsilon_relations_rdms2}):
\begin{align}\label{eq:rho2jkjkexact2}
\rho^{(2)}_{j\ssp k,\,j\ssp k}
=\frac{\lambda_j\lambda_k}{\NFB}
\frac{\partial^2\NFB}{\partial\lambda_j\,\partial \lambda_k}
=\frac{\lambda_j}{\NFB}\frac{\partial}{\partial\lambda_j}\big(\NFB\rho^{(1)}_{kk}\big)
=\rho^{(1)}_{jj}\rho^{(1)}_{kk}+\lambda_j\frac{\partial\rho^{(1)}_{kk}}{\partial\lambda_j}
=\rho^{(1)}_{jj}\rho^{(1)}_{kk}+2(n+s_j)\frac{\partial\rho^{(1)}_{kk}}{\partial\pcss_j}\;.
\end{align}
These relations allow us to determine $\rho_{j\ssp k,\,j\ssp k}^{(2)}$, for $j\neq k$, from the 1RDM, and then $\rho_{j\ssp j,\,j\ssp j}^{(2)}$ is determined by the marginalization condition,
\begin{align}\label{eq:rho2jjjjexact}
 \rho^{(2)}_{j j,\,j j}=(2n-1)\rho_{jj}^{(1)}-\sum_{k\neq j} \rho^{(2)}_{j\ssp k,\,j\ssp k}\;.
\end{align}
Equation~(\ref{eq:difference_rho2}) can then be used to calculate the off-diagonal elements~$\rho^{(2)}_{kk,jj}$:
\begin{align}\label{eq:rho2kkjjexact}
\rho^{(2)}_{kk,jj}
=\frac{\lambda_j\lambda_k}{\lambda_j^2-\lambda_k^2}\big(\rho_{jj}^{(1)}-\rho_{kk}^{(1)}\big)\;,
\quad j\ne k.
\end{align}
Equations~(\ref{eq:rho2jkjkexact})--(\ref{eq:rho2kkjjexact}) provide a recipe for calculating the 2RDM from the 1RDM.

We close this subsection by discussing a subtle point regarding formulas like Eqs.~(\ref{eq:difference_xi}) and~(\ref{eq:rho2jkjkexact}), which involve rewriting exact results that depend on the Schmidt coefficients $\lambda_j=\sqrt{1+\pcss_j/n}$ in terms of the PCS parameters~$\pcss_j$.  Rescaling the Schmidt coefficients by a factor $r$ rescales the normalization factor~$\NFB$ by a factor $r^{2n}$; this factor disappears from all the RDMs, however, and this property means that we can choose any normalization for the Schmidt coefficients.  We also know that in the large-$N$ limit, adding a constant $\delta$ to all the \PCS~parameters $\pcss_j$, rescales $\Upsilon(\vec s)$ by a factor $e^\delta$ and thus does not affect the large-$N$ RDMs.  What we have to be careful about in exact expressions is that adding $\delta$ to all the \PCS~parameters is not a rescaling of the Schmidt coefficients.  Instead, we rescale the Schmidt coefficients by a factor $r=\sqrt{1+\delta/n}$ by defining new \PCS~parameters $\pcss'_j=(1+\delta/n)\pcss_j+\delta$ and thus new Schmidt coefficients $\lambda'_j=\lambda_j\sqrt{1+\delta/n}$. In the large-$N$ limit, the term $\delta/n$ in the primed \PCS~parameters should be ignored, as it is of the size of terms neglected in the large-$N$ limit; this takes us back to simply adding a constant to the PCS parameter.  Exact expressions for RDMs are invariant under the rescaling by $r=\sqrt{1+\delta/n}$, but because the rescaling depends on $n$, when one expands in powers of $1/n$ by introducing the \PCS~parameters, the terms in the expansion are generally not invariant under the rescaling.  In Eq.~(\ref{eq:difference_xi}), this does not cause a problem, because the rescaling only introduces terms of higher order than those kept; thus we can derive the Bogoliubov and post-Bogoliubov corrections in Eq.~(\ref{eq:difference_xi}).  In contrast, in Eq.~(\ref{eq:rho2jkjkexact}), the rescaling mixes the leading-order term with the Bogoliubov correction; this shows up as a difficulty in identifying the Bogoliubov correction to the large-$N$ contribution.

\subsection{2RDM examples in the large-\textit{N} limit}
\label{sec:2RDMexamples}

In the case $\rank=2$, we can solve for the 2RDM exactly in the large-$N$ limit. Plugging the results~(\ref{eq:two_particle_rank2_a}), (\ref{eq:two_particle_rank2_b}), and~(\ref{eq:two_particle_rank2_c}) into Eq.~(\ref{eq:two_particle_rdm_c}), we have the following expression for the 2RDM in the Schmidt basis, $\{\,\ket{11}\,$,  $\,\ket{12}\,$,$\,\ket{21}\,$,$\,\ket{22}\,\}$,
\begin{align}\label{eq:two_particle_rdm_matrix_form}
 \rho^{(2)}\simeq \frac{n^2}{2I_0}
 \scalebox{0.9}{$
 \begin{pmatrix}
  3I_0+4I_1+I_2  &0        &0         & I_0-I_2\\[3pt]
  0              &I_0-I_2  & I_0-I_2  & 0 \\
  0              &I_0-I_2  & I_0-I_2  & 0 \\
  I_0-I_2        & 0       &0         & 3 I_0-4 I_1+ I_2
 \end{pmatrix}$
 }\;,
\end{align}
where $I_0$, $I_1$, and $I_2$ are the zeroth, first, and second order modified Bessel functions with argument $\pcss_-=(\pcss_1-\pcss_2)/2$. Equivalently, we can write $\rho^{(2)}$ in the Pauli basis,
\begin{align}\label{eq:two_particle_rdm_Pauli_form}
 \rho^{(2)}
 &\simeq n^2 \bigg(\identity\otimes \identity +\frac{I_1}{I_0}\: \big(\identity\otimes Z+Z\otimes \identity\big)
 +\frac{I_0+I_2}{2\, I_0}\: Z\otimes Z+\frac{I_0-I_2}{2\, I_0}\:  X\otimes X\bigg)\;.
\end{align}
The two-particle state~(\ref{eq:two_particle_rdm_Pauli_form}) is not entangled; i.e., it has zero concurrence.  It is known that all pairwise entanglement vanishes in large bosonic systems due to the monogamy of entanglement~\cite{koashi_entangled_2000}.

Another case that can be solved analytically in the large-$N$ limit is the totally degenerate case, i.e.,  $\pcss_1=\pcss_2=\cdots=\pcss_\rank$. Using Eqs.~(\ref{eq:diagonal_elements_degenerated}) and (\ref{eq:two_particle_rdm_c}), we have
\begin{align}
 \rho_{k_1 k_2,\,j_1 j_2}^{(2)} &\simeq \frac{(2n)^2}{\rank(\rank+2)} \Bigl(\delta_{j_1 j_2}\delta_{k_1 k_2}+ \delta_{j_1 k_1}\delta_{j_2\ssp k_2}+\delta_{j_1 k_2}\delta_{j_2 k_1}\Bigr)\;.
\end{align}

\subsection{Off-diagonal elements of \textit{q}RDMs in the large-\textit{N} limit}
\label{sec:higher_rdms}

We already know how to calculate the diagonal elements of the $q$RDMs in the large-$N$ limit, by using Eq.~(\ref{eq:upsilon_diagonal_elements_rdm}), but we have not yet discussed how to derive the off-diagonal elements for $q>2$.  It turns out that the methods we developed for the 2RDM are also useful for the $q$RDM, and in the large-$N$ limit, the result is very simple.

First, we establish how the $q$RDM matrix element~(\ref{eq:qRDM_PCS}) changes under exchange of $j_1$ and $k_1$ to leading order, i.e., to $\bigO(n^q)$:
\begin{align}
\begin{split}
 \lambda_{j_1}\rho_{k_1\cdots k_q,\,j_1 \cdots j_q}^{(q)}
  &=\frac{2n\lambda_{j_1}\lambda_{k_1}}{\NFB}\,\brab{\vac}\sA^{n}\, \a^\dagger_{j_1}\cdots \a^\dagger_{j_q} \a_{k_q}\cdots \a_{k_2} \a_{k_1}^\dagger \ssp\big(\sA^\dagger\big)^{n-1}\ketb{\vac}\\
 &=\frac{2 n\lambda_{j_1}\lambda_{k_1}}{\NFB}\,
 \brab{\vac}\sA^{n}\,
 \a_{k_1}^\dagger\a^\dagger_{j_2}\cdots\a^\dagger_{j_q} \a_{k_q}\cdots \a_{k_2}\a_{j_1}^\dagger\ssp\big(\sA^\dagger\big)^{n-1}\ketb{\vac}+ \bigO(n^{q-1})\,\\
 &=\frac{\lambda_{k_1}}{\NFB}\,
 \brab{\vac}\sA^{n}\,
 \a_{k_1}^\dagger\a^\dagger_{j_2}\cdots\a^\dagger_{j_q} \a_{k_q}\cdots \a_{k_2}\a_{j_1}\ssp\big(\sA^\dagger\big)^{n}\ketb{\vac}+ \bigO(n^{q-1})\\
  %&=\frac{1}{\NFB}\: \brab{\vac}\sA^{n}\, \a_{k_1}^\dagger  \a^\dagger_{j_2} \a_{k_2}  \a_{j_1}\ssp \big(\sA^\dagger\big)^{n}\ketb{\vac}+ \bigO(n)\nonumber\\[2pt]
 &=\lambda_{k_1}\rho_{j_1 k_2\cdots k_q,\,k_1 j_2 \cdots j_q}^{(q)}+ \bigO(n^{q-1})
 \;.
\end{split}
\end{align}
In the large-$N$ limit, $\lambda_{j_1}/\lambda_{k_1} = 1 +\bigO(1/n)$, so at leading order, we have
\begin{align}\label{eq:invariant}
 \rho_{k_1 k_2\cdots k_q,\,j_1 j_2 \cdots j_q}^{(q)} = \rho_{j_1 k_2\cdots k_q,\,k_1 j_2 \cdots j_q}^{(q)} + \bigO(n^{q-1})\;,
\end{align}
Because of the particle-exchange symmetry, $\rho_{k_1\cdots k_q,\,j_1 \cdots j_q}^{(q)}$ is invariant under any permutation of $\{j_1, \ldots, j_q\}$ and any permutation of $\{k_1, \ldots, k_q\}$.  Thus  Eq.~(\ref{eq:invariant}) implies that $\rho_{k_1\cdots k_q,\,j_1 \cdots j_q}^{(q)}$ remains the same at leading order under any permutation of all the indices, $\{j_1, \ldots, j_q,\,k_1, \ldots, k_q\}$.

Start now with any nonzero matrix element~$\rho_{k_1 k_2\cdots k_q,\,j_1 j_2 \cdots j_q}^{(q)}$.  Recalling from Eq.~(\ref{eq:qj}) that any particular index occurs an even number of times, those occurrences can be permuted to positions where that index is paired up on the row and column sides of the matrix element, thus giving a diagonal matrix element.  Thus the original nonzero matrix element is equal at leading order to a diagonal matrix element, which can be obtained from Eq.~(\ref{eq:upsilon_diagonal_elements_rdm}).  This relates any off-diagonal element of $\rho^{(q)}$ to a corresponding diagonal element at leading order, i.e., $\bigO(n^q)$.  Exact relations, not assuming that $N$ is large, but of increasing complexity as $q$ increases, could be worked out using the techniques developed in this section.

\section{Two-Site Bose-Hubbard Model}
\label{sec:two_site_bose_hubbard_model}

To illustrate the utility of the PCS formalism, we investigate the performance of the PCS \emph{Ansatz\/} in approximating ground-state properties and RDMs of the two-site Bose-Hubbard (BH) model, given by the Hamiltonian
\begin{align}\label{BH_hamil}
\sH_\mathrm{tbh} = -J\ssp (\b_1^\dagger \b_2 + \b_2^\dagger \b_1) + \frac{U}{2}\ssp \Big(\b_1^\dagger \b_1^\dagger \b_1 \b_1 + \b_2^\dagger \b_2^\dagger \b_2 \b_2\Big)\;,
\end{align}
which describes the hopping of $N$ interacting bosons between two identical lattice sites. Here $\b_k (\b_k^\dag)$ represents the annihilation (creation) operator for a boson on site $k=1,2$, $J$ sets the strength of tunneling between sites, and $U>0$ determines the strength of the on-site repulsive interaction between two particles.  Despite its simplicity, the two-site BH Hamiltonian~(\ref{BH_hamil}) describes the rich physics of a quantum phase transition from a superfluid to a Mott insulator~\cite{fisher_89} with increasing ratio $U/(n J)$, for integer-filling $n$.  In this section we focus on situations where $n\gg 1$.  For very weak interactions, $(nU)/J\ll1$, the ground state of the BH model is a superfluid with almost all particles occupying a single (condensate) mode that is symmetrically delocalized between the two sites; this regime is well described by GP mean-field theory.  Depletion of the single condensate mode increases as the interaction strength increases, with the transition from a superfluid to a Mott insulator occurring when $U/(nJ)\sim 1$.  For strong interactions, $U/(nJ)\gg 1$, the condensate fragments into two uncorrelated components at the two sites.  Notice that in terms of powers of $1/n$, there is another regime of relatively weak interaction strength, $U/J\sim1$, between the transition to mean-field GP at $(nU)/J\sim 1$ and the transition to a Mott insulator at $U/(nJ)\sim1$.  In this intermediate regime, the ground state is still close to a single-mode condensate; we discuss how the PCS formalism treats this intermediate regime in the following.  Even though the \PCS\ \emph{Ansatz\/} and the number-conserving Bogoliubov approximation share the same form~(\ref{eq:pcs_Sacha}), the latter only works for small corrections about a single-mode condensate and fails when there are two or more well-populated modes.  The \PCS\ approach is nonperturbative and does not have such a constraint.

Because the size of the Hilbert space grows linearly with $N$, we can directly diagonalize Eq.~(\ref{BH_hamil}) numerically for large system sizes ($N\gtrsim 100$) and, hence, benchmark the PCS \emph{Ansatz\/} against the exact BH ground state for a wide range of parameters. Specifically, we focus on the comparison of the ground-state energy and on the reconstruction of the one- and two-particle RDMs for the ground state.  As appropriate, we include comparisons with the ground state of the conventional Bogoliubov approximation to the two-site BH Hamiltonian~\cite{paraoanu_josephson_2001}; the predictions of the conventional Bogoliubov approximation are developed in App.~\ref{sec:bogoliubov_approximation_to_the_two_site_bh_model}.  Lastly, we briefly explore the potential of the PCS \emph{Ansatz\/} by showing that it can faithfully represent the time evolution of the two-site BH model~\cite{JiangPhD}, which will be discussed in detail elsewhere.

We start here with the exact $N$-particle ground state $\ket{\varPsi_\mathrm{ex}}$ and ground-state energy $E$ for parameters ranging across the transition region from very weak interactions, $(nU)/J \ll 1$, to strong interactions, $U/(nJ) \gg 1$.  We then calculate the exact one- and two-particle RDMs,
\begin{align}
	\exrho^{(1)}_{kj}&=\bra{\varPsi_\mathrm{ex}} \b_j^\dagger \b_k \ket{\varPsi_\mathrm{ex}}\;,\label{1RDM_BH}\\
	\exrho^{(2)}_{k_1k_2,\,j_1j_2}&=\bra{\varPsi_\mathrm{ex}} \b_{j_1}^\dagger \b_{j_2}^\dagger \b_{k_2}\b_{k_1} \ket{\varPsi_\mathrm{ex}}\;.\label{2RDM_BH}
\end{align}
Note that the RDMs~(\ref{1RDM_BH}) and (\ref{2RDM_BH}) are written in the physical basis that defines sites~1 and~2 and, in turn, determine directly important physical properties, such as the on-site number fluctuation,
\begin{align}\label{num_fl_BH}
	\Delta N_j = \sqrt{\exrho^{(2)}_{jj,\,jj}-\exrho^{(1)}_{jj}\!\left(\exrho^{(1)}_{jj}-1\right)}\;,
\end{align}
the (normalized) second-order site correlation functions,
\begin{subequations}\label{CG2_BH1}
\begin{align}
	C^{(2)}_{12}&=\frac{\exrho^{(2)}_{12,\,12}}{\exrho^{(1)}_{11}\, \exrho^{(1)}_{22}}\;,\\
    G^{(2)}_{12}&=\frac{\exrho^{(2)}_{11,\,22}}{\exrho^{(1)}_{11}\, \exrho^{(1)}_{22}}\;,
\end{align}
\end{subequations}
and the ground-state energy,
\begin{align}\label{E0_BH1}
	E=
-J \left(\exrho^{(1)}_{12}+\exrho^{(1)}_{21}\right)
+\frac{U}{2}\left(\exrho^{(2)}_{11,\,11}+\exrho^{(2)}_{22,\,22} \right)\;.
\end{align}

Before going further, it is useful to spell out the symmetries of the ground-state 2RDM $\exrho^{(2)}$.  The 2RDM is, of course, Hermitian.  Time-reversal invariance of the Hamiltonian~(\ref{BH_hamil}) implies that $\exrho^{(2)}$ has real matrix elements; together with Hermiticity, this means that $\exrho^{(2)}$ is a real, symmetric matrix.  The boson symmetry implies that the two row labels or the two column labels can be interchanged without changing the value of the matrix elements.  These symmetries imply that the 2RDM can be parameterized as
\begin{align}\label{eq:exact_2rdm_BH_without_site_label_symmetry}
 \exrho^{(2)}=\half
 \begin{pmatrix}
  \alpha-\delta_1 &\beta        &\beta         & \alpha-\delta-4\gamma\\
  \beta              &\alpha+\delta & \alpha+\delta  & \beta \\
  \beta              &\alpha+\delta & \alpha+\delta & \beta \\
  \alpha-\delta-4\gamma        & \beta       &\beta         & \alpha-\delta_2
 \end{pmatrix}\;,
\end{align}
where all the parameters are real and $\delta=(\delta_1+\delta_2)/2$.  We use the basis ordering $\{\ket{11}, \ket{12},\ket{21}, \ket{22}\}$.  Normalization implies that
\begin{align}\label{eq:alpha}
\alpha=\frac12\tr\exrho^{(2)}=\frac12 N(N-1)=n(2n-1)\;.
\end{align}
Notice that the boson symmetry means that the matrix~(\ref{eq:exact_2rdm_BH_without_site_label_symmetry}) is symmetric under interchange of the middle two rows or the middle two columns.  There is one further symmetry we can apply.  The Bose-Hubbard Hamiltonian~(\ref{BH_hamil}) is symmetric under interchange of the site labels.  The ground state inherits this symmetry, which implies that the matrix $\sigma^{(2)}$ is invariant under an inversion through its center.  The one further identification this gives us is that $\exrho^{(2)}_{11,\,11}=\exrho^{(2)}_{22,\,22}$, i.e., $\delta_1=\delta_2$, which leaves the 2RDM in the form
\begin{align}\label{eq:exact_2rdm_BH}
 \exrho^{(2)}=\half
 \begin{pmatrix}
  \alpha-\delta &\beta        &\beta         & \alpha-\delta-4\gamma\\
  \beta              &\alpha+\delta & \alpha+\delta  & \beta \\
  \beta              &\alpha+\delta & \alpha+\delta & \beta \\
  \alpha-\delta-4\gamma        & \beta       &\beta         & \alpha-\delta
 \end{pmatrix},
\end{align}
The 1RDM in the site basis takes the form
\begin{align}\label{1RDM_BH_ex}
\exrho^{(1)} = \frac{1}{N-1}
 \begin{pmatrix}
  \alpha&\beta    \\
  \beta &\alpha
 \end{pmatrix}\;,
\end{align}
and the second-order correlation functions~(\ref{CG2_BH1}) and the ground-state energy~(\ref{E0_BH1}) become
\begin{gather}
	C^{(2)}_{12}=\frac{\exrho^{(2)}_{12,\,12}}{\big(\exrho^{(1)}_{11}\big)^2}
    =1-\frac{1}{N}+\frac{2\delta}{N^2}\;,\label{C2_BH}\\
    G^{(2)}_{12}=\frac{\exrho^{(2)}_{11,\,22}}{\big(\exrho^{(1)}_{11}\big)^2}
    =1-\frac{1}{N}-\frac{2\delta}{N^2}-\frac{8\gamma}{N^2}\;,\label{G2_BH}\\
	E=-2J\exrho^{(1)}_{12}+U\exrho^{(2)}_{11,\,11}
    =-\frac{2J\beta}{N-1}+\frac{U}{2}\!\left(\frac{N(N-1)}{2}-\delta \right)\;.\label{E0_BH}
\end{gather}

The connection to the PCS~\emph{Ansatz\/} is made in the Schmidt basis, in which the 1RDM has the diagonal form
\begin{align}\label{1RDM_BH_Sch}
\rho^{(1)} =
 \begin{pmatrix}
  \exrho^{(1)}_{11} + \exrho^{(1)}_{12}& 0    \\
  0   & \exrho^{(1)}_{11} - \exrho^{(1)}_{12}
 \end{pmatrix}
 =\frac{1}{N-1}
 \begin{pmatrix}
  \alpha+\beta&0    \\
  0&\alpha-\beta
 \end{pmatrix}
\equiv
  \begin{pmatrix}
   \varrho_1   & 0    \\
   0   & \varrho_2
  \end{pmatrix}
 \,.
\end{align}
This diagonal form is obtained via the transformation
\begin{align}\label{Sch_transf}
    \begin{pmatrix}
     \a_1^\dagger   \\[2pt]
     e^{i\mu}\a_2^\dagger
    \end{pmatrix}=\frac{1}{\sqrt{2}}
    \begin{pmatrix}
     1   & 1    \\[4pt]
     1   & -1
    \end{pmatrix} \begin{pmatrix}
  \b_1^\dagger    \\[2pt]
  \b_2^\dagger
 \end{pmatrix}\;.
\end{align}
Here $e^{i\mu}$ is the one phase (after removing the global phase freedom) that is left arbitrary in the two Schmidt orbitals.  We show below that this phase is fixed (to within an irrelevant sign) by choosing it to minimize the \PCS\ ground-state energy, and this leads to the choice $\mu= \pi/2$, i.e., $e^{i\mu}=i$.

\begin{figure}[ht]
	\centering
	\includegraphics[width=.85\textwidth]{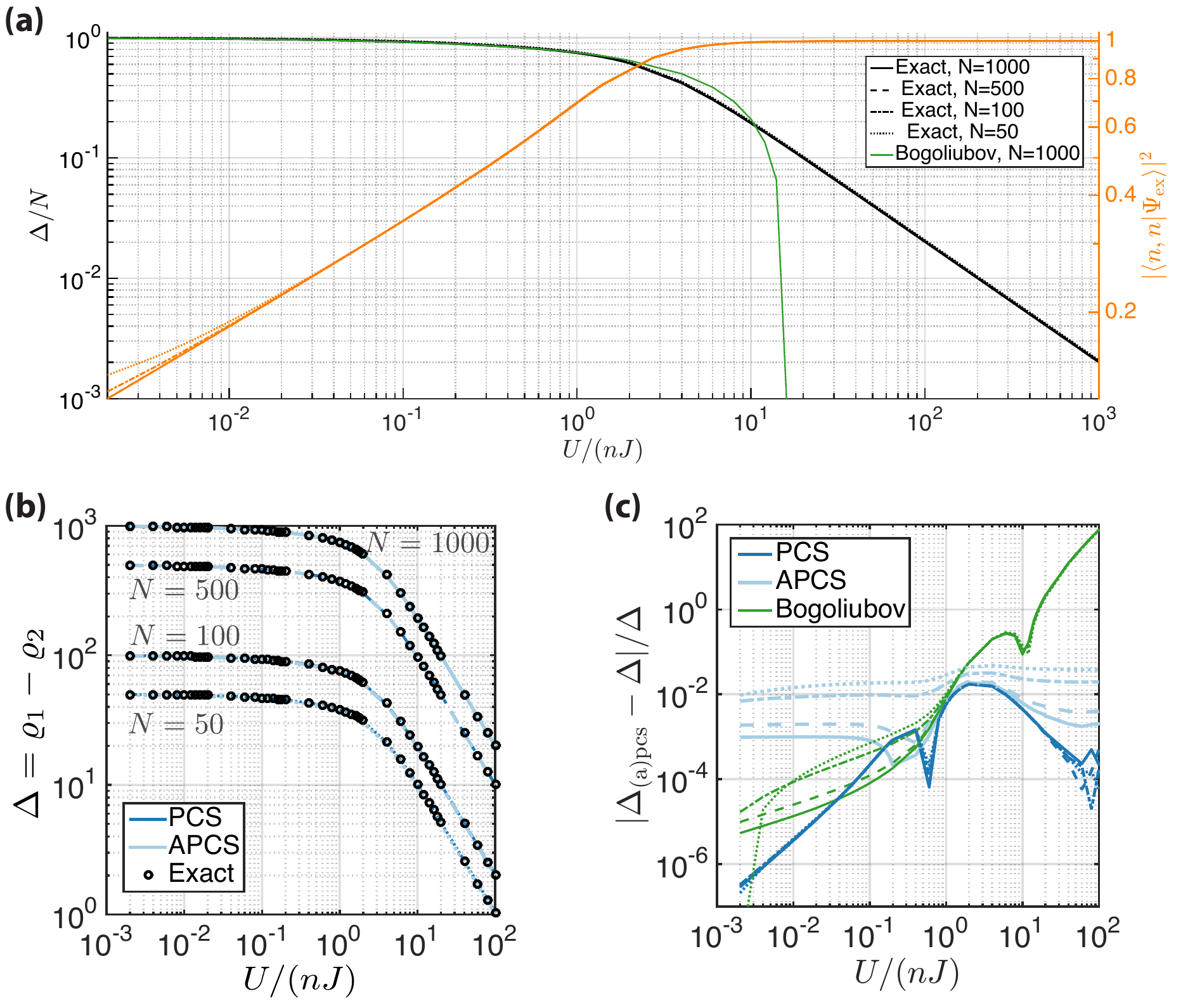}
% 	\subfloat[]{
% 	 \includegraphics[width=.85\textwidth]{diff_lam.pdf}
% 	\label{fig:diff_lam}
% 	}	
% 	\vspace{-2em}
% 	\subfloat[]{
% 	\includegraphics[width=.4\linewidth]{diff_lam_pcs.pdf}
% 	\label{fig:diff_lam_pcs}
% 	}		
% 	\hspace{.5em}		
% 	\subfloat[]{
% 	\includegraphics[width=.4\linewidth]{diff_lam_pcs_err.pdf}	
% 	\label{fig:diff_lam_pcs_err}
% 	}
\caption{(Color online) (a)~Population imbalance $\Delta$ of the Schmidt modes for the exact ground-state solution (black lines, left axis), given by Eq.~(\ref{eq:Deltabeta}), as a function of $U/(nJ)$ and for number of bosons $N = $ 50 (dotted lines), 100 (dot-dashed lines), 500 (dashed lines), 1\,000 (solid lines).  Note that by plotting $\Delta/N$, we remove the trivial dependence on particle number coming from the 1RDM normalization and, as a result, all cases lie on top of each other.  This plot evidences the transition between the superfluid/Mott-insulator regimes of the BH model.  For weak interactions $U/(nJ)\ll 1$, the ground state corresponds to a superfluid with a near-unity occupation of a single Schmidt mode ($\Delta/N \sim \varrho_1/N \sim 1$), whose spatial wave function is symmetrically delocalized between the two sites. As interactions increase, so does the depletion of the dominant Schmidt mode; for strong interactions $U/(nJ)\gg 1$, the superfluid fragments into uncorrelated components on the two sites, with both 1RDM eigenvalues approaching the same order of magnitude (i.e., $\Delta/N \ll 1$).  In the limit $U/(nJ)\rightarrow \infty$, $\Delta/N \rightarrow 0$ and the ground state becomes the Mott-insulator state $\ket{n,n}$, a product state of $n=N/2$ particles at each site.  This is demonstrated on the right axis of~(a) by the fidelity $|\braket{n,n}{\varPsi_{\rm ex}}|^2$ (orange lines).  For reference, we also plot the population imbalance $\Delta_\mathrm{bog}$ given by the conventional Bogoliubov approximation for $N=1\,000$ [thin, green (gray) line, left axis] (see App.~\ref{sec:bogoliubov_approximation_to_the_two_site_bh_model} for details).  As long as interactions remain weak, i.e., $U/(nJ)\alt 1$, the Bogoliubov approximation provides a good description of the condensate depletion.  In the transition region from a superfluid to a Mott insulator, $U/(nJ)\sim1$, the Bogoliubov approximation breaks down, and as interactions become sufficiently strong and the system transitions to a Mott insulator, the Bogoliubov approximation fails entirely, unable even to proceed past $U/(nJ)\simeq64$.  (b)~Comparison of the exact population imbalance $\Delta$ (circles) against the PCS and APCS predictions~(\ref{eq:occu_diff_s}) and~(\ref{eq:occu_diff_s_F}) [blue and light blue (gray and light gray) lines] for $N=$ 50 (dotted lines), 100 (dot-dashed lines), 500 (dashed lines), 1\,000 (solid lines).  These plots confirm that the ground state is close to a single condensate when $U/(nJ)\alt1/n$.  (c)~Relative error between the exact expressions and the PCS and APCS predictions [blue and light blue (gray and light gray) lines] for $N=50, 100, 500,$ and $1\,000$; plotted for comparison are the Bogoliubov predictions $\Delta_\mathrm{bog}$ of Eq.~(\ref{eq:Delta_bog_2}) for the same values of $N$ [thin, green (gray) lines].  The performance of the exact PCS approximation is remarkably good, nearly matching the exact solution over the whole parameter space and for all particle numbers considered.  As expected, the error of the exact PCS goes to zero away from the transition region and increases for $U/(nJ)\sim 1$.  In this transition region, the APCS nearly matches the exact PCS, with the accuracy of the large-$N$ approximation increasing with $N$.  The fluctuating relative errors of the exact PCS in the Mott-insulator regime come from the numerical minimization of Eq.~(\ref{eq:EPCSexact}), whose minimum gets shallower with increasing $U/(nJ)$.  The Bogoliubov approximation breaks down through the transition region and fails entirely in the Mott-insulator regime.  For $U/(nJ)\alt 10^{-1}$, the exact PCS approximation is generally more accurate than the Bogoliubov approximation, whereas APCS is less accurate, reflecting the fact that APCS cannot model precisely a single-mode condensate.}
\label{fig:rho1}
\end{figure}

Since $\tr\rho^{(1)} = N$, Eq.~(\ref{1RDM_BH_Sch}) is uniquely defined by the population imbalance of the Schmidt modes,
\begin{align}\label{eq:Deltabeta}
\Delta = \varrho_1 - \varrho_2=2\exrho^{(1)}_{12}=\frac{2\beta}{N-1}\;,
\end{align}
which we plot in Fig.~\ref{fig:rho1}(a) for the exact ground state as a function of $U/(nJ)$.  It is important to note that $\Delta$ contains the same information as the off-site correlation function $\exrho^{(1)}_{12}$.  For weak interactions, $U/(nJ)\ll 1$, the ground state of the BH model is a superfluid with almost all particles occupying the (condensate) Schmidt mode 1 ($\Delta \sim \varrho_1\sim N$), which is symmetrically delocalized between the two sites.  As $U/(nJ)$ increases, the dominant Schmidt mode becomes depleted, and the transition to a Mott insulator occurs around $U/(nJ)\sim 1$.  For strong interactions, $U/(nJ)\gg 1$, the two Schmidt modes are nearly equally occupied, i.e., $\Delta\sim0$, and the superfluid fragments into two uncorrelated components at the two sites.  In the limit $U/(nJ)\rightarrow \infty$, the ground state becomes the Mott-insulator state $\ket{n,n}$, i.e., a product state of $n$ particles at each site, as shown by the near-unity fidelity $|\braket{n,n}{\varPsi_{\rm ex}}|^2$ in Fig.~\ref{fig:rho1}(a) (right axis).

Using the transformation~(\ref{Sch_transf}), we can write the Hamiltonian~(\ref{BH_hamil}) as
\begin{align}\label{eq:tbh_schmidt}
\begin{split}
 \sH_\mathrm{tbh} &= -J\ssp (\a_1^\dagger \a_1 - \a_2^\dagger \a_2) + \frac{U}{4}\ssp \Big(\a_1^\dagger \a_1^\dagger \a_1\a_1+ \a_2^\dagger \a_2^\dagger \a_2\a_2+4\a_1^\dagger \a_2^\dagger \a_2\a_1+e^{-2i\mu}\a_1^\dagger \a_1^\dagger \a_2\a_2+e^{2i\mu}\a_2^\dagger \a_2^\dagger \a_1\a_1\Big)\\
 &= -J\ssp (\a_1^\dagger \a_1 - \a_2^\dagger \a_2) + \frac{U}{4}\ssp \Big(2\a_1^\dagger \a_2^\dagger \a_2\a_1+e^{-2i\mu}\a_1^\dagger \a_1^\dagger \a_2\a_2+e^{2i\mu}\a_2^\dagger \a_2^\dagger \a_1\a_1\Big)+\frac{N(N-1)U}{4}\;.
 \end{split}
\end{align}
In this basis the expectation value of the ground-state energy takes the form
\begin{align}\label{eq:pcs_energy_initial}
E &= -J\big(\rho^{(1)}_{11}-\rho^{(1)}_{22}\big)
+ \frac{U}{4} \Big(2\rho^{(2)}_{12,\,12}+e^{-2i\mu}\rho^{(2)}_{22,\,11}+e^{2i\mu}\rho^{(2)}_{11,\,22}\Big)+ \frac{N(N-1)U}{4}\;,
\end{align}
where $\rho^{(2)}_{k_1 k_2,\, j_1 j_2} = \brab{\varPsi_\mathrm{ex}}\,\a_{j_1}^\dagger \a_{j_2}^\dagger \a_{k_2} \a_{k_1}\, \ketb{\varPsi_\mathrm{ex}}$.  We remind the reader that the subscripts on $\rho$ index Schmidt orbitals, as in all previous sections of the paper, and not sites, as in the subscripts on $\exrho$.  Now we use the \PCS\ \emph{Ansatz\/} to determine a particular Schmidt basis.  By taking the expectation value of Eq.~(\ref{eq:tbh_schmidt}) over the \PCS~\emph{Ansatz}, one obtains an expression of the same form as Eq.~(\ref{eq:pcs_energy_initial}), but with the values of the matrix elements given by the \PCS~formalism. Because the Schmidt-basis matrix elements $\rho^{(2)}_{22,\,11}$ and $\rho^{(2)}_{11,\,22}$ in the \PCS~formalism are always nonnegative [see discussion following Eq.~(\ref{eq:two_particle_rdm_diagonal})], the \PCS\ ground-state energy is minimized, as promised, by the phase choice $e^{i\mu}=i$.  This specifies the Schmidt basis for the \PCS~formalism.  In the Schmidt basis, the ground-state energy then has the form
\begin{align}\label{eq:pcs_energy}
E &= -J\big(\rho^{(1)}_{11}-\rho^{(1)}_{22}\big)
+ \frac{U}{4} \Big(2\rho^{(2)}_{12,\,12}-\rho^{(2)}_{22,\,11}-\rho^{(2)}_{11,\,22}\Big)+ \frac{N(N-1)U}{4}\;,
\end{align}
The transition from superfluid to Mott insulator is described by the competition between the first and second terms on the right-hand side of Eq.~(\ref{eq:pcs_energy}).  In the Schmidt basis, the two-site Bose-Hubbard Hamiltonian in the Bogoliubov approximation now becomes
\begin{align}\label{eq:tbh_schmidt_final}
 \sH_\mathrm{tbh} &= -J\ssp (\a_1^\dagger \a_1 - \a_2^\dagger \a_2) + \frac{U}{4}\ssp \Big(2\a_1^\dagger \a_2^\dagger \a_2\a_1-\a_1^\dagger \a_1^\dagger \a_2\a_2-\a_2^\dagger \a_2^\dagger \a_1\a_1\Big)+\frac{N(N-1)U}{4}\;.
\end{align}

We can transform the 2RDM~(\ref{eq:exact_2rdm_BH}) to the Schmidt basis using the transformation~(\ref{Sch_transf}) with $e^{i\mu}=i$, obtaining
\begin{align}\label{eq:exact_2rdm_Sch}
 \rho^{(2)}=
 \begin{pmatrix}
  \alpha+\beta-\gamma &0        &0         & \gamma+\delta\\
  0              &\gamma & \gamma  & 0 \\
  0              &\gamma & \gamma & 0 \\
  \gamma+\delta        & 0       &0         & \alpha-\beta-\gamma
 \end{pmatrix}\;.
\end{align}
The exact ground-state 2RDM~(\ref{eq:exact_2rdm_Sch}) has the same general form~(\ref{eq:two_particle_rdm_b}) as a PCS state.  The only use of the PCS \emph{Ansatz\/} in deriving the ground-state energy~$(\ref{eq:pcs_energy_initial})$ and the 2RDM~$(\ref{eq:exact_2rdm_Sch})$ is to determine a particular Schmidt basis.  Once that determination is made, the ground-state energy and the 2RDM have the given forms, both for the exact results and for the PCS \emph{Ansatz}; what changes between the two are the values of the matrix elements in the expressions.  The four parameters in the 2RDM are given by Eqs.~(\ref{eq:alpha}) and~(\ref{eq:Deltabeta}) and by
\begin{subequations}\label{eq:gammadelta}
\begin{align}
\gamma&=\rho_{12,12}^{(2)}=\frac{N(N-1)}{4}-\frac{N^2}{8}\big(C_{12}^{(2)}+G_{12}^{(2)}\big)\;,\\
\delta&=\rho_{11,22}^{(2)}-\rho_{12,12}^{(2)}=-\frac{N^2}{2}\big(1-C_{12}^{(2)}\big)+\frac{N}{2}\;.
\end{align}
\end{subequations}

For the present case of Schmidt rank $\rank=2$, we have calculated the PCS normalization factor exactly in Eq.~(\ref{NFB_hypergeo}), and we use that expression to derive exact PCS RDMs in App.~\ref{sec:pcs_2rdm_for_schmidt_rank_two}.  In App.~\ref{app:Iterative_Relations}, we show how to evaluate the PCS 1RDM for $\rank\geq 2$ in time $\bigO(n)$ based on a simple iterative method, but we do not use that method here.  We center our discussion in this section around the use of the large-$N$ results derived in the previous sections to determine the RDMs approximately.  We relegate the calculation of the exact PCS to App.~\ref{sec:pcs_2rdm_for_schmidt_rank_two}, but include its predictions in our analysis in order to investigate how well the PCS \emph{Ansatz\/} does before any approximations are made to it.  To avoid confusion when presenting both predictions, we refer for the remainder of this section---and in Fig.~\ref{fig:rho1} and App.~\ref{sec:pcs_2rdm_for_schmidt_rank_two}---to the approximate \PCS~\emph{Ansatz\/} as APCS and reserve~\PCS\ for results of the PCS~\emph{Ansatz\/} without approximations.

We expect the exact PCS to do well for weak interactions, $U/(nJ)\ll 1$, where we get a single condensate by choosing Schmidt coefficients $\lambda_1=1$ and $\lambda_2=0$; the resulting pair creation operator, $\sA^\dagger=(\a_1^\dagger)^2=(\b_1^\dagger+\b_2^\dagger)^2/2$, creates pairs of particles in the mode $\a_1$.   We also expect the exact PCS to do well for strong interactions, $U/(nJ)\gg 1$, where we get $n$ particles at each site by choosing $\lambda_1^2=\lambda_2^2=1/2$; the resulting pair creation operator, $\sA^\dagger=[(\a_1^\dagger)^2+(\a_2^\dagger)^2]/\sqrt2=\sqrt{2}\b_1^\dagger\b_2^\dagger$, creates particles pairwise at the two sites.  The purpose of the numerics for the exact PCS is thus to test how well the PCS~\emph{Ansatz\/} does in the transition region, $U/(nJ)\sim1$, and what we find below is that it does a credible job throughout all of $U/(nJ)$ parameter space, even for relatively small particle numbers.

The situation is different for the approximate PCS.  Recall from our previous discussions of two Schmidt modes that all physical quantities in the APCS are functions of the parameter $\pcss_-=(\pcss_1-\pcss_2)/2$ and that the approximations made in the large-$N$ limit require that $\pcss_-\ll n$.  The purpose of the APCS predictions is really to see how well the APCS approximates the exact \PCS.  In the transition region between a superfluid and a Mott insulator, $U/(nJ)\sim1$, we expect $\pcss_-\sim 1$, and in the strong-interaction Mott regime, $U/(nJ)\gg1$, we expect $\pcss_-\ll1$, so throughout these two regimes, we expect the APCS to match the exact PCS well even for fairly small particle numbers.  In the mean-field GP regime of very weak interactions, $(nU)/J\agt1$, the APCS attempts to model a single condensate by having $\pcss_-$ formally become of order $N$ or larger; this, however, contradicts the definition of \PCS\ parameters in Eq.~(\ref{eq:lambda_zeta}).  We cannot expect the APCS predictions to do well or even to be physical when $\pcss_-$ gets as big as or bigger than $n$.  The reason for this problem, of course, is that the approximations made in the APCS are inconsistent with giving an accurate description of a single condensate mode.  What we find below, however, is that there is an intermediate regime of weak interactions, $U/J\sim1$, where $\pcss_-\sim\sqrt N$, and in this regime, the APCS does a credible job of modeling a single-mode condensate, with corrections of order $1/\sqrt N$.  The numerics are roughly consistent with all of this, except that the APCS does worse than expected in the strong-interaction Mott regime, for reasons that we discuss further below.

Comparison of the 2RDM~(\ref{eq:exact_2rdm_Sch}) with the large-$N$ approximation~(\ref{eq:two_particle_rdm_matrix_form}) gives the following APCS values:
\begin{subequations}\label{eq:APCSparameters}
\begin{align}
\alpha_\mathrm{apcs}&=2n^2+\bigO(N)\;,\\
\beta_\mathrm{apcs}&=2n^2\chi_1(\pcss_-)+\bigO(N)\;,\\
\gamma_\mathrm{apcs}&=\textstyle{\frac12}n^2\big[1-\chi_2(\pcss_-)\big]+\bigO(N)\;,\\
\delta_\mathrm{apcs}&=0+\bigO(N)\;.\label{eq:APCSdelta}
\end{align}
\end{subequations}
Here $\pcss_-=(\pcss_1-\pcss_2)/2$ is the APCS parameter, and $\chi_j(\pcss_-)=I_j(\pcss_-)/I_0(\pcss_-)$, with $\BesselI_j$ denoting the $j$th modified Bessel function.

To find the APCS predictions for the ground-state energy, however, we need to go beyond the parameters~(\ref{eq:APCSparameters}).  A glance at the 2RDM~(\ref{eq:exact_2rdm_Sch}) and at the ground-state energy~(\ref{eq:pcs_energy}) shows that if $\delta=0$, then there is no competition between tunneling and the on-site interaction, so we do not capture the transition from a single-mode condensate to a Mott insulator.  To model this transition, we hybridize the large-$N$ results with the exact PCS relations derived in Sec.~\ref{sec:2rdms}.  We start by using Eq.~(\ref{eq:difference_xi}) to get a more accurate value of $\delta$,
\begin{align}\label{eq:rho2_terms_pcs_energy}
-2\delta=2\rho^{(2)}_{12,\,12}- \rho^{(2)}_{22,\,11}-\rho^{(2)}_{11,\,22} = -N\,
+\frac{\pcss_-}{N}
\big(\rho^{(1)}_{11}-\rho^{(1)}_{22}\big)+\bigO(1/N)\;.
\end{align}
This then reduces the energy~(\ref{eq:pcs_energy}) to
\begin{align}\label{eq:pcs_energy_b}
E_\mathrm{apcs} &= \bigg(\frac{Us_-}{4N}-J\bigg)\big(\rho^{(1)}_{11}-\rho^{(1)}_{22}\big)
+  \frac{N(N-2)U}{4}+\bigO(U/N)\;.
\end{align}

We reiterate here some terminology that we employed in previous sections.  The large-$N$ approximation that we developed in Sec.~\ref{sec:diagonal_large_N} is a generalization of the mean-field description to multimode condensates.  Corrections to the large-$N$ approximation, such as those derived for exact 2RDMs in Sec.~\ref{sec:2rdms}, occur as increasing powers of $1/N$.  Thus, as we pointed out in Sec.~\ref{sec:2rdms}, we refer to the first term, $-N$, on the right-hand side of Eq.~(\ref{eq:rho2_terms_pcs_energy}) as a Bogoliubov-order correction to the large-$N$ 2RDM and the second term as a post-Bogoliubov correction.  It is the post-Bogoliubov correction that gives rise to the quantity $U\pcss_-/(4N)$ in Eq.~(\ref{eq:pcs_energy_b}).  In the transition region, where the on-site interaction competes with the tunneling splitting, the competition is expressed by the two quantities $U\pcss_-/(4N)$ and $J$, which are of the same order in the transition region.  This stands in marked contrast with the mean-field GP regime of a single condensate mode, which is nominally defined by $(nU)/J\ll 1$ [$U/(nJ)\ll1/n^2$]; i.e., the magnitude of the tunneling splitting, $nJ$, far exceeds the on-site interaction energy, $Un^2$.  Because the Bogoliubov correction in Eq.~(\ref{eq:rho2_terms_pcs_energy}) is a constant, however, the single-condensate behavior of the mean-field regime extends, at least approximately, to values of $J$ one order smaller in $1/N$, i.e., to the intermediate regime of weak interaction strength, $U/J\sim1$, as is evident from the plots in Fig.~\ref{fig:rho1}(a).

Using the relations~(\ref{eq:upsilon_s=2_1rdm_a}) and (\ref{eq:upsilon_s=2_1rdm_b}), we have
\begin{align}\label{eq:occu_diff_s}
\Delta_\mathrm{apcs}=\rho^{(1)}_{11}-\rho^{(1)}_{22} = N\chi_1(\pcss_-)+\bigO(1)\;.
\end{align}
The APCS energy can thus be written as a function of the parameter $\pcss_-$ alone,
\begin{align}\label{eq:pcs_energy_c}
E_\mathrm{apcs} &= \bigg(\frac{U\pcss_-}{4}-JN\bigg)\chi_1(\pcss_-)
+\frac{N(N-2)U}{4}+ \bigO(U/N)\;.
\end{align}
The APCS ground-state energy for the two-site Bose-Hubbard model is found by minimizing this expression with respect to $\pcss_-$; the optimal value of $\pcss_-$ is determined by
\begin{align}\label{eq:mod_PCS_grd}
\chi_1( \pcss_-)-2\Big(1+\chi_2( \pcss_-)-2\ssp\chi_1^2( \pcss_-)\Big) \Big(\frac{NJ}{U}-\frac{\pcss_-}{4}\Big)=0\;.
\end{align}
The optimal value of $\pcss_-$ (solved numerically) is then inserted into the parameters~(\ref{eq:APCSparameters}) to give the APCS 1RDM and 2RDM and into Eqs.~(\ref{eq:occu_diff_s}) and~(\ref{eq:pcs_energy_c}) to find the APCS population difference and ground-state energy.  We stress that Eq.~(\ref{eq:rho2_terms_pcs_energy}) has a precision of $\bigO(1/N)$ and Eq.~(\ref{eq:pcs_energy_c}) has a precision of $\bigO(1)$, both two orders beyond the precision of the large-$N$ approximation discussed in Sec.~\ref{sec:diagonal_large_N} and one order beyond the Bogoliubov corrections.  The ability of the \PCS~\emph{Ansatz\/} to include selectively such crucial post-Bogoliubov corrections is the primary strength of the formalism, but does indicate that using the formalism requires care if one wants to find physically meaningful results.

We can use the large-argument asymptotic expansions of the modified Bessel functions to find that for $\pcss_-\gg1$, the ground-state value determined by Eq.~(\ref{eq:mod_PCS_grd}) is
\begin{align}\label{eq:sminusintermediate}
\pcss_-=\sqrt{\frac{2NJ}{U}}+\bigO\bigg(\sqrt{\frac{U}{NJ}}\bigg)\;.
\end{align}
In the single-condensate, mean-field regime, where $(Un)/J\alt1$, this gives $\pcss_-\agt n$, which lies outside the region of validity of the large-$N$ approximation, reflecting the expected fact that the large-$N$ approximation cannot accommodate a single-mode condensate.  In the transition region, where $U/(nJ)\sim1$, this gives $\pcss_-\sim 1$, as expected, but meaning that the asymptotic expansions used to find Eq.~(\ref{eq:sminusintermediate}) are not valid.  The sweet spot for Eq.~(\ref{eq:sminusintermediate}) is in the intermediate regime of weak interaction strength, where $U/J\sim1$, which gives $\pcss_-\sim\sqrt N$; the population difference and ground-state energy in this intermediate regime are
\begin{align}\label{eq:Deltaapcs_large_s}
\Delta_\mathrm{apcs}&=N\Bigg(1-\sqrt{\frac{U}{8NJ}}+\bigO\bigg(\frac{U}{NJ}\bigg)\Bigg)\;,\\
E_\mathrm{apcs}&=-JN\bigg(1-\sqrt{\frac{U}{2NJ}}\;\bigg)+\frac{N(N-2)U}{4}+\bigO(U)\;.
\label{eq:Eapcs_large_s}
\end{align}
both of which are close to the values for a single-mode condensate.  Notice that the limits of validity of the approximate APCS predictions~(\ref{eq:Deltaapcs_large_s}) and~(\ref{eq:Eapcs_large_s}), i.e., $1\ll\pcss_-=\sqrt{2NJ/U}\ll N$, are equivalent to being in the intermediate regime of weak interaction strength, $1/N\ll U/J\ll N$.

We can also investigate the behavior of the APCS in the Mott regime by considering the power-series expansions of the modified Bessel functions for small $\pcss_-$.  What one obtains are corrections to the Mott population difference and the Mott ground-state energy that are nominally as small as or smaller than the terms neglected in Eqs.~(\ref{eq:occu_diff_s}) and~(\ref{eq:pcs_energy_c}).  This suggests that though the APCS gives an essentially correct description of the Mott regime for large $N$, it is missing details that are contained in the exact ground state and in the exact PCS ground state; the plots in Fig.~\ref{fig:rho1}(c) are consistent with this suggestion.  Moreover,
the difficulty in finding the APCS ground-state zero of Eq.~(\ref{eq:pcs_energy_c}) is a hint that it might be difficult to find the exact PCS ground-state energy in the Mott-insulator regime from Eq.~(\ref{eq:EPCSexact}).  Indeed, we find that Eq.~(\ref{eq:EPCSexact}) has a very shallow minimum in the Mott regime; this accounts for the noisy data for the exact PCS in the Mott-insulator regime in Fig.~\ref{fig:rho1}(c).

In Fig.~\ref{fig:rho1}(b), we plot the APCS population difference~$\Delta_\mathrm{apcs}$ from Eq.~(\ref{eq:occu_diff_s}), the exact PCS population difference $\Delta_\mathrm{pcs}$ from Eq.~(\ref{eq:occu_diff_s_F}), and also the exact solution $\Delta$ as functions of $U/(nJ)$ for $N=50, 100, 500, 1\,000$. The corresponding relative error is shown in Fig.~\ref{fig:rho1}(c), which gives a first indication of the accuracy of the PCS formalism.  The agreement of the PCS approximation is remarkably good over the whole parameter space and for all particle numbers considered.  In fact, the APCS gives good predictions across the entire parameter space, improving with increasing $N$, but fairly good even for $N=50$.  In the transition region, the APCS matches the exact PCS for $N\agt500$, but is not as good as the exact PCS, for any value of $N$, in the regimes of weak or strong interaction strength~$U$.

\begin{figure}[htbp]
  \centering
    \includegraphics[width=\textwidth]{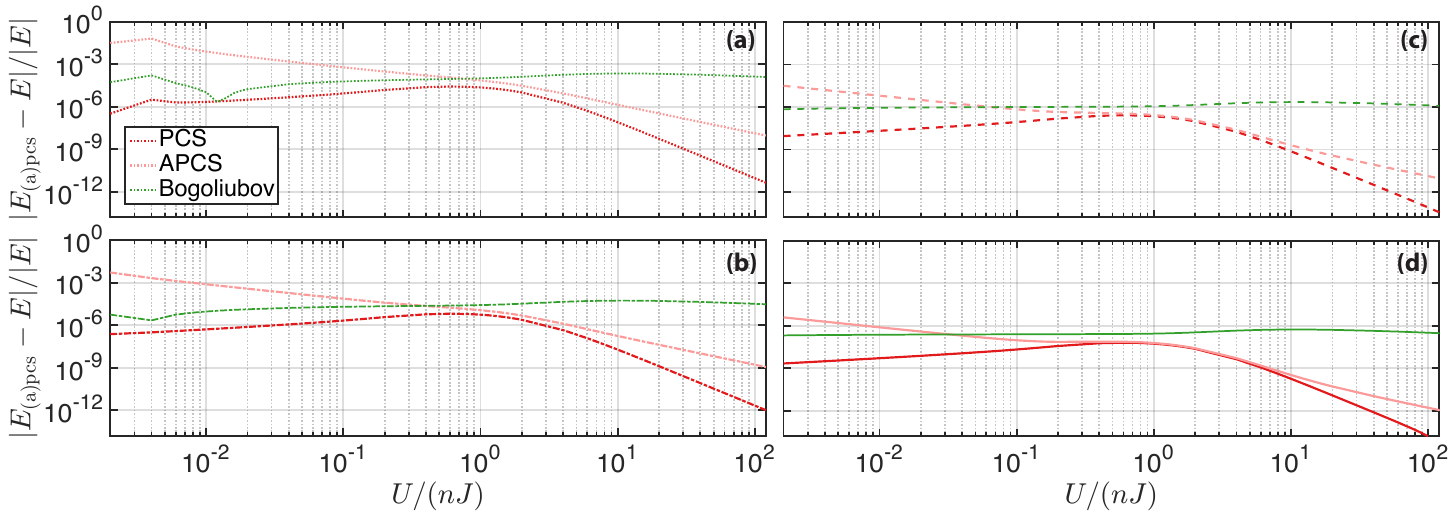}
  \caption{(Color online) Relative error of the PCS [dark red (dark gray) lines] and APCS [light coral (light gray) lines] approximations to the exact ground-state energy of the two-site BH model as a function of $U/(nJ)$, for (a)~$N = $ 50 (dotted lines), (b)~100 (dot-dashed lines), (c)~500 (dashed lines), and (d)~1\,000 (solid lines).  Plotted for comparison are the relative errors of the energy of the Bogoliubov ground state [green (gray) lines], calculated using Eq.~(\ref{eq:E_bog}).  The agreement of the (A)PCS approximation is very good over the whole parameter space, with the accuracy of the approximation increasing with $N$.  As $U/(nJ)$ becomes smaller than 1, the accuracy of the APCS energy is degraded, reflecting the inability of the large-$N$ approximation to describe a single-mode condensate. For $U/(nJ)\gtrsim 1$, the APCS predictions are quite accurate even for $N=50$.  The exact PCS prediction is significantly more accurate than the conventional Bogoliubov approximation for all parameters considered, providing a compelling demonstration of the potential of the PCS formalism.  Not only does it outperform the Bogoliubov approximation for weak interactions, it also provides a means for an accurate description across the transition into the Mott-insulator phase.  The APCS prediction, on the other hand, does not outperform the Bogoliubov approximation in the weak-interaction regime, because the approximations made for the APCS are inconsistent with giving an accurate description of a single condensate mode.  Yet, around $U/(nJ)\sim 1$, APCS becomes more accurate than Bogoliubov and is able to capture successfully the transition to the Mott-insulator phase.}
  \label{fig:gs_energy}
\end{figure}

\begin{figure}[htbp]
  \centering
    \includegraphics[width=.9\textwidth]{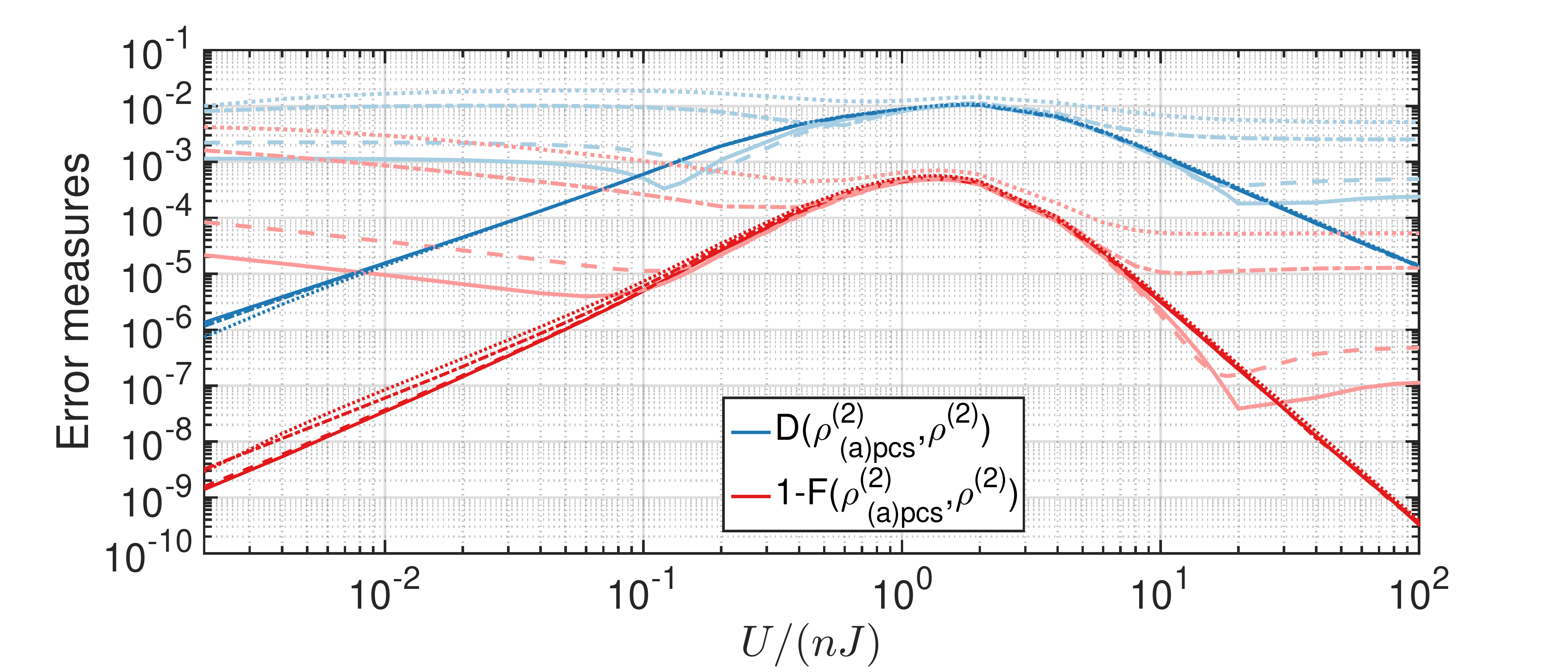}
  \caption{(Color online) Measures of the error of the PCS (darker colors) and APCS (lighter colors) 2RDMs, $\rho_\mathrm{pcs}^{(2)}$ and $\rho_\mathrm{apcs}^{(2)}$, relative to the exact 2RDM $\rho^{(2)}$, as quantified by the trace distance~(\ref{trace_dist_2RDM}) [blue (upper) lines] and infidelity~(\ref{infidelity_2RDM}) [red (lower) lines] as a function of $U/(nJ)$, for $N = $ 50 (dotted lines), 100 (dot-dashed lines), 500 (dashed lines), 1\,000 (solid lines). Both measures vanish if and only if the 2RDMs being compared are equal. A nonzero value quantifies the overall error (bounded by 1) of the (A)PCS 2RDM.  Both error metrics reveal a very good agreement between $\rho_\mathrm{(a)pcs}^{(2)}$ and $\rho^{(2)}$ over the entire parameter space, from small ($N=50$) to large ($N=1000$) particle numbers.  For $\rho^{(2)}_\mathrm{pcs}$, the agreement is nearly exact in the limits $U/(nJ)\rightarrow 0$ and $U/(nJ)\rightarrow \infty$. The PCS accuracy is worse near the phase transition, $U/(nJ)\sim 1$, where the trace distance (infidelity) shows an imprecision of the order of $10^{-2}$ ($10^{-3}$).  The agreement between $\rho_\mathrm{apcs}^{(2)}$ and $\rho^{(2)}$ is nearly as good near the transition, but saturates to a nonzero value for $U/(nJ)\ll 1$ and $U/(nJ)\gg 1$.  Not surprisingly, the accuracy of the APCS approximation increases with increasing particle number.
  }
  \label{fig:error_measures}
\end{figure}

In Fig.~\ref{fig:gs_energy}, we compare the APCS ground-state energy~(\ref{eq:pcs_energy_c}) and the PCS ground-state energy~(\ref{eq:EPCSexact}) to the exact ground-state energy, by computing their relative error as a function of $U/(nJ)$, for $N=50, 100, 500, 1\,000$. The error of the PCS approximation is impressively small for all parameters considered.  Note that the largest discrepancy (of only a few percent) of the APCS approximation happens for $U/(nJ)\sim0$ and $N=50$, a regime where we do not expect the APCS to do well, because it cannot describe a single-mode condensate, and where, in any case, the ground state, as a single-mode condensate, is trivial and the PCS \emph{Ansatz\/} is unnecessary.  Just as for the population imbalance, the APCS matches the exact PCS in the transition region for $N\agt500$, but is not as good as the exact PCS, for any value of $N$, in the regimes of weak or strong interaction strength~$U$.

\begin{figure}[htbp]
  \centering
	% \subcaptionbox*{\label{fig:beta}}{\includegraphics[width=.3\textwidth]{beta.pdf}}\hspace{0em}
	\subfloat[]{\label{fig:gamma} \includegraphics[width=.43\textwidth]{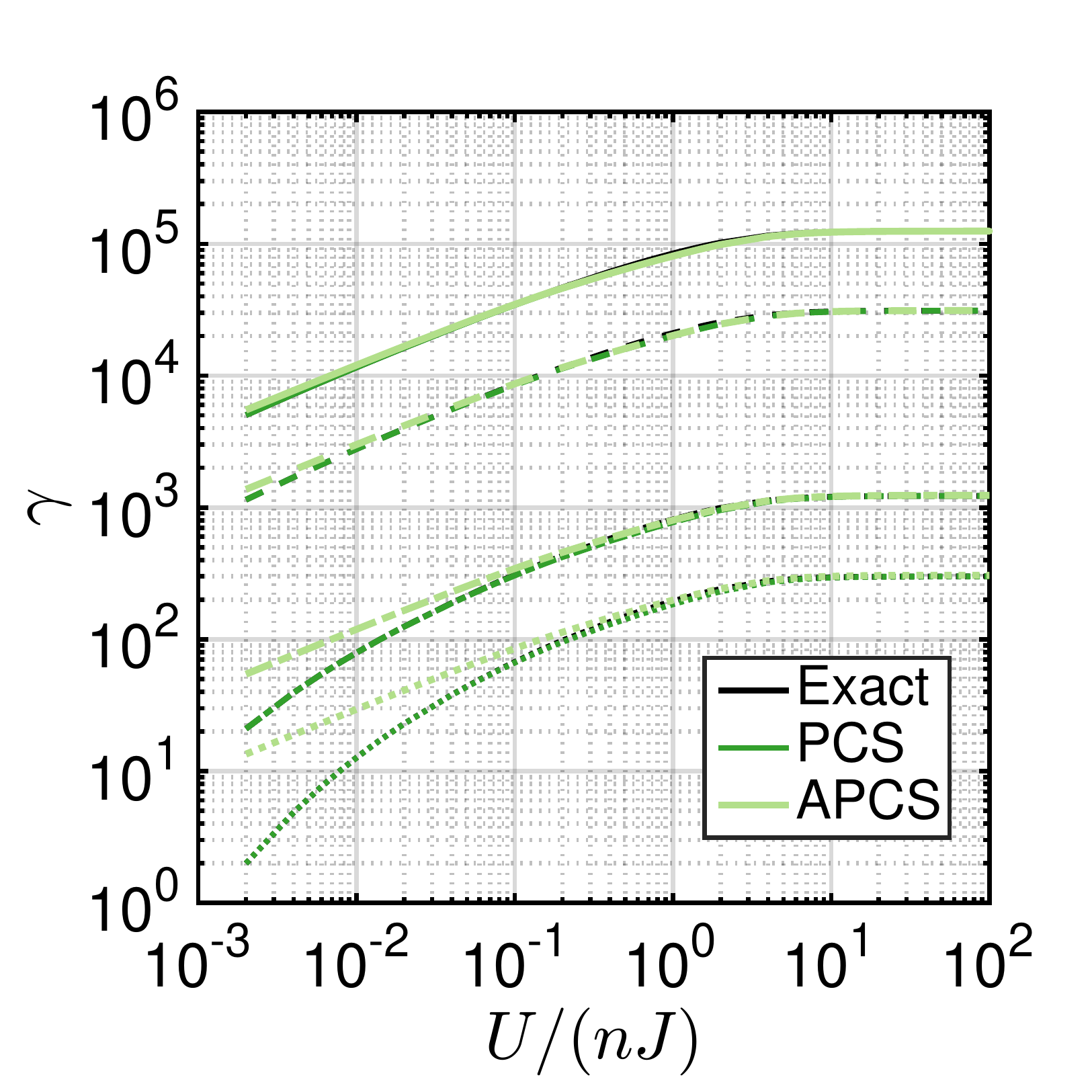}}\hspace{.7em}
  	\subfloat[]{\label{fig:delta} \includegraphics[width=.43\textwidth]{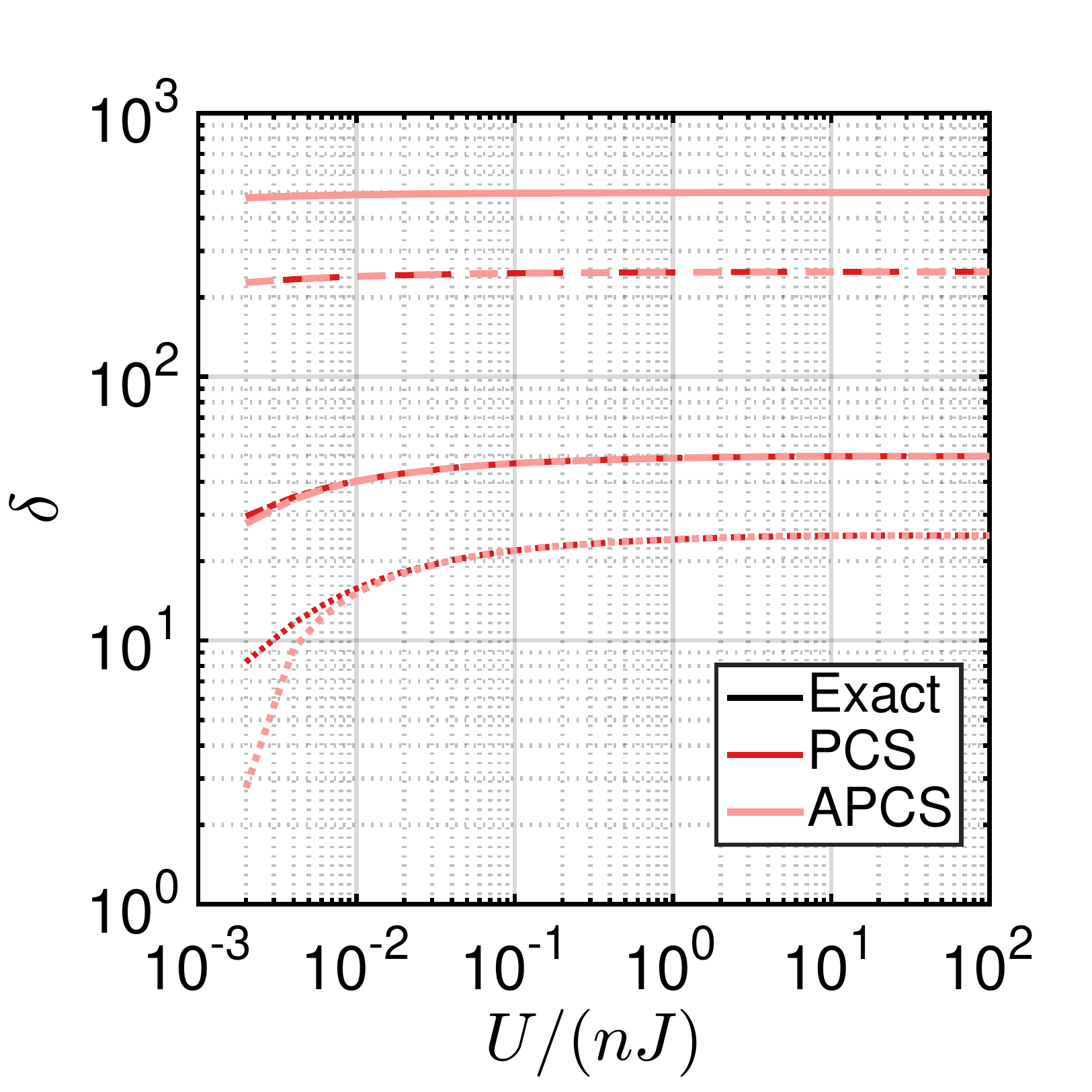}}\vspace{-.3em}\\
	% \subcaptionbox*{\label{fig:beta_err}}{\includegraphics[width=.31\textwidth]{beta_err.pdf}}\hspace{-.3em}
	\subfloat[]{\label{fig:gamma_err} \includegraphics[width=.43\textwidth]{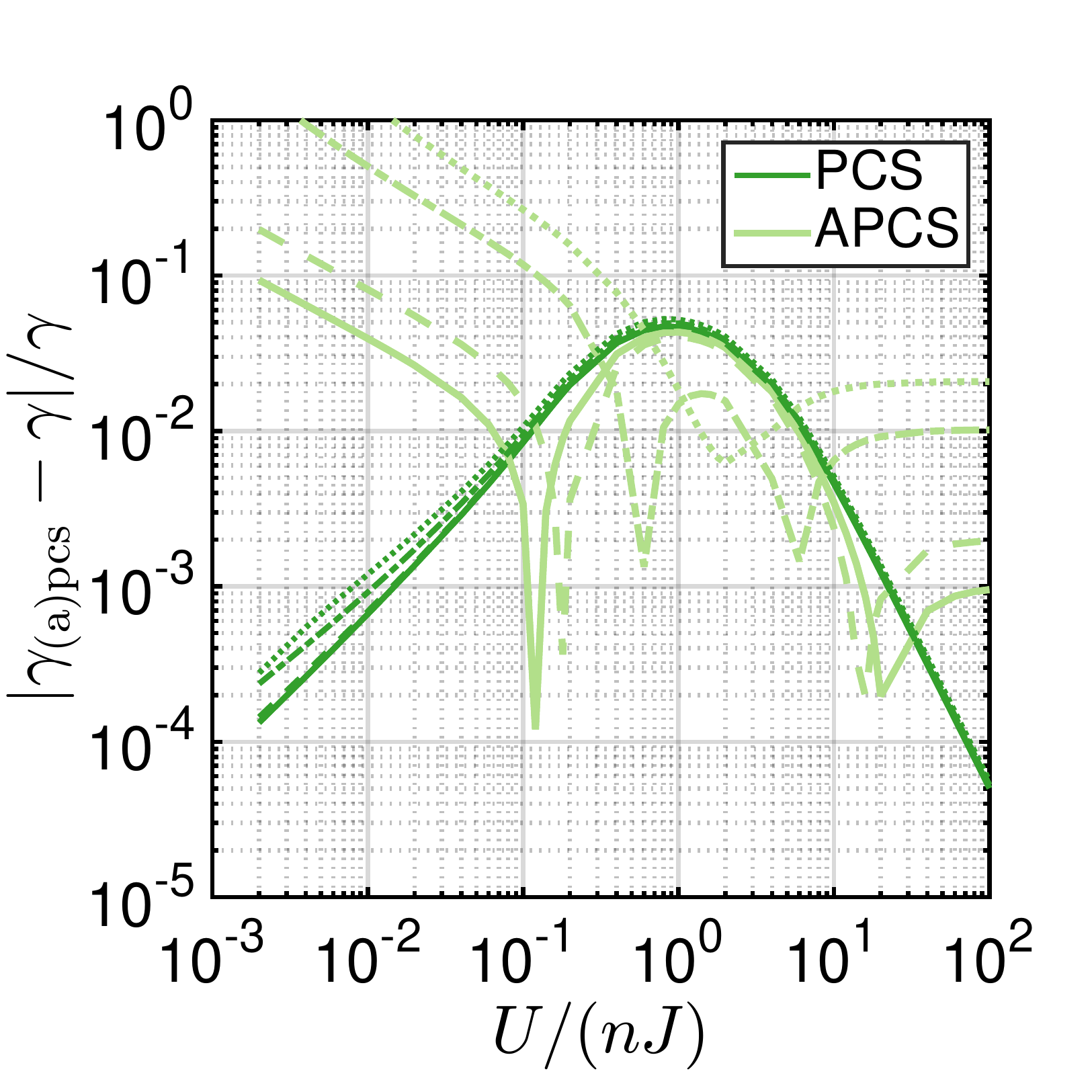}}\hspace{.7em}
  	\subfloat[]{\label{fig:delta_err} \includegraphics[width=.43\textwidth]{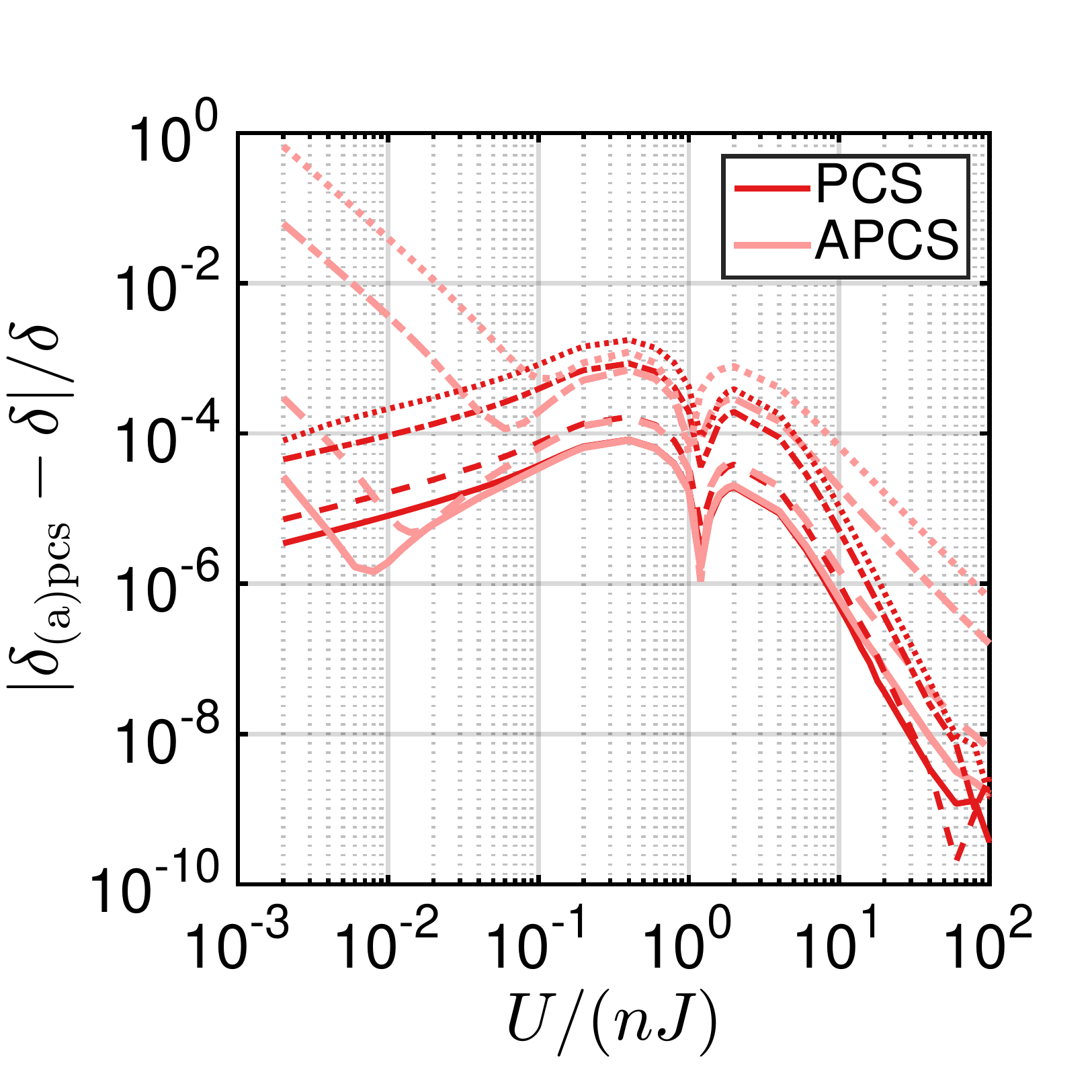}}
  \caption{(a) and (b)~Comparison of the PCS (darker colors) and APCS (lighter colors) 2RDM parameters $\gamma_\mathrm{(a)pcs}$ and $\delta_\mathrm{(a)pcs}$ to the exact ones $\gamma$ and $\delta$ [see Eqs.~(\ref{eq:gammadelta}a) and (\ref{eq:gammadelta}b)], with their respective relative errors (c) and (d), as a function of $U/(nJ)$ for $N=50~\text{(dotted)}, 100~\text{(dot-dash)}, 500~\text{(dashed)}, 1\,000~\text{(solid)}$.  The behavior here is generally consistent with that of the error measures plotted in Fig.~\ref{fig:error_measures}; the exact PCS predictions well approximate the exact 2RDM parameters, with both lines lying on top of each other. } %The error in the PCS approximation of $\gamma$ shows similar behaviour seen in the error captured by the measures~(\ref{trace_dist_2RDM}) and (\ref{infidelity_2RDM}): small for $U/(nJ)\ll 1$ and $U/(nJ)\gg 1$, and maximum when $U/(nJ)\sim 1$. Interestingly, $\gamma_{\rm apcs}\simeq \gamma_{\rm pcs}$ for all $N$ and $U/(nJ)$ values considered, thus revealing that $\gamma$ is well captured by the large-$N$ approximation, regardless of the value of $N$. The error increases when the structure of the PCS \emph{Ansatz\/} is no longer sufficient to approximate $\ket{\varPsi_{\rm ex}}$ with high precision. This in no longer true for the approximation of $\delta$, which explains the mismatch between the PCS and APCS 2RDMs for $U/(nJ)\ll 1$ and $U/(nJ)\gg 1$, also seen in Fig.~\ref{fig:gs_energy} for the ground-state energy.}
  \label{fig:params_2RDM}
\end{figure}

\begin{figure}
   \centering
   \includegraphics[width=0.75\textwidth]{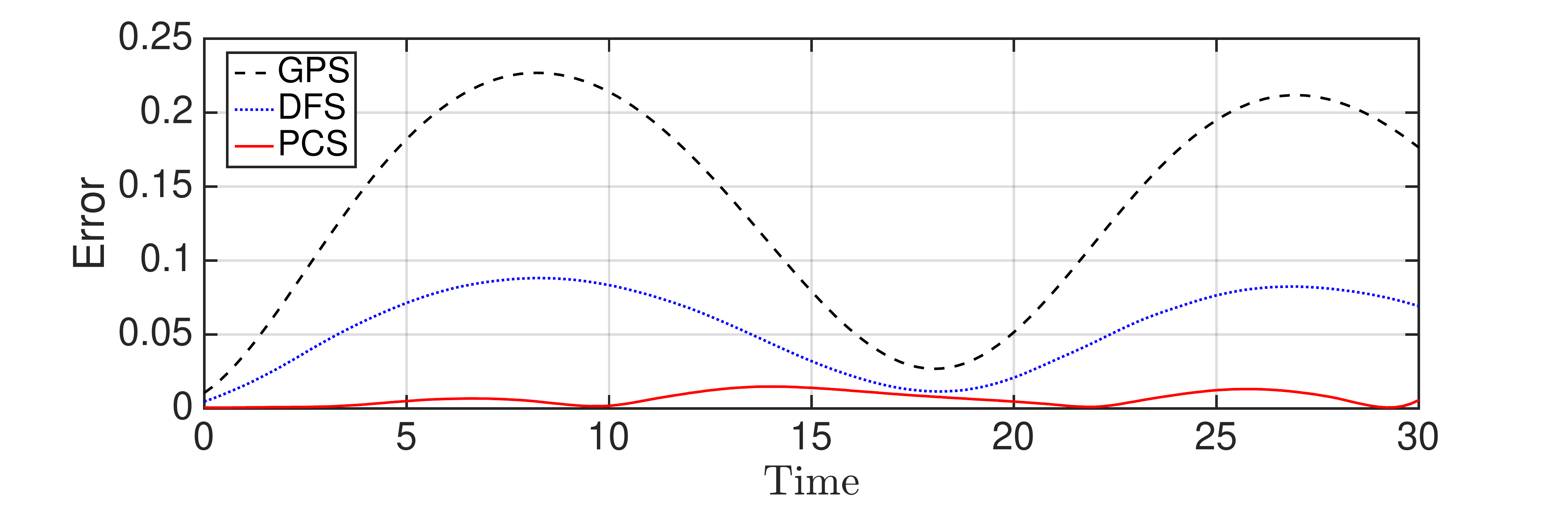}
   \caption[Errors to the 2RDMs of PCS: Time Evolution]{ Time evolution of the two-site Bose-Hubbard model of $N= 1\,000$ atoms.  Initially, all the atoms sit in the ground state of the noninteracting Hamiltonian, and then an on-site  interaction with strength $U=0.8 J$ is suddenly turned on.  The plots show the trace distances (normalized to 1) of the exact 2RDM to the closest 2RDMs given by PCS, the double-Fock state (DFS), and the Gross-Pitaevskii state (GPS) as functions of evolution time (in units of $\hbar/J$).}
   \label{fig:pcs_dfs_td_err}
\end{figure}

We turn our attention now to the PCS construction of the two-particle RDM for the ground state. In the Schmidt basis~(\ref{Sch_transf}), the exact 2RDM $\rho^{(2)}$ and the PCS 2RDM, in its exact form~$\rho_\mathrm{pcs}^{(2)}$, given in App.~\ref{sec:pcs_2rdm_for_schmidt_rank_two}, and its approximate, large-$N$ form $\rho_\mathrm{apcs}^{(2)}$, constructed from the parameters~(\ref{eq:APCSparameters}), have the general form~(\ref{eq:exact_2rdm_Sch}).  Thus we can assess the accuracy of the PCS 2RDMs by comparing them directly to the exact 2RDM.  First, to serve as overall error measures, we compute the trace distance and the infidelity between the PCS RDMs, $\rho_\mathrm{pcs}^{(2)}$ and $\rho_\mathrm{apcs}^{(2)}$, and the exact 2RDM $\rho^{(2)}$.  These measures are given by
\begin{align}
	D\!\left(\rho_\mathrm{(a)pcs}^{(2)},\rho^{(2)}\right)
&= \half\!\tr \big|\rho_\mathrm{(a)pcs}^{(2)}-\rho^{(2)}\big|\;,\label{trace_dist_2RDM}\\
	1-F\!\left(\rho_\mathrm{(a)pcs}^{(2)},\rho^{(2)}\right)
&= 1-\!\left(\tr\sqrt{ \sqrt{\rho^{(2)}}\,\rho_\mathrm{(a)pcs}^{(2)} \sqrt{\rho^{(2)}}}\,\right)^2\;,
\label{infidelity_2RDM}
\end{align}
with all density matrices now normalized to unity.  Both quantities vanish if and only if $\rho_\mathrm{(a)pcs}^{(2)}=\rho^{(2)}$; hence a nonzero value quantifies the overall error (bounded by 1) of the PCS 2RDM.  Figure~\ref{fig:error_measures} shows the very good agreement between the PCS 2RDMs and the exact 2RDM, as measured by the metrics~(\ref{trace_dist_2RDM}) and (\ref{infidelity_2RDM}), over the entire $U/(nJ)$ parameter space, from small ($N=50$) to large ($N=1000$) particle numbers.  As shown by both metrics, $\rho_\mathrm{pcs}^{(2)}$ is very close to $\rho^{(2)}$ for all parameters considered.  It is worth pointing out that in the extreme limits of zero and hard-core interactions [$U/(nJ)\rightarrow 0$ and $U/(nJ)\rightarrow \infty$], the exact PCS \emph{Ansatz\/} gives a nearly exact prediction, but reaches its worst point in the transition region $U/(nJ)\sim 1$.  In the transition region, the APCS performs just as well, with the trace distance (infidelity) showing imprecision of the order of $10^{-2}$ ($10^{-3}$); this region of good agreement gets wider as $N$ increases, but the APCS deviates from the PCS predictions for weak and strong interactions.

To understand the origin of the disagreements among the 2RDMs revealed in Fig.~\ref{fig:error_measures}, we compare the parameters that define the exact and (A)PCS 2RDM matrix elements [see Eqs.~(\ref{eq:exact_2rdm_BH}) and~(\ref{eq:exact_2rdm_Sch})].  Note that for this purpose, we revert to setting the normalization of all 2RDMs equal to $N(N-1)$, as opposed to the 1 in Fig.~\ref{fig:error_measures}.  The PCS values of these parameters are given in Eqs.~(\ref{eq:occu_diff_s_F}), (\ref{eq:F1111}), and~(\ref{eq:F1122}), and the APCS values by Eqs.~(\ref{eq:APCSparameters}).  The physical role played by these parameters is clear from looking at how they express the correlations in the site-label basis.  From Eq.~(\ref{eq:Deltabeta}) the parameter $\beta$ expresses the first-order correlation function between the two sites; since it also gives the population difference of the Schmidt modes, we have already considered it in Fig.~\ref{fig:rho1}, so do not consider it further now.  As one sees from Eqs.~(\ref{eq:gammadelta}), the parameters $\gamma$ and $\delta$ give the second-order site correlation functions.  We show in Fig.~\ref{fig:params_2RDM} a direct comparison between the parameters that define the exact and the PCS 2RDM matrix elements, together with their respective relative errors.  We already have some evidence of the behavior of $\delta$ from its role in the ground-state energy~(\ref{E0_BH}), where it describes how the interaction strength $U$ competes with the tunneling $J$ through the transition from the mean-field regime to a Mott insulator. Indeed, it was the need for this contribution that motivated us to develop the more accurate, hybrid estimate of $\delta$ in Eq.~(\ref{eq:rho2_terms_pcs_energy}).  In the plots in Fig.~\ref{fig:params_2RDM}, we use the more accurate, hybrid estimate,
\begin{align}
\delta_\mathrm{apcs}=\frac{N}{2}-\frac{\pcss_-}{2}\chi_1(\pcss_-)+\bigO(1/N)\;,
\end{align}
as the APCS result, in place of the large-$N$ estimate of zero from Eq.~(\ref{eq:APCSdelta}).  That $\gamma$ has the large-$N$ behavior proportional to $N^2$, whereas the leading-order behavior of $\delta$ is the Bogoliubov correction proportional to $N$ is immediately evident from the difference in scale of the ordinates in Fig.~\ref{fig:params_2RDM}.

Finally, as a teaser to motivate using the PCS formalism to tackle many-body dynamics, we briefly investigate how faithfully PCS can represent the time evolution of the two-site BH model~(\ref{BH_hamil}) compared to other \emph{ans\"atze}.  We consider a system of $N=1\,000$ particles initially in the ground state of the Hamiltonian~(\ref{BH_hamil}) with $U/J = 0$; this ground state is a single condensate mode that is symmetrically delocalized between the two sites.  Then we suddenly quench the system by turning on an on-site repulsion $U/J=0.8$ [$U/(nJ)=0.0016$], which is in the intermediate regime of weak interactions.  We calculate the resulting evolution of the exact state vector numerically using the Crank-Nicolson method; the exact 2RDM is derived from the state vector.  We then optimize the parameter spaces of PCS, double-Fock state (DFS), and Gross-Pitaevskii state (GPS), i.e., a single-mode condensate, to minimize their trace distance to the exact~\hbox{2RDM}.  The errors of the 2RDMs, as measured by the trace distances normalized to 1, given by PCS, DFS, and GPS, are plotted in Fig.~\ref{fig:pcs_dfs_td_err} as functions of evolution time (in units of $\hbar/J$).  The oscillation of the error given by GPS is a consequence of the collapse and revival of phase~\cite{greiner_collapse_2002, schachenmayer_atomic_2011}; i.e., the purity of the 1RDM oscillates.  These numerical results suggest that PCS might be useful in describing the dynamics in the strongly interacting regime.

\section{Conclusion and Future Work}
\label{sec:conclusions}

In this paper we introduce the bosonic particle-correlated state (BPCS), a state of $N=l\times n$ bosons that is derived by symmetrizing the $n$-fold tensor product of an arbitrary $l$-boson (pure or mixed) state $\sigma^{(l)}$.  We analyze in detail the pure-state case for $l=2$, i.e., $\sigma^{(2)}=\ket{\varPsi^{(2)}}\bra{\varPsi^{(2)}}$, which we call a pair-correlated state (PCS).  When there is just one Schmidt coefficient in the two-particle wave function, PCS reproduces the mean-field, Gross-Pitaevskii description of a single-mode condensate.  When there is one dominant Schmidt coefficient in $\ket{\varPsi^{(2)}}$, the leading-order corrections to just one Schmidt mode reproduce the particle-number-conserving Bogoliubov approximation.  The allure of the PCS \emph{Ansatz\/} lies in the case of many Schmidt coefficients of nearly the same size; in this situation, PCS describes a fragmented state that has large two-particle quantum correlations.  At leading order in the large-$N$ limit, this PCS description is a sort of mean-field theory for a multimode condensate; the corrections to leading order describe Bogoliubov and higher-order modifications to the behavior of the multimode condensate.  We provide methods for calculating the one- and two-particle reduced density matrices of PCS; from these RDMs come the predictions for physical observables such as mode populations and correlation functions.

As a test of the \PCS\ \emph{Ansatz}, we consider the two-site Bose-Hubbard model for the case of two identical lattice sites and large particle numbers.  We calculate the ground-state energy of this model and analyze in detail the one- and two-particle reduced density matrices of the ground state.  By comparing to exact results, we find that the PCS description, both in its exact form and in its large-$N$ approximate form, provides a remarkably good account of these ground-state properties.  The main lesson of this analysis concerns the transition from a mean-field, single-mode condensate, in the regime where tunneling between sites dominates the on-site interaction, to a Mott insulator, in the regime where the on-site interaction dominates.  This transition is well described in the large-$N$ limit by the \PCS\ formalism.  For this large-$N$ approximate \PCS\ to give an accurate modeling, however, requires including in the approximate \PCS\ ground-state energy not just Bogoliubov-order corrections, but also the first correction beyond Bogoliubov order.  As a teaser, we also present results for how well the \PCS\ \emph{Ansatz\/} can match the dynamics of the symmetric two-site Bose-Hubbard model when it is suddenly quenched from being a single-mode condensate; our results suggest that the \PCS\ formalism has the potential to describe this process quite well.

The success of the PCS formalism for the symmetric, two-site Bose-Hubbard model motivates further work.  One question involves the ground state of an asymmetric Bose-Hubbard model, where comparison of Eqs.~(\ref{eq:exact_2rdm_BH_without_site_label_symmetry}) and~(\ref{eq:exact_2rdm_BH}) shows that in the asymmetric case, the exact ground-state two-particle RDM does not have \PCS\ form.  Does the \PCS\ formalism continue to provide a good description in this situation?  A second question involves generalizing to the ground state of multi-site Bose-Hubbard Hamiltonians.  The \PCS\ description can incorporate exactly the strong-interaction Mott-insulator state of disconnected condensates at each site in the case of two sites, but it cannot do so for three or more sites.  Does the \PCS\ formalism continue to provide an accurate description of the transition from single condensate to Mott insulator for three or more sites?  The question here might be put more generally as whether the correlations built into the two-particle RDM in the \PCS\ description are the dominant correlations when there can be correlations among more than two sites.

The ultimate goal of our work on the \PCS\ \emph{Ansatz\/} is to analyze the dynamics of a Bose-Einstein condensate as one or more barriers is raised and lowered within it.  In this situation, one must model how the ``sites'' within the BEC are evolving, i.e., how some appropriate set of single-particle wave functions are changing in time, and at the same time model how these sites are populated.  The \PCS\ formalism has the potential to capture this situation by providing a prescription for how the Schmidt orbitals are changing in time and by using the evolving two-particle \PCS\ correlations to describe how the Schmidt orbitals are populated.  Developing this description of \PCS\ dynamics is the main goal of our future work.

\begin{acknowledgments}
This work was supported by National Science Foundation Grants No.~PHY-1314763 and No.~PHY-1521016 and by Office of Naval Research Grant No.~N00014-15-1-2167. A.B.T. acknowledges support from AFOSR Grant No.~FA9550-12-1-0057. Z.J. acknowledges support from the NASA Advanced Exploration Systems program and NASA Ames Research Center.
\end{acknowledgments}

%###################################################################################
% \bibliographystyle{apsrev4-1_with_title}
% \bibliography{ref_notes}

\begin{thebibliography}{56}%
\makeatletter
\providecommand \@ifxundefined [1]{%
 \@ifx{#1\undefined}
}%
\providecommand \@ifnum [1]{%
 \ifnum #1\expandafter \@firstoftwo
 \else \expandafter \@secondoftwo
 \fi
}%
\providecommand \@ifx [1]{%
 \ifx #1\expandafter \@firstoftwo
 \else \expandafter \@secondoftwo
 \fi
}%
\providecommand \natexlab [1]{#1}%
\providecommand \enquote  [1]{``#1''}%
\providecommand \bibnamefont  [1]{#1}%
\providecommand \bibfnamefont [1]{#1}%
\providecommand \citenamefont [1]{#1}%
\providecommand \href@noop [0]{\@secondoftwo}%
\providecommand \href [0]{\begingroup \@sanitize@url \@href}%
\providecommand \@href[1]{\@@startlink{#1}\@@href}%
\providecommand \@@href[1]{\endgroup#1\@@endlink}%
\providecommand \@sanitize@url [0]{\catcode `\\12\catcode `\$12\catcode
  `\&12\catcode `\#12\catcode `\^12\catcode `\_12\catcode `\%12\relax}%
\providecommand \@@startlink[1]{}%
\providecommand \@@endlink[0]{}%
\providecommand \url  [0]{\begingroup\@sanitize@url \@url }%
\providecommand \@url [1]{\endgroup\@href {#1}{\urlprefix }}%
\providecommand \urlprefix  [0]{URL }%
\providecommand \Eprint [0]{\href }%
\providecommand \doibase [0]{http://dx.doi.org/}%
\providecommand \selectlanguage [0]{\@gobble}%
\providecommand \bibinfo  [0]{\@secondoftwo}%
\providecommand \bibfield  [0]{\@secondoftwo}%
\providecommand \translation [1]{[#1]}%
\providecommand \BibitemOpen [0]{}%
\providecommand \bibitemStop [0]{}%
\providecommand \bibitemNoStop [0]{.\EOS\space}%
\providecommand \EOS [0]{\spacefactor3000\relax}%
\providecommand \BibitemShut  [1]{\csname bibitem#1\endcsname}%
\let\auto@bib@innerbib\@empty
%</preamble>
\bibitem [{\citenamefont {Gross}(1961)}]{gross_structure_1961}%
  \BibitemOpen
  \bibfield  {author} {\bibinfo {author} {\bibfnamefont {E.~P.}\ \bibnamefont
  {Gross}},\ }\bibfield  {title} {\enquote {\bibinfo {title} {Structure of a
  quantized vortex in boson systems},}\ }\href {\doibase 10.1007/BF02731494}
  {\bibfield  {journal} {\bibinfo  {journal} {Il Nuovo Cimento Series 10}\
  }\textbf {\bibinfo {volume} {20}},\ \bibinfo {pages} {454} (\bibinfo {year}
  {1961})}\BibitemShut {NoStop}%
\bibitem [{\citenamefont {Pitaevskii}(1961)}]{pitaevsk_vortex_1961}%
  \BibitemOpen
  \bibfield  {author} {\bibinfo {author} {\bibfnamefont {L.}~\bibnamefont
  {Pitaevskii}},\ }\bibfield  {title} {\enquote {\bibinfo {title} {Vortex lines
  in an imperfect {B}ose gas},}\ }\href@noop {} {\bibfield  {journal} {\bibinfo
   {journal} {Soviet Physics {JETP}}\ }\textbf {\bibinfo {volume} {13}},\
  \bibinfo {pages} {451} (\bibinfo {year} {1961})}\BibitemShut {NoStop}%
\bibitem [{\citenamefont {Jaksch}\ \emph {et~al.}(1998)\citenamefont {Jaksch},
  \citenamefont {Bruder}, \citenamefont {Cirac}, \citenamefont {Gardiner},\
  and\ \citenamefont {Zoller}}]{jaksch_cold_1998}%
  \BibitemOpen
  \bibfield  {author} {\bibinfo {author} {\bibfnamefont {D.}~\bibnamefont
  {Jaksch}}, \bibinfo {author} {\bibfnamefont {C.}~\bibnamefont {Bruder}},
  \bibinfo {author} {\bibfnamefont {J.~I.}\ \bibnamefont {Cirac}}, \bibinfo
  {author} {\bibfnamefont {C.~W.}\ \bibnamefont {Gardiner}}, \ and\ \bibinfo
  {author} {\bibfnamefont {P.}~\bibnamefont {Zoller}},\ }\bibfield  {title}
  {\enquote {\bibinfo {title} {Cold bosonic atoms in optical lattices},}\
  }\href {\doibase 10.1103/PhysRevLett.81.3108} {\bibfield  {journal} {\bibinfo
   {journal} {Physical Review Letters}\ }\textbf {\bibinfo {volume} {81}},\
  \bibinfo {pages} {3108} (\bibinfo {year} {1998})}\BibitemShut {NoStop}%
\bibitem [{\citenamefont {Greiner}\ \emph
  {et~al.}(2002{\natexlab{a}})\citenamefont {Greiner}, \citenamefont {Mandel},
  \citenamefont {Esslinger}, \citenamefont {H\"{a}nsch},\ and\ \citenamefont
  {Bloch}}]{greiner_quantum_2002}%
  \BibitemOpen
  \bibfield  {author} {\bibinfo {author} {\bibfnamefont {M.}~\bibnamefont
  {Greiner}}, \bibinfo {author} {\bibfnamefont {O.}~\bibnamefont {Mandel}},
  \bibinfo {author} {\bibfnamefont {T.}~\bibnamefont {Esslinger}}, \bibinfo
  {author} {\bibfnamefont {T.~W.}\ \bibnamefont {H\"{a}nsch}}, \ and\ \bibinfo
  {author} {\bibfnamefont {I.}~\bibnamefont {Bloch}},\ }\bibfield  {title}
  {\enquote {\bibinfo {title} {Quantum phase transition from a superfluid to a
  mott insulator in a gas of ultracold atoms},}\ }\href {\doibase
  10.1038/415039a} {\bibfield  {journal} {\bibinfo  {journal} {Nature}\
  }\textbf {\bibinfo {volume} {415}},\ \bibinfo {pages} {39} (\bibinfo {year}
  {2002}{\natexlab{a}})}\BibitemShut {NoStop}%
\bibitem [{\citenamefont {Bogoliubov}(1947)}]{bogoliubov_theory_1947}%
  \BibitemOpen
  \bibfield  {author} {\bibinfo {author} {\bibfnamefont {N.~N.}\ \bibnamefont
  {Bogoliubov}},\ }\bibfield  {title} {\enquote {\bibinfo {title} {On the
  theory of superfluidity},}\ }\href@noop {} {\bibfield  {journal} {\bibinfo
  {journal} {Journal of Physics (USSR)}\ }\textbf {\bibinfo {volume} {11}},\
  \bibinfo {pages} {23} (\bibinfo {year} {1947})}\BibitemShut {NoStop}%
\bibitem [{\citenamefont {Fetter}(1972)}]{fetter_nonuniform_1972}%
  \BibitemOpen
  \bibfield  {author} {\bibinfo {author} {\bibfnamefont {A.~L.}\ \bibnamefont
  {Fetter}},\ }\bibfield  {title} {\enquote {\bibinfo {title} {Nonuniform
  states of an imperfect {Bose} gas},}\ }\href {\doibase
  10.1016/0003-4916(72)90330-2} {\bibfield  {journal} {\bibinfo  {journal}
  {Annals of Physics}\ }\textbf {\bibinfo {volume} {70}},\ \bibinfo {pages}
  {67} (\bibinfo {year} {1972})}\BibitemShut {NoStop}%
\bibitem [{\citenamefont {Huang}(1987)}]{huang_statistical_1987}%
  \BibitemOpen
  \bibfield  {author} {\bibinfo {author} {\bibfnamefont {K.}~\bibnamefont
  {Huang}},\ }\href@noop {} {\emph {\bibinfo {title} {Statistical Mechanics,
  {\rm 2nd Edition}}}}\ (\bibinfo  {publisher} {Wiley},\ \bibinfo {year}
  {1987})\BibitemShut {NoStop}%
\bibitem [{\citenamefont {Penrose}\ and\ \citenamefont
  {Onsager}(1956)}]{penrose_bose-einstein_1956}%
  \BibitemOpen
  \bibfield  {author} {\bibinfo {author} {\bibfnamefont {O.}~\bibnamefont
  {Penrose}}\ and\ \bibinfo {author} {\bibfnamefont {L.}~\bibnamefont
  {Onsager}},\ }\bibfield  {title} {\enquote {\bibinfo {title}
  {{B}ose-{E}instein condensation and liquid helium},}\ }\href {\doibase
  10.1103/PhysRev.104.576} {\bibfield  {journal} {\bibinfo  {journal} {Physical
  Review}\ }\textbf {\bibinfo {volume} {104}},\ \bibinfo {pages} {576}
  (\bibinfo {year} {1956})}\BibitemShut {NoStop}%
\bibitem [{\citenamefont {Streltsov}\ \emph {et~al.}(2004)\citenamefont
  {Streltsov}, \citenamefont {Cederbaum},\ and\ \citenamefont
  {Moiseyev}}]{streltsov_ground-state_2004}%
  \BibitemOpen
  \bibfield  {author} {\bibinfo {author} {\bibfnamefont {A.~I.}\ \bibnamefont
  {Streltsov}}, \bibinfo {author} {\bibfnamefont {L.~S.}\ \bibnamefont
  {Cederbaum}}, \ and\ \bibinfo {author} {\bibfnamefont {N.}~\bibnamefont
  {Moiseyev}},\ }\bibfield  {title} {\enquote {\bibinfo {title} {Ground-state
  fragmentation of repulsive {Bose-Einstein} condensates in double-trap
  potentials},}\ }\href {\doibase 10.1103/PhysRevA.70.053607} {\bibfield
  {journal} {\bibinfo  {journal} {Physical Review A}\ }\textbf {\bibinfo
  {volume} {70}},\ \bibinfo {pages} {053607} (\bibinfo {year}
  {2004})}\BibitemShut {NoStop}%
\bibitem [{\citenamefont {Mueller}\ \emph {et~al.}(2006)\citenamefont
  {Mueller}, \citenamefont {Ho}, \citenamefont {Ueda},\ and\ \citenamefont
  {Baym}}]{mueller_fragmentation_2006}%
  \BibitemOpen
  \bibfield  {author} {\bibinfo {author} {\bibfnamefont {E.~J.}\ \bibnamefont
  {Mueller}}, \bibinfo {author} {\bibfnamefont {T.-L.}\ \bibnamefont {Ho}},
  \bibinfo {author} {\bibfnamefont {M.}~\bibnamefont {Ueda}}, \ and\ \bibinfo
  {author} {\bibfnamefont {G.}~\bibnamefont {Baym}},\ }\bibfield  {title}
  {\enquote {\bibinfo {title} {Fragmentation of {Bose-Einstein} condensates},}\
  }\href {\doibase 10.1103/PhysRevA.74.033612} {\bibfield  {journal} {\bibinfo
  {journal} {Physical Review A}\ }\textbf {\bibinfo {volume} {74}},\ \bibinfo
  {pages} {033612} (\bibinfo {year} {2006})}\BibitemShut {NoStop}%
\bibitem [{\citenamefont {Alon}\ \emph {et~al.}(2007)\citenamefont {Alon},
  \citenamefont {Streltsov},\ and\ \citenamefont
  {Cederbaum}}]{alon_time-dependent_2007}%
  \BibitemOpen
  \bibfield  {author} {\bibinfo {author} {\bibfnamefont {O.~E.}\ \bibnamefont
  {Alon}}, \bibinfo {author} {\bibfnamefont {A.~I.}\ \bibnamefont {Streltsov}},
  \ and\ \bibinfo {author} {\bibfnamefont {L.~S.}\ \bibnamefont {Cederbaum}},\
  }\bibfield  {title} {\enquote {\bibinfo {title} {Time-dependent multi-orbital
  mean-field for fragmented {Bose-Einstein} condensates},}\ }\href {\doibase
  10.1016/j.physleta.2006.10.048} {\bibfield  {journal} {\bibinfo  {journal}
  {Physics Letters A}\ }\textbf {\bibinfo {volume} {362}},\ \bibinfo {pages}
  {453} (\bibinfo {year} {2007})}\BibitemShut {NoStop}%
\bibitem [{\citenamefont {Vardi}\ and\ \citenamefont
  {Anglin}(2001)}]{vardi_bose-einstein_2001}%
  \BibitemOpen
  \bibfield  {author} {\bibinfo {author} {\bibfnamefont {A.}~\bibnamefont
  {Vardi}}\ and\ \bibinfo {author} {\bibfnamefont {J.~R.}\ \bibnamefont
  {Anglin}},\ }\bibfield  {title} {\enquote {\bibinfo {title} {{Bose-Einstein}
  condensates beyond mean field theory: Quantum backreaction as decoherence},}\
  }\href {\doibase 10.1103/PhysRevLett.86.568} {\bibfield  {journal} {\bibinfo
  {journal} {Physical Review Letters}\ }\textbf {\bibinfo {volume} {86}},\
  \bibinfo {pages} {568} (\bibinfo {year} {2001})}\BibitemShut {NoStop}%
\bibitem [{\citenamefont {Tikhonenkov}\ \emph {et~al.}(2007)\citenamefont
  {Tikhonenkov}, \citenamefont {Anglin},\ and\ \citenamefont
  {Vardi}}]{tikhonenkov_quantum_2007}%
  \BibitemOpen
  \bibfield  {author} {\bibinfo {author} {\bibfnamefont {I.}~\bibnamefont
  {Tikhonenkov}}, \bibinfo {author} {\bibfnamefont {J.~R.}\ \bibnamefont
  {Anglin}}, \ and\ \bibinfo {author} {\bibfnamefont {A.}~\bibnamefont
  {Vardi}},\ }\bibfield  {title} {\enquote {\bibinfo {title} {Quantum dynamics
  of {Bose-Hubbard} hamiltonians beyond the {Hartree-Fock-Bogoliubov}
  approximation: The {Bogoliubov} back-reaction approximation},}\ }\href
  {\doibase 10.1103/PhysRevA.75.013613} {\bibfield  {journal} {\bibinfo
  {journal} {Physical Review A}\ }\textbf {\bibinfo {volume} {75}},\ \bibinfo
  {pages} {013613} (\bibinfo {year} {2007})}\BibitemShut {NoStop}%
\bibitem [{\citenamefont {Laughlin}(1983)}]{laughlin_anomalous_1983}%
  \BibitemOpen
  \bibfield  {author} {\bibinfo {author} {\bibfnamefont {R.~B.}\ \bibnamefont
  {Laughlin}},\ }\bibfield  {title} {\enquote {\bibinfo {title} {Anomalous
  quantum hall effect: An incompressible quantum fluid with fractionally
  charged excitations},}\ }\href {\doibase 10.1103/PhysRevLett.50.1395}
  {\bibfield  {journal} {\bibinfo  {journal} {Physical Review Letters}\
  }\textbf {\bibinfo {volume} {50}},\ \bibinfo {pages} {1395} (\bibinfo {year}
  {1983})}\BibitemShut {NoStop}%
\bibitem [{\citenamefont {Gutzwiller}(1963)}]{gutzwiller_effect_1963}%
  \BibitemOpen
  \bibfield  {author} {\bibinfo {author} {\bibfnamefont {M.~C.}\ \bibnamefont
  {Gutzwiller}},\ }\bibfield  {title} {\enquote {\bibinfo {title} {Effect of
  correlation on the ferromagnetism of transition metals},}\ }\href {\doibase
  10.1103/PhysRevLett.10.159} {\bibfield  {journal} {\bibinfo  {journal}
  {Physical Review Letters}\ }\textbf {\bibinfo {volume} {10}},\ \bibinfo
  {pages} {159} (\bibinfo {year} {1963})}\BibitemShut {NoStop}%
\bibitem [{\citenamefont {Rokhsar}\ and\ \citenamefont
  {Kotliar}(1991)}]{rokhsar_gutzwiller_1991}%
  \BibitemOpen
  \bibfield  {author} {\bibinfo {author} {\bibfnamefont {D.~S.}\ \bibnamefont
  {Rokhsar}}\ and\ \bibinfo {author} {\bibfnamefont {B.~G.}\ \bibnamefont
  {Kotliar}},\ }\bibfield  {title} {\enquote {\bibinfo {title} {{Gutzwiller}
  projection for bosons},}\ }\href {\doibase 10.1103/PhysRevB.44.10328}
  {\bibfield  {journal} {\bibinfo  {journal} {Physical Review B}\ }\textbf
  {\bibinfo {volume} {44}},\ \bibinfo {pages} {10328} (\bibinfo {year}
  {1991})}\BibitemShut {NoStop}%
\bibitem [{\citenamefont {Umrigar}\ \emph {et~al.}(1988)\citenamefont
  {Umrigar}, \citenamefont {Wilson},\ and\ \citenamefont
  {Wilkins}}]{umrigar_optimized_1988}%
  \BibitemOpen
  \bibfield  {author} {\bibinfo {author} {\bibfnamefont {C.~J.}\ \bibnamefont
  {Umrigar}}, \bibinfo {author} {\bibfnamefont {K.~G.}\ \bibnamefont {Wilson}},
  \ and\ \bibinfo {author} {\bibfnamefont {J.~W.}\ \bibnamefont {Wilkins}},\
  }\bibfield  {title} {\enquote {\bibinfo {title} {Optimized trial wave
  functions for quantum monte carlo calculations},}\ }\href {\doibase
  10.1103/PhysRevLett.60.1719} {\bibfield  {journal} {\bibinfo  {journal}
  {Physical Review Letters}\ }\textbf {\bibinfo {volume} {60}},\ \bibinfo
  {pages} {1719} (\bibinfo {year} {1988})}\BibitemShut {NoStop}%
\bibitem [{\citenamefont {Reatto}(1969)}]{reatto_bose-einstein_1969}%
  \BibitemOpen
  \bibfield  {author} {\bibinfo {author} {\bibfnamefont {L.}~\bibnamefont
  {Reatto}},\ }\bibfield  {title} {\enquote {\bibinfo {title} {{Bose-Einstein}
  condensation for a class of wave functions},}\ }\href {\doibase
  10.1103/PhysRev.183.334} {\bibfield  {journal} {\bibinfo  {journal} {Physical
  Review}\ }\textbf {\bibinfo {volume} {183}},\ \bibinfo {pages} {334}
  (\bibinfo {year} {1969})}\BibitemShut {NoStop}%
\bibitem [{\citenamefont {Fabrocini}\ and\ \citenamefont
  {Polls}(1999)}]{fabrocini_beyond_1999}%
  \BibitemOpen
  \bibfield  {author} {\bibinfo {author} {\bibfnamefont {A.}~\bibnamefont
  {Fabrocini}}\ and\ \bibinfo {author} {\bibfnamefont {A.}~\bibnamefont
  {Polls}},\ }\bibfield  {title} {\enquote {\bibinfo {title} {Beyond the
  {Gross-Pitaevskii} approximation: Local density versus correlated basis
  approach for trapped bosons},}\ }\href {\doibase 10.1103/PhysRevA.60.2319}
  {\bibfield  {journal} {\bibinfo  {journal} {Physical Review A}\ }\textbf
  {\bibinfo {volume} {60}},\ \bibinfo {pages} {2319} (\bibinfo {year}
  {1999})}\BibitemShut {NoStop}%
\bibitem [{\citenamefont {{DuBois}}\ and\ \citenamefont
  {Glyde}(2001)}]{dubois_bose-einstein_2001}%
  \BibitemOpen
  \bibfield  {author} {\bibinfo {author} {\bibfnamefont {J.~L.}\ \bibnamefont
  {{DuBois}}}\ and\ \bibinfo {author} {\bibfnamefont {H.~R.}\ \bibnamefont
  {Glyde}},\ }\bibfield  {title} {\enquote {\bibinfo {title} {{Bose-Einstein}
  condensation in trapped bosons: A variational {Monte} {Carlo} analysis},}\
  }\href {\doibase 10.1103/PhysRevA.63.023602} {\bibfield  {journal} {\bibinfo
  {journal} {Physical Review A}\ }\textbf {\bibinfo {volume} {63}},\ \bibinfo
  {pages} {023602} (\bibinfo {year} {2001})}\BibitemShut {NoStop}%
\bibitem [{\citenamefont {Cowell}\ \emph {et~al.}(2002)\citenamefont {Cowell},
  \citenamefont {Heiselberg}, \citenamefont {Mazets}, \citenamefont {Morales},
  \citenamefont {Pandharipande},\ and\ \citenamefont
  {Pethick}}]{cowell_cold_2002}%
  \BibitemOpen
  \bibfield  {author} {\bibinfo {author} {\bibfnamefont {S.}~\bibnamefont
  {Cowell}}, \bibinfo {author} {\bibfnamefont {H.}~\bibnamefont {Heiselberg}},
  \bibinfo {author} {\bibfnamefont {I.~E.}\ \bibnamefont {Mazets}}, \bibinfo
  {author} {\bibfnamefont {J.}~\bibnamefont {Morales}}, \bibinfo {author}
  {\bibfnamefont {V.~R.}\ \bibnamefont {Pandharipande}}, \ and\ \bibinfo
  {author} {\bibfnamefont {C.~J.}\ \bibnamefont {Pethick}},\ }\bibfield
  {title} {\enquote {\bibinfo {title} {Cold {Bose} gases with large scattering
  lengths},}\ }\href {\doibase 10.1103/PhysRevLett.88.210403} {\bibfield
  {journal} {\bibinfo  {journal} {Physical Review Letters}\ }\textbf {\bibinfo
  {volume} {88}},\ \bibinfo {pages} {210403} (\bibinfo {year}
  {2002})}\BibitemShut {NoStop}%
\bibitem [{\citenamefont {Leggett}(2001)}]{leggett_bose-einstein_2001}%
  \BibitemOpen
  \bibfield  {author} {\bibinfo {author} {\bibfnamefont {A.~J.}\ \bibnamefont
  {Leggett}},\ }\bibfield  {title} {\enquote {\bibinfo {title} {{Bose-Einstein}
  condensation in the alkali gases: Some fundamental concepts},}\ }\href
  {\doibase 10.1103/RevModPhys.73.307} {\bibfield  {journal} {\bibinfo
  {journal} {Reviews of Modern Physics}\ }\textbf {\bibinfo {volume} {73}},\
  \bibinfo {pages} {307} (\bibinfo {year} {2001})}\BibitemShut {NoStop}%
\bibitem [{\citenamefont {Yang}(1962)}]{yang_concept_1962}%
  \BibitemOpen
  \bibfield  {author} {\bibinfo {author} {\bibfnamefont {C.~N.}\ \bibnamefont
  {Yang}},\ }\bibfield  {title} {\enquote {\bibinfo {title} {Concept of
  off-diagonal long-range order and the quantum phases of liquid {He} and of
  superconductors},}\ }\href {\doibase 10.1103/RevModPhys.34.694} {\bibfield
  {journal} {\bibinfo  {journal} {Reviews of Modern Physics}\ }\textbf
  {\bibinfo {volume} {34}},\ \bibinfo {pages} {694} (\bibinfo {year}
  {1962})}\BibitemShut {NoStop}%
\bibitem [{\citenamefont {Jiang}\ and\ \citenamefont {Caves}()}]{pcs2}%
  \BibitemOpen
  \bibfield  {author} {\bibinfo {author} {\bibfnamefont {Z.}~\bibnamefont
  {Jiang}}\ and\ \bibinfo {author} {\bibfnamefont {C.~M.}\ \bibnamefont
  {Caves}},\ }\href@noop {} {}\bibinfo {note} {In preparation}\BibitemShut
  {NoStop}%
\bibitem [{\citenamefont {Jiang}(2014)}]{JiangPhD}%
  \BibitemOpen
  \bibfield  {author} {\bibinfo {author} {\bibfnamefont {Z.}~\bibnamefont
  {Jiang}},\ }\emph {\bibinfo {title} {Particle Correlations in {Bose-Einstein}
  Condensates}},\ \href {http://digitalrepository.unm.edu/phyc_etds/29/} {Ph.D.
  thesis},\ \bibinfo  {school} {University of New Mexico} (\bibinfo {year}
  {2014})\BibitemShut {NoStop}%
\bibitem [{\citenamefont {Pa\v{s}kauskas}\ and\ \citenamefont
  {You}(2001)}]{paskauskas_quantum_2001}%
  \BibitemOpen
  \bibfield  {author} {\bibinfo {author} {\bibfnamefont {R.}~\bibnamefont
  {Pa\v{s}kauskas}}\ and\ \bibinfo {author} {\bibfnamefont {L.}~\bibnamefont
  {You}},\ }\bibfield  {title} {\enquote {\bibinfo {title} {Quantum
  correlations in two-boson wave functions},}\ }\href {\doibase
  10.1103/PhysRevA.64.042310} {\bibfield  {journal} {\bibinfo  {journal}
  {Physical Review A}\ }\textbf {\bibinfo {volume} {64}},\ \bibinfo {pages}
  {042310} (\bibinfo {year} {2001})}\BibitemShut {NoStop}%
\bibitem [{\citenamefont {Horn}\ and\ \citenamefont
  {Johnson}(2013)}]{horn_matrix_2013}%
  \BibitemOpen
  \bibfield  {author} {\bibinfo {author} {\bibfnamefont {R.~A.}\ \bibnamefont
  {Horn}}\ and\ \bibinfo {author} {\bibfnamefont {C.~R.}\ \bibnamefont
  {Johnson}},\ }\href@noop {} {\emph {\bibinfo {title} {Matrix analysis (2nd
  edition)}}}\ (\bibinfo  {publisher} {Cambridge University Press},\ \bibinfo
  {address} {Cambridge},\ \bibinfo {year} {2013})\BibitemShut {NoStop}%
\bibitem [{\citenamefont {Bardeen}\ \emph
  {et~al.}(1957{\natexlab{a}})\citenamefont {Bardeen}, \citenamefont {Cooper},\
  and\ \citenamefont {Schrieffer}}]{bardeen_microscopic_1957}%
  \BibitemOpen
  \bibfield  {author} {\bibinfo {author} {\bibfnamefont {J.}~\bibnamefont
  {Bardeen}}, \bibinfo {author} {\bibfnamefont {L.~N.}\ \bibnamefont {Cooper}},
  \ and\ \bibinfo {author} {\bibfnamefont {J.~R.}\ \bibnamefont {Schrieffer}},\
  }\bibfield  {title} {\enquote {\bibinfo {title} {Microscopic theory of
  superconductivity},}\ }\href {\doibase 10.1103/PhysRev.106.162} {\bibfield
  {journal} {\bibinfo  {journal} {Physical Review}\ }\textbf {\bibinfo {volume}
  {106}},\ \bibinfo {pages} {162} (\bibinfo {year}
  {1957}{\natexlab{a}})}\BibitemShut {NoStop}%
\bibitem [{\citenamefont {Bardeen}\ \emph
  {et~al.}(1957{\natexlab{b}})\citenamefont {Bardeen}, \citenamefont {Cooper},\
  and\ \citenamefont {Schrieffer}}]{bardeen_theory_1957}%
  \BibitemOpen
  \bibfield  {author} {\bibinfo {author} {\bibfnamefont {J.}~\bibnamefont
  {Bardeen}}, \bibinfo {author} {\bibfnamefont {L.~N.}\ \bibnamefont {Cooper}},
  \ and\ \bibinfo {author} {\bibfnamefont {J.~R.}\ \bibnamefont {Schrieffer}},\
  }\bibfield  {title} {\enquote {\bibinfo {title} {Theory of
  superconductivity},}\ }\href {\doibase 10.1103/PhysRev.108.1175} {\bibfield
  {journal} {\bibinfo  {journal} {Physical Review}\ }\textbf {\bibinfo {volume}
  {108}},\ \bibinfo {pages} {1175} (\bibinfo {year}
  {1957}{\natexlab{b}})}\BibitemShut {NoStop}%
\bibitem [{\citenamefont {Valatin}\ and\ \citenamefont
  {Butler}(1958)}]{valatin_collective_1958}%
  \BibitemOpen
  \bibfield  {author} {\bibinfo {author} {\bibfnamefont {J.~G.}\ \bibnamefont
  {Valatin}}\ and\ \bibinfo {author} {\bibfnamefont {D.}~\bibnamefont
  {Butler}},\ }\bibfield  {title} {\enquote {\bibinfo {title} {On the
  collective properties of a boson system},}\ }\href {\doibase
  10.1007/BF02859603} {\bibfield  {journal} {\bibinfo  {journal} {Il Nuovo
  Cimento}\ }\textbf {\bibinfo {volume} {10}},\ \bibinfo {pages} {37} (\bibinfo
  {year} {1958})}\BibitemShut {NoStop}%
\bibitem [{\citenamefont {Coniglio}\ and\ \citenamefont
  {Marinaro}(1967)}]{coniglio_condensation_1967}%
  \BibitemOpen
  \bibfield  {author} {\bibinfo {author} {\bibfnamefont {A.}~\bibnamefont
  {Coniglio}}\ and\ \bibinfo {author} {\bibfnamefont {M.}~\bibnamefont
  {Marinaro}},\ }\bibfield  {title} {\enquote {\bibinfo {title} {On
  condensation for an interacting boson system},}\ }\href {\doibase
  10.1007/BF02712190} {\bibfield  {journal} {\bibinfo  {journal} {Il Nuovo
  Cimento B Series 10}\ }\textbf {\bibinfo {volume} {48}},\ \bibinfo {pages}
  {249} (\bibinfo {year} {1967})}\BibitemShut {NoStop}%
\bibitem [{\citenamefont {Evans}\ and\ \citenamefont
  {Imry}(1969)}]{evans_pairing_1969}%
  \BibitemOpen
  \bibfield  {author} {\bibinfo {author} {\bibfnamefont {W.~A.~B.}\
  \bibnamefont {Evans}}\ and\ \bibinfo {author} {\bibfnamefont
  {Y.}~\bibnamefont {Imry}},\ }\bibfield  {title} {\enquote {\bibinfo {title}
  {On the pairing theory of the {Bose} superfluid},}\ }\href {\doibase
  10.1007/BF02711051} {\bibfield  {journal} {\bibinfo  {journal} {Il Nuovo
  Cimento B Series 10}\ }\textbf {\bibinfo {volume} {63}},\ \bibinfo {pages}
  {155} (\bibinfo {year} {1969})}\BibitemShut {NoStop}%
\bibitem [{\citenamefont {Nozi\'eres}\ and\ \citenamefont
  {Saint~James}(1982)}]{nozieres_particle_1982}%
  \BibitemOpen
  \bibfield  {author} {\bibinfo {author} {\bibfnamefont {P.}~\bibnamefont
  {Nozi\'eres}}\ and\ \bibinfo {author} {\bibfnamefont {D.}~\bibnamefont
  {Saint~James}},\ }\bibfield  {title} {\enquote {\bibinfo {title} {Particle
  vs. pair condensation in attractive {Bose} liquids},}\ }\href {\doibase
  10.1051/jphys:019820043070113300} {\bibfield  {journal} {\bibinfo  {journal}
  {Journal de Physique}\ }\textbf {\bibinfo {volume} {43}},\ \bibinfo {pages}
  {1133} (\bibinfo {year} {1982})}\BibitemShut {NoStop}%
\bibitem [{\citenamefont {Leggett}(2003)}]{leggett_relation_2003}%
  \BibitemOpen
  \bibfield  {author} {\bibinfo {author} {\bibfnamefont {A.~J.}\ \bibnamefont
  {Leggett}},\ }\bibfield  {title} {\enquote {\bibinfo {title} {The relation
  between the {Gross-Pitaevskii} and {Bogoliubov} descriptions of a dilute
  {Bose} gas},}\ }\href {\doibase 10.1088/1367-2630/5/1/103} {\bibfield
  {journal} {\bibinfo  {journal} {New Journal of Physics}\ }\textbf {\bibinfo
  {volume} {5}},\ \bibinfo {pages} {103} (\bibinfo {year} {2003})}\BibitemShut
  {NoStop}%
\bibitem [{\citenamefont {Ho}\ and\ \citenamefont
  {Yip}(2000)}]{ho_fragmented_2000}%
  \BibitemOpen
  \bibfield  {author} {\bibinfo {author} {\bibfnamefont {T.-L.}\ \bibnamefont
  {Ho}}\ and\ \bibinfo {author} {\bibfnamefont {S.~K.}\ \bibnamefont {Yip}},\
  }\bibfield  {title} {\enquote {\bibinfo {title} {Fragmented and single
  condensate ground states of {spin-1} {Bose} gas},}\ }\href {\doibase
  10.1103/PhysRevLett.84.4031} {\bibfield  {journal} {\bibinfo  {journal}
  {Physical Review Letters}\ }\textbf {\bibinfo {volume} {84}},\ \bibinfo
  {pages} {4031} (\bibinfo {year} {2000})}\BibitemShut {NoStop}%
\bibitem [{\citenamefont {Leggett}(1980)}]{leggett_diatomic_1980}%
  \BibitemOpen
  \bibfield  {author} {\bibinfo {author} {\bibfnamefont {A.~J.}\ \bibnamefont
  {Leggett}},\ }\bibfield  {title} {\enquote {\bibinfo {title} {Diatomic
  molecules and cooper pairs},}\ }in\ \href
  {http://link.springer.com/chapter/10.1007/BFb0120125} {\emph {\bibinfo
  {booktitle} {Modern Trends in the Theory of Condensed Matter}}},\ \bibinfo
  {series and number} {\bibinfo {series} {Lecture Notes in Physics}\ No.\
  \bibinfo {number} {115}},\ \bibinfo {editor} {edited by\ \bibinfo {editor}
  {\bibfnamefont {A.}~\bibnamefont {P\c{e}kalski}}\ and\ \bibinfo {editor}
  {\bibfnamefont {J.~A.}\ \bibnamefont {Przystawa}}}\ (\bibinfo  {publisher}
  {Springer Berlin Heidelberg},\ \bibinfo {year} {1980})\ pp.\ \bibinfo {pages}
  {13--27}\BibitemShut {NoStop}%
\bibitem [{\citenamefont {Leggett}\ and\ \citenamefont
  {Zhang}(2012)}]{leggett_becbcs_2012}%
  \BibitemOpen
  \bibfield  {author} {\bibinfo {author} {\bibfnamefont {A.~J.}\ \bibnamefont
  {Leggett}}\ and\ \bibinfo {author} {\bibfnamefont {S.}~\bibnamefont
  {Zhang}},\ }\bibfield  {title} {\enquote {\bibinfo {title} {The {BEC-BCS}
  crossover: Some history and some general observations},}\ }in\ \href
  {http://link.springer.com/chapter/10.1007/978-3-642-21978-8_2} {\emph
  {\bibinfo {booktitle} {The {BCS-BEC} Crossover and the Unitary Fermi Gas}}},\
  \bibinfo {series and number} {\bibinfo {series} {Lecture Notes in Physics}\
  No.\ \bibinfo {number} {836}},\ \bibinfo {editor} {edited by\ \bibinfo
  {editor} {\bibfnamefont {W.}~\bibnamefont {Zwerger}}}\ (\bibinfo  {publisher}
  {Springer Berlin Heidelberg},\ \bibinfo {year} {2012})\ pp.\ \bibinfo {pages}
  {33--47}\BibitemShut {NoStop}%
\bibitem [{\citenamefont {Combescot}\ \emph {et~al.}(2003)\citenamefont
  {Combescot}, \citenamefont {Leyronas},\ and\ \citenamefont
  {Tanguy}}]{combescot_n-exciton_2003}%
  \BibitemOpen
  \bibfield  {author} {\bibinfo {author} {\bibfnamefont {M.}~\bibnamefont
  {Combescot}}, \bibinfo {author} {\bibfnamefont {X.}~\bibnamefont {Leyronas}},
  \ and\ \bibinfo {author} {\bibfnamefont {C.}~\bibnamefont {Tanguy}},\
  }\bibfield  {title} {\enquote {\bibinfo {title} {On the \mbox{$N$}-exciton
  normalization factor},}\ }\href {\doibase 10.1140/epjb/e2003-00003-1}
  {\bibfield  {journal} {\bibinfo  {journal} {The European Physical Journal B -
  Condensed Matter and Complex Systems}\ }\textbf {\bibinfo {volume} {31}},\
  \bibinfo {pages} {17} (\bibinfo {year} {2003})}\BibitemShut {NoStop}%
\bibitem [{\citenamefont {Law}(2005)}]{law_quantum_2005}%
  \BibitemOpen
  \bibfield  {author} {\bibinfo {author} {\bibfnamefont {C.~K.}\ \bibnamefont
  {Law}},\ }\bibfield  {title} {\enquote {\bibinfo {title} {Quantum
  entanglement as an interpretation of bosonic character in composite
  two-particle systems},}\ }\href {\doibase 10.1103/PhysRevA.71.034306}
  {\bibfield  {journal} {\bibinfo  {journal} {Physical Review A}\ }\textbf
  {\bibinfo {volume} {71}},\ \bibinfo {pages} {034306} (\bibinfo {year}
  {2005})}\BibitemShut {NoStop}%
\bibitem [{\citenamefont {Combescot}\ \emph {et~al.}(2008)\citenamefont
  {Combescot}, \citenamefont {Betbeder-Matibet},\ and\ \citenamefont
  {Dubin}}]{combescot_many-body_2008}%
  \BibitemOpen
  \bibfield  {author} {\bibinfo {author} {\bibfnamefont {M.}~\bibnamefont
  {Combescot}}, \bibinfo {author} {\bibfnamefont {O.}~\bibnamefont
  {Betbeder-Matibet}}, \ and\ \bibinfo {author} {\bibfnamefont
  {F.}~\bibnamefont {Dubin}},\ }\bibfield  {title} {\enquote {\bibinfo {title}
  {The many-body physics of composite bosons},}\ }\href {\doibase
  10.1016/j.physrep.2007.11.003} {\bibfield  {journal} {\bibinfo  {journal}
  {Physics Reports}\ }\textbf {\bibinfo {volume} {463}},\ \bibinfo {pages}
  {215} (\bibinfo {year} {2008})}\BibitemShut {NoStop}%
\bibitem [{\citenamefont {Tichy}\ \emph {et~al.}(2014)\citenamefont {Tichy},
  \citenamefont {Bouvrie},\ and\ \citenamefont {M{\o}lmer}}]{tichy_how_2014}%
  \BibitemOpen
  \bibfield  {author} {\bibinfo {author} {\bibfnamefont {M.~C.}\ \bibnamefont
  {Tichy}}, \bibinfo {author} {\bibfnamefont {P.~A.}\ \bibnamefont {Bouvrie}},
  \ and\ \bibinfo {author} {\bibfnamefont {K.}~\bibnamefont {M{\o}lmer}},\
  }\bibfield  {title} {\enquote {\bibinfo {title} {How bosonic is a pair of
  fermions?}}\ }\href {\doibase 10.1007/s00340-014-5819-9} {\bibfield
  {journal} {\bibinfo  {journal} {Applied Physics B}\ }\textbf {\bibinfo
  {volume} {117}},\ \bibinfo {pages} {785} (\bibinfo {year}
  {2014})}\BibitemShut {NoStop}%
\bibitem [{\citenamefont {Dziarmaga}\ and\ \citenamefont
  {Sacha}(2003)}]{dziarmaga_bogoliubov_2003}%
  \BibitemOpen
  \bibfield  {author} {\bibinfo {author} {\bibfnamefont {J.}~\bibnamefont
  {Dziarmaga}}\ and\ \bibinfo {author} {\bibfnamefont {K.}~\bibnamefont
  {Sacha}},\ }\bibfield  {title} {\enquote {\bibinfo {title} {Bogoliubov theory
  of a {Bose}-{Einstein} condensate in the particle representation},}\ }\href
  {\doibase 10.1103/PhysRevA.67.033608} {\bibfield  {journal} {\bibinfo
  {journal} {Physical Review A}\ }\textbf {\bibinfo {volume} {67}},\ \bibinfo
  {pages} {033608} (\bibinfo {year} {2003})}\BibitemShut {NoStop}%
\bibitem [{\citenamefont {Dziarmaga}\ and\ \citenamefont
  {Sacha}(2006)}]{dziarmaga_images_2006}%
  \BibitemOpen
  \bibfield  {author} {\bibinfo {author} {\bibfnamefont {J.}~\bibnamefont
  {Dziarmaga}}\ and\ \bibinfo {author} {\bibfnamefont {K.}~\bibnamefont
  {Sacha}},\ }\bibfield  {title} {\enquote {\bibinfo {title} {Images of a
  {Bose}-{Einstein} condensate: diagonal dynamical {Bogoliubov} vacuum},}\
  }\href {\doibase 10.1088/0953-4075/39/1/007} {\bibfield  {journal} {\bibinfo
  {journal} {Journal of Physics B: Atomic, Molecular and Optical Physics}\
  }\textbf {\bibinfo {volume} {39}},\ \bibinfo {pages} {57} (\bibinfo {year}
  {2006})}\BibitemShut {NoStop}%
\bibitem [{\citenamefont {Girardeau}\ and\ \citenamefont
  {Arnowitt}(1959)}]{girardeau_theory_1959}%
  \BibitemOpen
  \bibfield  {author} {\bibinfo {author} {\bibfnamefont {M.}~\bibnamefont
  {Girardeau}}\ and\ \bibinfo {author} {\bibfnamefont {R.}~\bibnamefont
  {Arnowitt}},\ }\bibfield  {title} {\enquote {\bibinfo {title} {Theory of
  many-boson systems: Pair theory},}\ }\href {\doibase 10.1103/PhysRev.113.755}
  {\bibfield  {journal} {\bibinfo  {journal} {Physical Review}\ }\textbf
  {\bibinfo {volume} {113}},\ \bibinfo {pages} {755} (\bibinfo {year}
  {1959})}\BibitemShut {NoStop}%
\bibitem [{\citenamefont
  {Gardiner}(1997)}]{gardiner_particle-number-conserving_1997}%
  \BibitemOpen
  \bibfield  {author} {\bibinfo {author} {\bibfnamefont {C.~W.}\ \bibnamefont
  {Gardiner}},\ }\bibfield  {title} {\enquote {\bibinfo {title}
  {Particle-number-conserving {B}ogoliubov method which demonstrates the
  validity of the time-dependent {G}ross-{P}itaevskii equation for a highly
  condensed {B}ose gas},}\ }\href {\doibase 10.1103/PhysRevA.56.1414}
  {\bibfield  {journal} {\bibinfo  {journal} {Physical Review A}\ }\textbf
  {\bibinfo {volume} {56}},\ \bibinfo {pages} {1414} (\bibinfo {year}
  {1997})}\BibitemShut {NoStop}%
\bibitem [{\citenamefont {Castin}\ and\ \citenamefont
  {Dum}(1998)}]{castin_low-temperature_1998}%
  \BibitemOpen
  \bibfield  {author} {\bibinfo {author} {\bibfnamefont {Y.}~\bibnamefont
  {Castin}}\ and\ \bibinfo {author} {\bibfnamefont {R.}~\bibnamefont {Dum}},\
  }\bibfield  {title} {\enquote {\bibinfo {title} {Low-temperature
  {B}ose-{E}instein condensates in time-dependent traps: Beyond the
  \mbox{$U$}(1) symmetry-breaking approach},}\ }\href {\doibase
  10.1103/PhysRevA.57.3008} {\bibfield  {journal} {\bibinfo  {journal}
  {Physical Review A}\ }\textbf {\bibinfo {volume} {57}},\ \bibinfo {pages}
  {3008} (\bibinfo {year} {1998})}\BibitemShut {NoStop}%
\bibitem [{\citenamefont {Bloch}\ and\ \citenamefont
  {Messiah}(1962)}]{bloch_canonical_1962}%
  \BibitemOpen
  \bibfield  {author} {\bibinfo {author} {\bibfnamefont {C.}~\bibnamefont
  {Bloch}}\ and\ \bibinfo {author} {\bibfnamefont {A.}~\bibnamefont
  {Messiah}},\ }\bibfield  {title} {\enquote {\bibinfo {title} {The canonical
  form of an antisymmetric tensor and its application to the theory of
  superconductivity},}\ }\href {\doibase 10.1016/0029-5582(62)90377-2}
  {\bibfield  {journal} {\bibinfo  {journal} {Nuclear Physics}\ }\textbf
  {\bibinfo {volume} {39}},\ \bibinfo {pages} {95} (\bibinfo {year}
  {1962})}\BibitemShut {NoStop}%
\bibitem [{\citenamefont {Perelomov}(1977)}]{perelomov_generalized_1977}%
  \BibitemOpen
  \bibfield  {author} {\bibinfo {author} {\bibfnamefont {A.~M.}\ \bibnamefont
  {Perelomov}},\ }\bibfield  {title} {\enquote {\bibinfo {title} {Generalized
  coherent states and some of their applications},}\ }\href {\doibase
  10.1070/PU1977v020n09ABEH005459} {\bibfield  {journal} {\bibinfo  {journal}
  {Soviet Physics Uspekhi}\ }\textbf {\bibinfo {volume} {20}},\ \bibinfo
  {pages} {703} (\bibinfo {year} {1977})}\BibitemShut {NoStop}%
\bibitem [{\citenamefont {Hollenhorst}(1979)}]{hollenhorst_quantum_1979}%
  \BibitemOpen
  \bibfield  {author} {\bibinfo {author} {\bibfnamefont {J.~N.}\ \bibnamefont
  {Hollenhorst}},\ }\bibfield  {title} {\enquote {\bibinfo {title} {Quantum
  limits on resonant-mass gravitational-radiation detectors},}\ }\href
  {\doibase 10.1103/PhysRevD.19.1669} {\bibfield  {journal} {\bibinfo
  {journal} {Physical Review D}\ }\textbf {\bibinfo {volume} {19}},\ \bibinfo
  {pages} {1669} (\bibinfo {year} {1979})}\BibitemShut {NoStop}%
\bibitem [{\citenamefont {Schumaker}(1986)}]{Schumaker1986a}%
  \BibitemOpen
  \bibfield  {author} {\bibinfo {author} {\bibfnamefont {B.~L.}\ \bibnamefont
  {Schumaker}},\ }\bibfield  {title} {\enquote {\bibinfo {title} {Quantum
  mechanical pure states with {Gaussian} wave functions},}\ }\href@noop {}
  {\bibfield  {journal} {\bibinfo  {journal} {Physics Reports}\ }\textbf
  {\bibinfo {volume} {135}},\ \bibinfo {pages} {317} (\bibinfo {year}
  {1986})}\BibitemShut {NoStop}%
\bibitem [{\citenamefont {Jiang}\ and\ \citenamefont {Caves}(2016)}]{Nconserv}%
  \BibitemOpen
  \bibfield  {author} {\bibinfo {author} {\bibfnamefont {Z.}~\bibnamefont
  {Jiang}}\ and\ \bibinfo {author} {\bibfnamefont {C.~M.}\ \bibnamefont
  {Caves}},\ }\bibfield  {title} {\enquote {\bibinfo {title}
  {Particle-number-conserving {Bogoliubov} approximation for {Bose-Einstin}
  condensates using extended catalytic states},}\ }\href {\doibase
  10.1103/PhysRevA.93.033623} {\bibfield  {journal} {\bibinfo  {journal}
  {Physical Review~A}\ }\textbf {\bibinfo {volume} {93}},\ \bibinfo {pages}
  {033623} (\bibinfo {year} {2016})}\BibitemShut {NoStop}%
\bibitem [{\citenamefont {Koashi}\ \emph {et~al.}(2000)\citenamefont {Koashi},
  \citenamefont {Bu\v{z}ek},\ and\ \citenamefont
  {Imoto}}]{koashi_entangled_2000}%
  \BibitemOpen
  \bibfield  {author} {\bibinfo {author} {\bibfnamefont {M.}~\bibnamefont
  {Koashi}}, \bibinfo {author} {\bibfnamefont {V.}~\bibnamefont {Bu\v{z}ek}}, \
  and\ \bibinfo {author} {\bibfnamefont {N.}~\bibnamefont {Imoto}},\ }\bibfield
   {title} {\enquote {\bibinfo {title} {Entangled webs: Tight bound for
  symmetric sharing of entanglement},}\ }\href {\doibase
  10.1103/PhysRevA.62.050302} {\bibfield  {journal} {\bibinfo  {journal}
  {Physical Review A}\ }\textbf {\bibinfo {volume} {62}},\ \bibinfo {pages}
  {050302} (\bibinfo {year} {2000})}\BibitemShut {NoStop}%
\bibitem [{\citenamefont {Fisher}\ \emph {et~al.}(1989)\citenamefont {Fisher},
  \citenamefont {Weichman}, \citenamefont {Grinstein},\ and\ \citenamefont
  {Fisher}}]{fisher_89}%
  \BibitemOpen
  \bibfield  {author} {\bibinfo {author} {\bibfnamefont {M.~P.~A.}\
  \bibnamefont {Fisher}}, \bibinfo {author} {\bibfnamefont {P.~B.}\
  \bibnamefont {Weichman}}, \bibinfo {author} {\bibfnamefont {G.}~\bibnamefont
  {Grinstein}}, \ and\ \bibinfo {author} {\bibfnamefont {D.~S.}\ \bibnamefont
  {Fisher}},\ }\bibfield  {title} {\enquote {\bibinfo {title} {Boson
  localization and the superfluid-insulator transition},}\ }\href
  {http://journals.aps.org/prb/pdf/10.1103/PhysRevB.40.546} {\bibfield
  {journal} {\bibinfo  {journal} {Physical Review B}\ }\textbf {\bibinfo
  {volume} {40}},\ \bibinfo {pages} {546} (\bibinfo {year} {1989})}\BibitemShut
  {NoStop}%
\bibitem [{\citenamefont {Paraoanu}\ \emph {et~al.}(2001)\citenamefont
  {Paraoanu}, \citenamefont {Kohler}, \citenamefont {Sols},\ and\ \citenamefont
  {Leggett}}]{paraoanu_josephson_2001}%
  \BibitemOpen
  \bibfield  {author} {\bibinfo {author} {\bibfnamefont {G.-S.}\ \bibnamefont
  {Paraoanu}}, \bibinfo {author} {\bibfnamefont {S.}~\bibnamefont {Kohler}},
  \bibinfo {author} {\bibfnamefont {F.}~\bibnamefont {Sols}}, \ and\ \bibinfo
  {author} {\bibfnamefont {A.~J.}\ \bibnamefont {Leggett}},\ }\bibfield
  {title} {\enquote {\bibinfo {title} {The {Josephson} plasmon as a
  {Bogoliubov} quasiparticle},}\ }\href {\doibase 10.1088/0953-4075/34/23/313}
  {\bibfield  {journal} {\bibinfo  {journal} {Journal of Physics B: Atomic,
  Molecular and Optical Physics}\ }\textbf {\bibinfo {volume} {34}},\ \bibinfo
  {pages} {4689} (\bibinfo {year} {2001})}\BibitemShut {NoStop}%
\bibitem [{\citenamefont {Greiner}\ \emph
  {et~al.}(2002{\natexlab{b}})\citenamefont {Greiner}, \citenamefont {Mandel},
  \citenamefont {H\"ansch},\ and\ \citenamefont
  {Bloch}}]{greiner_collapse_2002}%
  \BibitemOpen
  \bibfield  {author} {\bibinfo {author} {\bibfnamefont {M.}~\bibnamefont
  {Greiner}}, \bibinfo {author} {\bibfnamefont {O.}~\bibnamefont {Mandel}},
  \bibinfo {author} {\bibfnamefont {T.~W.}\ \bibnamefont {H\"ansch}}, \ and\
  \bibinfo {author} {\bibfnamefont {I.}~\bibnamefont {Bloch}},\ }\bibfield
  {title} {\enquote {\bibinfo {title} {Collapse and revival of the matter wave
  field of a {Bose-Einstein} condensate},}\ }\href {\doibase
  10.1038/nature00968} {\bibfield  {journal} {\bibinfo  {journal} {Nature}\
  }\textbf {\bibinfo {volume} {419}},\ \bibinfo {pages} {51} (\bibinfo {year}
  {2002}{\natexlab{b}})}\BibitemShut {NoStop}%
\bibitem [{\citenamefont {Schachenmayer}\ \emph {et~al.}(2011)\citenamefont
  {Schachenmayer}, \citenamefont {Daley},\ and\ \citenamefont
  {Zoller}}]{schachenmayer_atomic_2011}%
  \BibitemOpen
  \bibfield  {author} {\bibinfo {author} {\bibfnamefont {J.}~\bibnamefont
  {Schachenmayer}}, \bibinfo {author} {\bibfnamefont {A.~J.}\ \bibnamefont
  {Daley}}, \ and\ \bibinfo {author} {\bibfnamefont {P.}~\bibnamefont
  {Zoller}},\ }\bibfield  {title} {\enquote {\bibinfo {title} {Atomic
  matter-wave revivals with definite atom number in an optical lattice},}\
  }\href {\doibase 10.1103/PhysRevA.83.043614} {\bibfield  {journal} {\bibinfo
  {journal} {Physical Review A}\ }\textbf {\bibinfo {volume} {83}},\ \bibinfo
  {pages} {043614} (\bibinfo {year} {2011})}\BibitemShut {NoStop}%
\end{thebibliography}


\appendix

\section{Bogoliubov Approximation to the Two-Site Bose-Hubbard Model} % (fold)
\label{sec:bogoliubov_approximation_to_the_two_site_bh_model}

We start by considering the two-site Bose-Hubbard Hamiltonian in the Schmidt basis, which is given by Eq.~(\ref{eq:tbh_schmidt_final}).  In the absence of interactions ($U=0$), the ground state of $\sH_\mathrm{tbh}$ consists of a condensate, with all particles occupying the Schmidt mode $\a_1 = (\b_1+\b_2)/\sqrt{2}$; the ground-state energy is $-JN$.

To derive the Bogoliubov approximation to $\sH_\mathrm{tbh}$, we replace $\a_1$ and $\a_1^\dagger$ with $\sqrt N - \a_2^\dagger\a_2/2\sqrt N$ (so that $\a_1^\dagger\a_1$ is replaced by $N-\a_2^\dagger\a_2$) and keep terms up to quadratic order in $\a_2$ and $\a_2^\dagger$. This gives the Bogoliubov Hamiltonian, which in this case, is just a Hamiltonian for mode $\a_2$:
\begin{align}
\sH_\mathrm{bog} &=  -J (N+1)+\frac{N(N-2)U}{4}
+\Big(J +\frac{NU}{4}\Big)(\a_2^\dagger \a_2+ \a_2\a_2^\dagger)- \frac{NU}{4}\ssp \Big(\a_2\a_2 + \a_2^\dagger \a_2^\dagger\Big)\;.\label{eq:BH_hamil_2}
\end{align}
The Bogoliubov Hamiltonian~(\ref{eq:BH_hamil_2}) can be diagonalized by introducing the (squeezed) bosonic operators
\begin{align}
 \c &= \a_2 \cosh \gamma -\a_2^\dagger \sinh \gamma\;,\\
 \c^\dagger &= \a_2^\dagger \cosh \gamma -\a_2 \sinh \gamma\;,
\end{align}
from which we get
\begin{align}
 \c^\dagger \c +\c\ssp\c^\dagger
 &= (\a_2^\dagger\a_2 +\a_2\a_2^\dagger)\cosh2\gamma
 - (\a_2\a_2+\a_2^\dagger\a_2^\dagger)\sinh2\gamma\;.
\end{align}
With the choice
\begin{align}
 \tanh2\gamma = \frac{NU/4}{J +NU/4}\;,
\end{align}
Eq.~(\ref{eq:BH_hamil_2}) takes the form
\begin{align}
\sH_\mathrm{bog} &=  -J (N+1)+\frac{N(N-2)U}{4}
+\sqrt{J^2+\frac{NJU}{2}}\,\big(\c^\dagger \c +\c\ssp\c^\dagger\big)\;.
\end{align}
Thus the Bogoliubov ground state has energy
\begin{align}\label{eq:E_bog}
E_\mathrm{bog} &= -J(N+1)+\frac{N(N-2)U}{4}+\sqrt{J^2+\frac{NJU}{2}}
\end{align}
and a population in mode $\a_2$ equal to
\begin{align}\label{eq:depletion_bog}
 \langle \a_2^\dagger\a_2 \rangle = \sinh^2\!\gamma
 = \half\!\left(\frac{J+NU/4}{\sqrt{J^2+NJU/2}}-1\right)\;.
\end{align}
The population imbalance of the Schmidt modes is given by
\begin{align}\label{eq:Delta_bog}
\Delta_\mathrm{bog} = N - 2\langle \a_2^\dagger\a_2 \rangle
=N+1-\frac{J+NU/4}{\sqrt{J^2+NJU/2}}\;.
\end{align}

The Bogoliubov approximation provides a very good account in the regime of very weak interactions, $NU/J\ll1$, where the interaction energy is much smaller than the tunneling energy, but in this regime, the Bogoliubov corrections are really too tiny to worry about.  The approximation should remain valid over a larger range of interaction strengths. Indeed, it should give a reasonably good account as long as the depletion is small, i.e., $\langle\a_2^\dagger\a_2\rangle\ll N$; notice that this region of validity includes regions where $NU/J\gg1$.  Taking the Bogoliubov approximation at face value for all values of interaction strength and assuming $N\gg1$ and $NU/J\gg1$, we can put the Bogoliubov results~(\ref{eq:E_bog})--(\ref{eq:Delta_bog}) in the simpler, approximate forms
\begin{align}\label{eq:E_bog_2}
E_\mathrm{bog} &\simeq -JN\!\left(1-\sqrt{\frac{U}{2NJ}}\right)+\frac{N(N-2)U}{4}\;,\\
\frac{\langle \a_2^\dagger\a_2\rangle}{N}&\simeq \frac12\sqrt{\frac{U}{8NJ}}\;,\\
\Delta_\mathrm{bog}&\simeq N\!\left(1-\sqrt{\frac{U}{8NJ}}\right)\;.
\label{eq:Delta_bog_2}
\end{align}

As noted above, the Bogoliubov approximation should be valid as long as $\langle\a_2^\dagger\a_2\rangle/N\ll1$, i.e.,
$\sqrt{U/(NJ)}\ll1$, which means that we should be able to rely on the expressions~(\ref{eq:E_bog_2})--(\ref{eq:Delta_bog_2}) as long as $NU/J\gg1$ and $\sqrt{U/(NJ)}\ll1$.  This is the intermediate regime of weak interaction strength delimited by $1/N\ll U/J\ll N$.  Notice now that Eqs.~(\ref{eq:E_bog_2})--(\ref{eq:Delta_bog_2}) are identical to the APCS predictions contained in Eqs.~(\ref{eq:Deltaapcs_large_s}) and~(\ref{eq:Eapcs_large_s}), which apply in this same intermediate regime of weak interaction strength.  [In Figs.~\ref{fig:rho1}(c) and~\ref{fig:gs_energy}, the differences between APCS and the Bogoliubov approximation in this intermediate regime are accounted for by the fact that the differences are as small as the terms neglected in getting to Eqs.~(\ref{eq:Deltaapcs_large_s}) and~(\ref{eq:Eapcs_large_s}) and Eqs.~(\ref{eq:E_bog_2})--(\ref{eq:Delta_bog_2}).]

The crucial distinction between APCS and the Bogoliubov approximation is that APCS, by including post-Bogoliubov corrections, successfully navigates the transition from a superfluid to a Mott insulator as the interaction strength increases through $U/(NJ)\sim1$, whereas the Bogoliubov approximation continues to make the predictions~~(\ref{eq:E_bog_2})--(\ref{eq:Delta_bog_2}) as $U$ increases, failing to notice the transition.  The Bogoliubov approximation runs completely off the rails when the interaction strength increases to the point that  $\langle\a_2^\dagger\a_2\rangle=N$, i.e., $U/(NJ)=32$, and cannot proceed to stronger interactions (the Bogoliubov expressions yield values for stronger interaction strengths, but they are unphysical).  All these conclusions are satisfyingly consistent with our contention that the intermediate regime of weak interaction strength is where Bogoliubov corrections become important and to get beyond this regime and through the transition to a Mott insulator requires including post-Bogoliubov corrections, which the \PCS\ formalism can accommodate.

\section{Exact PCS 2RDM for Schmidt Rank Two} % (fold)
\label{sec:pcs_2rdm_for_schmidt_rank_two}

For Schmidt rank $\rank=2$, we have in Eq.~(\ref{NFB_hypergeo}) the exact PCS normalization factor $\NFB$ in terms of a Gauss hypergeometric function.  From this, we get, using Eq.~(\ref{eq:diagonal_elements_rdm_single_particle}),
\begin{align}\label{eq:occu_11}
\rho_{11}^{(1)}
&=\frac{\lambda_{1}}{\NFB}\:\frac{\partial \NFB}{\partial\lambda_{1}}
=2n+1-\sF\;,\\
\rho_{22}^{(1)}
&=\frac{\lambda_{2}}{\NFB}\:\frac{\partial \NFB}{\partial\lambda_{2}}
=-1+\sF\;,
\label{eq:occu_22}\\
\Delta_\mathrm{pcs}&=
\rho_{11}^{(1)}-\rho_{22}^{(1)}
=2n+2-2\sF
=\frac{2\beta_\mathrm{pcs}}{2n-1}\;,
\label{eq:occu_diff_s_F}
\end{align}
where
\begin{align}
\sF=\Fthreeoverone\;,\quad z\equiv\frac{\lambda_2^2}{\lambda_1^2}\;,
\end{align}
denotes a ratio of contiguous Gauss hypergeometric functions.  To get this result, we use, when taking the derivatives, the property
\begin{align}
z\frac{d\,\Fone/dz}{\Fone}=\frac12(\sF-1)\;,
\end{align}
which comes from the general identity,
\begin{align}
z\frac{dF(a,b;c;z)}{dz}=a\big[\,F(a+1,b;c;z)-F(a,b;c;z)\,\big]\;.
\end{align}

The population difference~(\ref{eq:occu_diff_s_F}) is sufficient to calculate the PCS ground-state energy.  In particular, from Eq.~(\ref{eq:difference_rho2}), we have
\begin{align}\label{eq:deltaPCSexact}
\delta_\mathrm{pcs}=\rho_{11,22}^{(2)}-\rho_{12,12}^{(2)}=
n-\frac12\frac{\lambda_1-\lambda_2}{\lambda_1+\lambda_2}\Delta_\mathrm{pcs}\;,
\end{align}
and plugging this into the expression~(\ref{E0_BH}) for the ground-state energy, we get
\begin{align}\label{eq:EPCSexact}
E_\mathrm{pcs}
=-J\Delta_\mathrm{pcs}+\frac{U}{2}\!\left(\frac{N(N-1)}{2}-\delta_\mathrm{pcs}\right)
=\left(-J+\frac{U}{4}\frac{\lambda_1-\lambda_2}{\lambda_1+\lambda_2}\right)\Delta_\mathrm{pcs}+\frac{N(N-2)U}{4}\;.
\end{align}
Equations~(\ref{eq:deltaPCSexact}) and~(\ref{eq:EPCSexact}) are the exact PCS analogues of the large-$N$, approximate expressions~(\ref{eq:rho2_terms_pcs_energy}) and ~(\ref{eq:pcs_energy_c}).  We find the exact PCS ground-state energy by minimizing $E_\mathrm{pcs}$ with respect to $z=\lambda_2^2/\lambda_1^2$; this minimum value of $z$ is then inserted in the expressions in this appendix to find the PCS ground-state 1RDM and 2RDM.

One can develop power-series expansions for the population difference~(\ref{eq:occu_diff_s_F}) and the ground-state energy~(\ref{eq:EPCSexact}) in the single-condensate, mean-field regime, $\lambda_2^2/\lambda_1^2\ll1$, and in the Mott regime, $|1-\lambda_2^2/\lambda_1^2|\ll1$; these are, however, of limited value since they only provide small corrections to the mean-field and Mott limiting behaviors, corrections valid only in the extreme mean-field regime and the extreme Mott regime.

The normalization factor~(\ref{NFB_hypergeo}) also allows us to calculate the nonzero matrix elements of the PCS 2RDM, $\rho^{(2)}$.  Using Eq.~(\ref{eq:diagonal_elements_rdm_a}), we find the following diagonal elements:
\begin{align}\label{eq:F1212}
\rho_{12,12}^{(2)}
&=\frac{\lambda_1\lambda_2}{\NFB}\frac{\partial^2\NFB}{\partial\lambda_1\,\partial\lambda_2}
=-n+\frac12\frac{1+z}{1-z}\Delta_\mathrm{pcs}=\gamma_\mathrm{pcs}\;,\\
\label{eq:F1111}
\rho_{11,11}^{(2)}
&=\frac{\lambda_1^2}{\NFB}\frac{\partial^2\NFB}{\partial\lambda_1^2}
=2n^2+\frac12\bigg(2n-1-\frac{1+z}{1-z}\bigg)\Delta_\mathrm{pcs}
=(2n-1)\rho_{11}^{(1)}-\rho_{12,12}^{(2)}
=n(2n-1)+\beta_\mathrm{pcs}-\gamma_\mathrm{pcs}\;,\\
\label{eq:F2222}
\rho_{22,22}^{(2)}
&=\frac{\lambda_2^2}{\NFB}\frac{\partial^2\NFB}{\partial\lambda_2^2}
=2n^2-\frac12\bigg(2n-1+\frac{1+z}{1-z}\bigg)\Delta_\mathrm{pcs}
=(2n-1)\rho_{22}^{(1)}-\rho_{12,12}^{(2)}
=n(2n-1)-\beta_\mathrm{pcs}-\gamma_\mathrm{pcs}\;.
\end{align}
When taking the derivatives, we use the property
\begin{align}
z^2\frac{d^2\Fone/dz^2}{\Fone}
=\frac14\!\left(\frac{-2n+1-3z}{1-z}+\frac{2n-1-(2n-3)z}{1-z}\sF\right)\;,
\end{align}
which comes from applying a relation between contiguous hypergeometric functions to the general identity,
\begin{align}
z^2\frac{d^2 F(a,b;c;z)}{dz^2}
=a(a+1)\big[\,F(a+2,b;c;z)-2F(a+1,b;c;z)+F(a,b;c;z)\,\big]\;.
\end{align}
There is, however, an easier way to find these 2RDM matrix elements than taking derivatives of the normalization factor.  The second-to-last formulas on the right in Eqs.~(\ref{eq:F1212})--(\ref{eq:F2222}) express the procedure developed in Sec.~\ref{sec:2rdms} for determining the diagonal elements of the 2RDM from the 1RDM.  In particular, these are Eqs.~(\ref{eq:rho2jkjkexact}) and~(\ref{eq:rho2jjjjexact}), and they can be used directly to find these 2RDM diagonal elements from the 1RDM.  The final formula on the right relates the 2RDM matrix elements to the parameters in the Eq.~(\ref{eq:exact_2rdm_Sch}).

Because $\rho_{12,12}^{(2)} = \rho_{21,21}^{(2)} = \rho_{12,21}^{(2)} = \rho_{21,12}^{(2)}$, the only nontrivial off-diagonal matrix elements remaining are $\rho^{(2)}_{11,22}=\rho^{(2)}_{22,11}$, which can be determined from Eq.~(\ref{eq:rho2kkjjexact}) to be
\begin{align}\label{eq:F1122}
\rho^{(2)}_{11,22}
=\frac{\sqrt z}{1-z}\Delta_\mathrm{pcs}
=\gamma_\mathrm{pcs}+\delta_\mathrm{pcs}\;.
\end{align}

Notice that in accordance with our general conclusions, the entire 2RDM in the Schmidt basis is determined by the 1RDM and, in particular, in this case of $\rank=2$, by the population imbalance $\Delta_\mathrm{pcs}$ or, equivalently, by the hypergeometric ratio~$\sF$.

\section{Iterative Relations}
\label{app:Iterative_Relations}

In App.~\ref{sec:pcs_2rdm_for_schmidt_rank_two}, we demonstrated that the exact 2RDM of the \PCS~\emph{Ansatz\/} for $\rank = 2$ can be expressed using hypergeometric functions.  It is very challenging, if not impossible, to generalize this analytical result to $\rank > 2$.  In this appendix, we derive an exact relation between the \PCS\ 1RDMs of $2n$ and $2n+2$ particles using Wick's theorem.  This leads to an iterative algorithm that gives the exact 1RDM for $\rank \geq 2$ in time $O(N)$.  Using Eqs.~(\ref{eq:rho2jkjkexact})--(\ref{eq:rho2kkjjexact}), one can calculate the 2RDM in time $O(1)$ from the 1RDM.  In the large-$N$ limit, this iterative algorithm for the 1RDMs can be turned into a differential equation; a solution based on aseries expansion is given here.  This procedure reduces the time complexity of the iterative algorithm for the 1RDMs to $O(1)$, at a price of introducing relative errors of order $O(1/N)$.

% The purposes of this Appendix are the following: (i)~to find an effective way of approximating the 1RDMs; (ii)~to gain physical intuition about the 1RDMs; and (iii)~to verify and extend results already obtained.
In the Schmidt basis, we have
\begin{align}\label{eq:iteration_1rdm}
\begin{split}
 \rho_{kj}^{(1)}(n+1)
 &=\frac{1}{\NFB_{n+1}}\, \brab{\vac}\sA^{n+1} \, a_j^\dagger a_k\,\big(\sA^\dagger\big)^{n+1}  \ketb{\vac}\\[3pt]
 &=\frac{4 (n+1)^2}{\NFB_{n+1}}\, \lambda_j\lambda_k\, \brab{\vac}\sA^{n}\, a_j a_k^\dagger\, \big(\sA^\dagger\big)^{n}\ketb{\vac}\\[3pt]
 &=\frac{4 (n+1)^2}{\NFB_{n+1}}\, \lambda_j\lambda_k \, \brab{\vac}\sA^{n}\, \big(a_k^\dagger a_j+\delta_{jk}\big) \,\big(\sA^\dagger\big)^{n}\ketb{\vac}\\[3pt]
 &=\frac{4 (n+1)^2\ssp \NFB_{n}}{\NFB_{n+1}}\,\Bigl( \lambda_j\lambda_k\, \rho_{jk}^{(1)}(n)+  \lambda_j^2\, \delta_{jk} \Bigr)\;.
\end{split}
\end{align}
This relation implies that $\rho^{(1)}(n+1)$ is diagonalized in the Schmidt basis provided that $\rho^{(1)}(n)$ is diagonalized.  Since $\rho^{(1)}(1)$ is diagonalized, mathematical induction allows us to conclude that $\rho^{(1)}(n)$ is diagonalized.

That $\rho^{(1)}(n)$ is diagonalized in the Schmidt basis can also be seen directly from the first line of Eq.~(\ref{eq:iteration_1rdm}): $\big(\sA^\dagger\big)^n\ketb{\vac}$ is a superposition of Fock states that have an even number of particles in each of the Schmidt single-particle states, so in the Fock-state superposition for $a_k\,\big(\sA^\dagger\big)^n\ketb{\vac}$, the single-particle state~$k$ always has an odd number of particles; thus $a_k\,\big(\sA^\dagger\big)^n\ketb{\vac}$ is orthogonal to $a_j\,\big(\sA^\dagger\big)^n\ketb{\vac}$ unless $j=k$.

For the diagonal elements $\rho^{(1)}_{jj}(n)\equiv\varrho_j(n)$, we have
\begin{align}\label{eq:iteration_diagonal_1rdm}
 \varrho_j(n+1)=\frac{4 (n+1)^2\ssp \NFB_{n}}{\NFB_{n+1}}\,\lambda_j^2\,\big[\varrho_j(n)+  1\big]\;.
\end{align}
Using the normalization condition, $\sum_j \varrho_j(n) = 2n$, we can write the ratio of the normalization factors, $\NFB_{n}/\NFB_{n+1}$, as
\begin{align}
\frac{\NFB_{n}}{\NFB_{n+1}}=\frac{1}{2(n+1)}\bigg(\sum_{k=1}^\rank\lambda_k^2\big[\varrho_k(n)+1\big]\bigg)^{-1}\;.
\end{align}
Plugging this into Eq.~(\ref{eq:iteration_diagonal_1rdm}), we get an iterative relation in which the normalization factors do not appear.
% \begin{align}
% \varrho_j(n)=
% \lambda_j^2\big[\varrho_j(n-1)+ 1\big]\,2n\bigg(\sum_{k=1}^\rank\lambda_k^2\big[\varrho_k(n-1)+1\big]\bigg)^{-1}\;.
% \end{align}
We can also defer normalization to the end of the process, instead of imposing it at each iteration.  The new iterative equation then reads
\begin{align}\label{eq:iteration_1rdm_defer}
 \varrho_j(n+1)=\lambda_j^2\,\bigg(\varrho_j(n)+ \frac{1}{2 n}\sum_{k=1}^\rank \varrho_k(n)\bigg)\;,
\end{align}
where we replace the 1 with $(1/2n)\,\sum_k\varrho_k(n)$ to deal with the fact that $\varrho_j(n)$ is not normalized. The number of steps required in each iteration is proportional to $\rank$ and is independent of $n$, so the time complexity of calculating $\rho^{(1)}(n)$ using Eq.~\eqref{eq:iteration_diagonal_1rdm} or \eqref{eq:iteration_1rdm_defer} is $O (\rank n)$. Because the \PCS\ 2RDM can be calculated from the 1RDM (up to relative phase factors of the orbitals), searching for the lowest energy state in the \PCS\ submanifold is exponentially faster than a brute-force approach considering the entire Hilbert space of $2n$ bosons.

We now let $n'$ denote the iteration variable, and we suppose that we wish to iterate from $n'=1$ to $n'=n$.
For sufficiently large $n$, all the probability concentrates on the dominant eigenvalues.  To get useful results, we assume, as in Sec.~\ref{sec:large_N}, that the differences of the $\lambda_j$s are small, and we use the parametrization~(\ref{eq:lambda_zeta}) for the endpoint of the iteration, i.e., $\lambda_j^2=1+\pcss_j/n$.  To turn the iterative equation into a continuous differential equation, we introduce the parameter $\tau=n'/n$.  The new iterative equation then reads as follows:
\begin{align}\label{eq:iteration_1rdm_large_n1}
 \varrho_j(n\tau+1)
 = \Big( 1+\frac{\pcss_j}{n}\,\Big)\bigg(\varrho_j(n\tau)+\frac{1}{2 n\tau}\sum_{k=1}^\rank \varrho_k(n\tau)\bigg)\;.
\end{align}
% where we replace the 1 with $(1/2n\tau)\,\sum_k\varrho_k$ to deal with the fact that $\varrho_j$ is not normalized.
We now can write
\begin{align}
\frac{\varrho_j(n\tau+1)-\varrho_j(n\tau)}{1/n}=
s_j\varrho_j(n\tau)+\Big(1+\frac{\pcss_j}{n}\Big)\frac{1}{2 \tau }\sum_{k=1}^\rank \varrho_k(n\tau)\;.
\end{align}
Taking the limit $n\rightarrow \infty$ yields the differential equation,
\begin{align}
 \frac{\dif \bar \varrho_j(\tau)}{\dif \tau}=\pcss_j\ssp \bar\varrho_j(\tau)+\frac{1}{2 \tau}\,  \sum_{k=1}^\rank \bar\varrho_k(\tau)\;,\label{eq:iteration_differential_a}
\end{align}
where $\bar\varrho_j(\tau)=\varrho_j(n\tau)$.  The problem with Eq.~(\ref{eq:iteration_differential_a}) is that it diverges at small $\tau$ due to the factor $1/2\tau$ unless $\bar\varrho_j(0)=0$ for all $j=1,2,\ldots,\rank$.  This divergence is a consequence of our decision to defer the normalization to the end; one remedy is to modify the differential equation to
\begin{align}\label{eq:iteration_differential_b}
 \frac{\dif \bar\varrho_j}{\dif \tau}=\Bigl(\pcss_j-\frac{\rank}{2 \tau }\Bigr)\,  \bar\varrho_j+\frac{1}{2 \tau}\,  \sum_{k=1}^\rank \bar\varrho_k\;,
\end{align}
where the extra term, which only introduces an overall factor, keeps $\bar\varrho_j$ from diverging for the initial condition $\bar\varrho_j(0)=1$, for $j = 1,2,\ldots,\rank$.  By mapping the iterative relation Eq.~\eqref{eq:iteration_1rdm_defer} to the differential equation~\eqref{eq:iteration_differential_b}, the \PCS\ 1RDM can be solved approximately within time independent of $N$.

For the case $\rank=2$, we assume $s_2 = -s_1$ without loss of generality. By introducing new variables, $\bar\varrho_\pm = (\bar\varrho_1 \pm \bar\varrho_2)/2$, we can write Eq.~\eqref{eq:iteration_differential_b} as
\begin{align}
 &\frac{\dif \bar\varrho_+}{\dif \tau} = s_- \bar\varrho_-\;,\label{eq:differ_eq_1}\\[2pt]
 &\frac{\dif \bar\varrho_-}{\dif \tau} = s_-\bar\varrho_+ - \frac{\bar\varrho_-}{\tau}\label{eq:differ_eq_2}\;,
\end{align}
where $s_- = (s_1-s_2)/2 =s_1$. Taking derivatives with respect to $\tau$ on both sides of Eqs.~\eqref{eq:differ_eq_1} and \eqref{eq:differ_eq_2} and manipulating the results, we have the decoupled second-order differential equations for $\bar\varrho_\pm$,
\begin{align}
&\frac{\dif^2 \bar\varrho_+}{\dif \tau^2} = s_-^2\bar\varrho_+ - \frac{1}{\tau}\frac{\dif \bar\varrho_+}{\dif \tau}\;,\\[2pt]
&\frac{\dif^2 \bar\varrho_-}{\dif \tau^2} = s_-^2 \bar\varrho_- + \frac{\bar\varrho_-}{\tau^2} - \frac{1}{\tau}\frac{\dif \bar\varrho_-}{\dif \tau}
\;.
\end{align}
These differential equations are solved by the zeroth- and first-order modified Bessel functions $\BesselI_0(\tau s_-)$ and $\BesselI_1(\tau s_-)$, respectively. Therefore, we recover our former results~(\ref{eq:upsilon_s=2_1rdm_a}) and~(\ref{eq:upsilon_s=2_1rdm_b}) using an entirely different approach.

For the general case $\rank\geq 2$, the solution to the linear differential~(\ref{eq:iteration_differential_b}) can be expressed using a Taylor series in~$\tau$.  This can be done most conveniently by introducing the transition matrix
$T(\tau)$, such that
\begin{align}
 \bar\varrho_j(\tau)&=\sum_{k=1}^\rank T_{jk}(\tau)\, \bar\varrho_k(0)\;,
 %=\sum_{k=1}^\rank T_{jk}(\tau)\;,
\end{align}
where $\bar\varrho_j(\tau)$, the diagonal elements of the 1RDM, form a vector.  Using the initial condition $\bar\varrho_j(0)=1$, for $j = 1,2,\ldots,\rank$, we have
\begin{align}
 \bar\varrho_j(\tau) =\sum_{k=1}^\rank T_{jk}(\tau)\;.
\end{align}
It is convenient to choose the initial value of the transition matrix as $T(0)=M/\rank$ [any matrix that stabilizes the vector $(1,\,1,\ldots,1)^T$ suffices], where $M_{jk}=1$ (for $j,k = 1,2,\ldots,\rank$) is the matrix of ones. The equation that governs the evolution of $T(\tau)$ can be derived from Eq.~(\ref{eq:iteration_differential_b}),
\begin{align}\label{eq:matrix_iteration_differential}
 \frac{\dif\ssp T(\tau)}{\dif \tau}=\Bigl(S-\frac{1}{2\tau}\, \bigl(\rank \identity-M\bigr) \Bigr)\, T(\tau)\;,
\end{align}
where $S=\diag(\pcss_1,\pcss_2,\ldots, \pcss_\rank)$, and $\identity$ is the identity matrix.  Suppose the transition matrix has a Taylor expansion,
\begin{align}\label{eq:taylor_transition_matrix}
T(\tau)=M/\rank+\sum_{\ell=1}^\infty \tau^\ell\, T_\ell\;,
\end{align}
where the matrices $T_\ell$, for $\ell=1,2,\ldots$, are to be determined. Putting Eq.~(\ref{eq:taylor_transition_matrix}) into Eq.~(\ref{eq:matrix_iteration_differential}), we get
\begin{align}
\sum_{\ell=1}^\infty \ell\, \tau^{\ell-1}\, T_\ell= \Bigl(S-\frac{1}{2\tau}\, \bigl(\rank\identity-M\bigr) \Bigr) M/\rank+\sum_{\ell=1}^\infty \, \Bigl(\tau^\ell S-\frac{\tau^{\ell-1}}{2}\, \bigl(\rank\identity-M\bigr) \Bigr)\, T_\ell\;.%-\frac{\rank}{2}\,  \tau^{m-1} \big(\identity-S\big)\, T_m\;.
\label{eq:taylor_comparison}
\end{align}
The term of order $\tau^{-1}$ on the right hand side of Eq.~(\ref{eq:taylor_comparison}) disappears with the choice $T(0)=M/\rank$. Comparing the coefficients of the terms $\tau^{\ell-1}$ on both sides of Eq.~(\ref{eq:taylor_comparison}), we have
\begin{align}\label{eq:taylor_solving_a}
 \Big( \ell\ssp \identity+\frac{1}{2}\, \big(\rank\identity-M\big)\Big)\, T_\ell =S\ssp T_{\ell-1}\,,\;\;\mbox{for $\ell\geq 1$}\;.
\end{align}
By inverting the matrix $\ell\ssp \identity+\big(\rank\identity-M\big)/2$, we have
\begin{align}\label{eq:taylor_solving_b}
 T_\ell=\frac{\identity+M/(2\ell)}{\ell+\rank/2}\, S\ssp T_{\ell-1}\;.
\end{align}
The $\ell$th order matrix $T_\ell$ can be solved iteratively by applying the above relation $\ell$ times to $T_0 = M/\rank$. To simply this procedure, we notice that
\begin{align}
M D M=\tr(D)\ssp M
\end{align}
holds for any diagonal matrix $D$, including powers of $S$. Therefore, the solution to Eq.~(\ref{eq:taylor_solving_b}) takes the form $T_\ell = P_\ell(S)\ssp M/\rank$, where $P_\ell(S)$ is some polynomial of $S$ of order $\ell$, e.g., $P_1(S)=S/(1+\rank/2)$ and $P_2(S) = [S^2+\tr(S^2)/4]/[1+\rank/2)(2+\rank/2)]$. The matrix norm of $T_\ell$ begins to fall quickly after $\ell > \pcss_1$, which gives an estimate on how many terms are needed in the expansion to get a desired precision.

In this appendix, we discussed an iterative algorithm to calculate the 1RDMs of the \PCS\ \emph{Ansatz\/} exactly in time $O(N)$. For large $N$, the iterative steps can be approximately mapped to a differential equation. This differential equation gives the same result as those derived previously for $\rank = 2$. For $\rank \geq 2$, the solution to the differential equations can be expressed as a Taylor expansion. Compared to our other approaches, the iterative approach is particularly suitable for numerical evaluations of the \PCS\ \emph{Ansatz\/}.

\end{document}